\documentclass[
	reprint,
	groupedaddress,
	amsmath,amssymb,
	prb,
	floatfix,
	superscriptaddress,
	twocolumn,
	aps,
]{revtex4-2}

\usepackage[utf8]{inputenc} %
\usepackage[T1]{fontenc}    %
\usepackage{helvet}         %
\usepackage{mathpazo}       %
\usepackage{microtype}      %
\usepackage{amsmath}
\usepackage{bm}
\usepackage{amsfonts}
\usepackage{amssymb}
\usepackage{amsthm}
\usepackage[caption=false]{subfig}
\usepackage{graphicx}
\usepackage{hyperref}
\usepackage{cleveref}
\usepackage[T1]{fontenc}
\usepackage{multirow}
\usepackage{lipsum} %
\usepackage{color,soul}
\usepackage[export]{adjustbox}
\usepackage{natbib}

\usepackage{todonotes}

\usepackage{xr}
\begin{document}
\title{Machine learning interatomic potentials for solid-state precipitation}
\author{Lorenzo Piersante}
\affiliation{Laboratory of materials design and simulation (MADES), Institute of Materials, \'{E}cole Polytechnique F\'{e}d\'{e}rale de Lausanne}
\author{Anirudh Raju Natarajan}
\email{anirudh.natarajan@epfl.ch}
\affiliation{Laboratory of materials design and simulation (MADES), Institute of Materials, \'{E}cole Polytechnique F\'{e}d\'{e}rale de Lausanne}
\affiliation{National Centre for Computational Design and Discovery of Novel Materials (MARVEL), \'{E}cole Polytechnique F\'{e}d\'{e}rale de Lausanne}

\begin{abstract}
    Machine learning interatomic potentials (MLIPs) are routinely used to model diverse atomistic phenomena, yet parameterizing them to accurately capture solid-state phase transformations remains difficult. We present error metrics and data-generation schemes designed to streamline the parameterization of MLIPs for modeling precipitation in multi-component alloys. We developed an algorithm that enumerates symmetrically distinct transformation pathways connecting chemical decorations on different parent crystal structures. Additionally, we introduce the weighted Kendall-$\tau$ coefficient and its semi-grand canonical generalization as metrics for quantifying MLIP accuracy in predicting low-temperature thermodynamics. We apply these approaches to parameterize an MLIP for a dilute Mg-Nd alloy. The resulting potential reproduces the complex early-stage precipitation behavior observed in experiment. Large-scale atomistic simulations reveal competition between order-disorder and structural transformations. Furthermore, these results suggest a continuous transition between high-symmetry hcp and bcc crystal structures during aging heat treatments.
\end{abstract}
\maketitle

\section{Introduction}

Precipitation or age hardening is the primary strengthening mechanism in many commercial engineering alloys. Age hardening relies on the formation of secondary phases embedded within a disordered matrix phase. These precipitates increase material strength by impeding dislocation motion. To achieve peak strength, age-hardening heat treatments must be tuned to optimize precipitate size, shape, and number density. Insufficient aging times produce small precipitates that dislocations can easily overcome. Conversely, overaging leads to precipitate coarsening, which reduces their number density and decreases the strengthening effect. Optimizing the aging process requires quantitative knowledge of thermodynamic and kinetic parameters, such as free energies and diffusion constants, as well as an understanding of nucleation mechanisms and rates within the matrix phase. While several studies have focused on developing techniques and models to estimate the thermodynamic and kinetic parameters of materials \cite{starikov_atomic_2025, marchand_foundation_2025, srinivasan_atomic_2025}, less is known about the nucleation mechanisms of precipitates.

The atomistic nucleation mechanism of precipitates is important for modeling the early stages of precipitation. As the time and length scales required to study nucleation are inaccessible to purely \textit{ab-initio} methods, machine learning interatomic potentials (MLIPs), trained on density functional theory (DFT) data, have emerged as reliable and computationally efficient alternatives. While the capabilities of MLIPs have advanced substantially in recent years, designing datasets and validation methods for specific atomistic processes is still challenging \cite{liu_discrepancies_2023, liu_assessing_2024}.

Precipitation in multicomponent alloys often proceeds through a series of phase transitions. The initial stage is typically an order-disorder transition, where precipitates form by the chemical ordering of elements on the same parent crystal structure as the disordered matrix phase. Longer aging promotes the formation of more stable precipitates, which can have crystal structures that are different from the matrix. This precipitation sequence has been proposed to arise from the lower nucleation barriers for precipitates formed through order-disorder phase transitions as compared to those formed through structural transformations. However, rigorous estimation of the barriers and driving forces for the competing transformations has remained elusive.

Multi-stage precipitation is common in many engineering alloys, including those based on aluminum, titanium, nickel, and magnesium. Rare-earth containing magnesium alloys are a prototypical example, where experiments reveal a complex precipitation sequence involving ordering and structural transformations\cite{nie_precipitation_2012, xu_shear_2014, liu_structure_2017, xie_diffusional-displacive_2021,natarajan_early_2016,natarajan_2017_unifieddescription}. For instance, in dilute Mg-Nd alloys with $x_{\text{Nd}} \le 0.5$ at.\%, the decomposition of a supersaturated solid solution (SSSS) results in several intermediate phases during aging. The precipitation process begins with the formation of Guinier-Preston (GP) zones, followed by the $\beta^{\prime\prime\prime}$ phase, which consists of rods of Nd atoms arranged in a zigzag pattern along the hcp $[01\bar{1}0]$ direction interwoven with hexagonal chains of solute atoms. The $\beta'''$ precipitates are comprised of a family of ordered phases with $\beta'$ occurring at a composition of $x_{\text{Nd}}$ = 0.125 and $\beta''$ at a composition of 0.25. The next precipitate, $\beta_1$, requires a structural transformation from hcp to bcc. Finally, more complex structural transitions result in the formation of a metastable $\beta$ and equilibrium $\beta_{e}$ phases. Peak strength is achieved through a fine distribution of $\beta^{\prime\prime\prime}$ and $\beta_1$ precipitates embedded within the disordered matrix \cite{nie_precipitation_2012,natarajan_early_2016}.

\begin{figure}[!ht]
    \centering
    \includegraphics[width=\columnwidth]{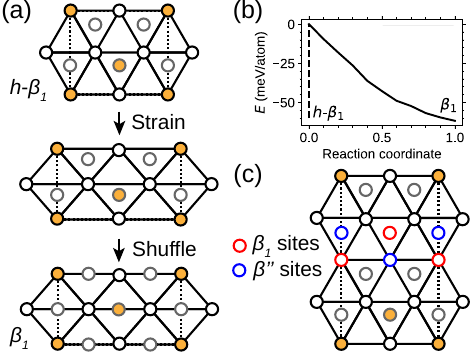}
    \caption{(a) Structural transformation from the ideal hcp decoration, $h$-$\beta_1$, to the bcc structure, $\beta_1$. (b) DFT energy landscape along the transformation path between $h$-$\beta_1$ and $\beta_1$. (c) Two possible sublattice decorations of a pre-existing $\beta'$ ordering; Nd substitution on the red sublattice yields $\beta_1$, while substitution on the blue sublattice creates local $\beta''$ ordering. Crystal structures are projected along the hcp [0001] direction. Orange and white circles represent Nd and Mg atoms, respectively. Atoms positioned between lattice sites reside in the layer above the basal plane.}
    \label{fig:burgers_transformation}
\end{figure}

The $\beta_1$ precipitate, one of the primary sources of strength in this alloy, forms through a structural phase transition from hcp to bcc. This hcp$\rightarrow$bcc transformation occurs through the Burgers mechanism which involves both a homogeneous deformation of the hcp lattice and a shuffle of atomic planes\cite{natarajan_2019_understandingdeformation}. As illustrated in \cref{fig:burgers_transformation}a, the basal plane of hcp is homogeneously strained to form the $\{110\}$ plane of bcc. Adjacent basal planes must then shuffle along the $\langle 01\bar{1}0 \rangle$ direction to attain the ideal bcc structure. The formation of the $\beta_1$ precipitate through the Burgers mechanism requires a specific arrangement of Mg and Nd atoms on the parent hcp structure, as shown in \cref{fig:burgers_transformation}c. DFT calculations in \cref{fig:burgers_transformation}b predict a barrierless transformation from the ordered hcp precursor ($h$-$\beta_1$) to the stable bcc precipitate ($\beta_1$). The existence of a barrierless pathway indicates the possibility of a continuous homogeneous nucleation mechanism for the formation of $\beta_1$.

A rigorous investigation of the nucleation mechanisms in this alloy requires an interatomic potential that accurately captures the hcp$\rightarrow$bcc transformation, thermodynamics, defect energetics and kinetic quantities such as vacancy migration barriers. An MLIP that faithfully models precipitation must predict the low-temperature phase stability. Models yielding zero-Kelvin ground states inconsistent with DFT or experiment, such as the modified embedded-atom method (MEAM) potential for Mg-Nd (\cref{supp-sup:meam_results} in the supplementary information\cite{supp_info}), have limited utility. To capture the multi-stage precipitation reactions, the MLIP must additionally describe both equilibrium and metastable thermodynamics. The potential also needs to reproduce salient features of structural transformations, such as the energy barriers that separate ordered phases on different parent lattices. Finally, estimating the effects of defects such as grain boundaries or dislocations requires training datasets with a large structural and chemical diversity.

This article describes our strategy for parameterizing an MLIP capable of reliably modeling precipitation in the Mg–Nd alloy. We begin in \cref{sect:MLIP_and_DFT} by introducing the MLIP formalism, summarizing the first-principles calculations, and describing the sampling methodology used to construct the training sets. The following two sections present systematic tools for dataset generation and validation beyond standard statistical errors, tailored to capture the phase transformations occurring during precipitation. Specifically, \cref{sect:dataset_construction} details our \emph{ab-initio} database and presents a crystal-symmetry based algorithm to systematically sample the potential energy surface of the hcp-to-bcc phase transition, while \cref{sect:validation} introduces the weighted Kendall–$\tau$ coefficient as an error metric for assessing phase stability in multi-component systems. We then demonstrate in \cref{sect:results} that these methodologies enabled parameterization of an MLIP that correctly captures the atomistic behavior of the Mg–Nd system. Finally, \cref{sect:discussion} applies the MLIP to model precipitation in dilute Mg–Nd alloys, revealing the atomistic nucleation mechanisms for both order-disorder and structural phase transformations.

\section{Interatomic potential and DFT calculations} \label{sect:MLIP_and_DFT}
The MLIP for the Mg-Nd alloy was parameterized with the atomic cluster expansion (ACE) framework \cite{drautz_atomic_2019, lysogorskiy_performant_2021, bochkarev_efficient_2022}. In the ACE formalism, the energy contribution of atom $i$ is expressed as:
\begin{equation}\label{eq:ace_energy}
    E_{i} =  \left ( \rho_{i}^{(1)} + \sqrt{\rho_{i}^{(2)} } \right )
\end{equation}
where $\rho_{i}^{(k)}$ is an atomic density expanded in a basis of cluster functions, $B$, centered on site $i$:
\begin{equation} \label{eq:ace_exp}
    \rho_{i}^{(k)} = \sum_{\alpha = 1}^{N_{max}} \sum_{ (\vec{\mu}_\alpha, \vec{n}_\alpha, \vec{l}_{\alpha})} c_{(\vec{\mu}_\alpha, \vec{n}_\alpha, \vec{l}_\alpha)}^{(k)} B_{i, (\vec{\mu}_\alpha, \vec{n}_\alpha, \vec{l}_\alpha)}
\end{equation}
The basis functions $B_{i, (\vec{\mu}_\alpha, \vec{n}_\alpha, \vec{l}_\alpha)}$ are rotationally invariant descriptors of the local chemical and geometrical environment surrounding atom $i$. These $B$-basis functions are constructed from radial basis functions and spherical harmonics. The multi-index $(\vec{\mu}_\alpha, \vec{n}_\alpha, \vec{l}_\alpha)$ specifies the chemical species ($\vec{\mu}_\alpha$), radial basis functions ($\vec{n}_\alpha$), and angular momenta ($\vec{l}_\alpha$) used to construct a descriptor for a cluster of $\alpha+1$ atoms. The expansion coefficients, $c^{(k)}$, are the model fitting parameters that are determined during training.

The accuracy and speed of an MLIP is determined by the number of descriptors or basis functions used in its construction. Consequently, hyperparameters such as the cutoff radius, the largest many-body interaction, and the number of basis functions must be carefully chosen. The Mg-Nd potential was constructed using 4246 cluster functions with an interaction distance of 7.8\AA. The basis set for pure Mg is identical to that used in \cite{ibrahim_atomic_2023}. A Ziegler-Biersack-Littmark (ZBL) potential was included to describe short-range ($r \lessapprox$ 1-2 \AA) repulsive interactions between atoms. \Cref{supp-sup:ace} provides a list of the basis function cutoffs and other model hyperparameters.

\begin{table}[!ht]
    \centering

    \begin{tabular}{|c | c | c | c | c | c |}
        \hline
        $n$/$l$ & Pair & Triplet & Quadruplet & Quintuplet & Total \\
        \hline
        Mg      & 15/0 & 5/4     & 4/3        & 3/2        & 694   \\
        Nd      & 15/0 & 8/8     & 4/4        & 1/1        & 890   \\
        Mg-Nd   & 15/0 & 8/8     & 3/2        & -          & 2662  \\ \hline
    \end{tabular}
    \caption{Basis function parameters. The table lists the maximum values for $n$ and $l$ allowed in the vectors $\vec{n}_{\alpha}$ and $\vec{l}_{\alpha}$ defined in \cref{eq:ace_exp}.}
    \label{tab:basis_functions}
\end{table}

The MLIP was trained using the \texttt{pacemaker} code \cite{lysogorskiy_performant_2021}. First, a coarse potential was parameterized, either through ladder training or by fitting all parameters to a subset of the training data. The resulting potential was then fine-tuned on the full training dataset. The unary interaction parameters were parameterized before fitting the binary interaction coefficients. To prevent overfitting $L_1$/$L_2$ regularization was employed, with the scaling factor set to $\approx 10^{-5}$ for Mg and Mg-Nd, and $10^{-8}$ for Nd. Energies and forces were weighted to ensure that low-energy structures were accurately reproduced. The weighting scheme described in \cite{bochkarev_efficient_2022} was applied to most structures. Additional weight was given to structures on or near the convex hull, as well as to those with neodymium compositions between 0 and 0.25.

\begin{figure}[!ht]
    \centering
    \includegraphics[width=\columnwidth ]{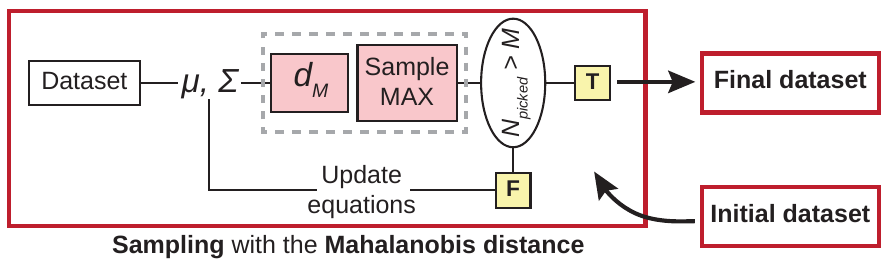}
    \caption{Sampling workflow for training dataset generation.}
    \label{fig:dataset_selection}
\end{figure}

We sampled a diverse training dataset from a pool of candidate structures using a selection metric based on the Mahalanobis distance \cite{mahalanobis_reprint_2018}:
\begin{equation} \label{eq:mahalanobis_distance}
    d(\vec{X})^2 = (\vec{X} - \vec{\mu})^T \Sigma^{-1} (\vec{X} - \vec{\mu})
  \end{equation}
where $d(\vec{X})$ is the Mahalanobis distance of an atomic environment, computed from its ACE descriptors $\vec{X}$. $\vec{\mu}$ and $\Sigma$ are the mean vector and covariance matrix computed from the ACE descriptors of all atomic environments in the current training set. The sampling process was iterative. We started with an initial subset of structures and then repeatedly added the candidate structure with the largest Mahalanobis distance to the training dataset. After adding each new structure, the mean vector and covariance matrix were recomputed with the update equations (\cref{supp-sup_eq:mean_update,supp-sup_eq:covariance_update}). This sampling method is schematically illustrated in \cref{fig:dataset_selection}. Following this approach, we created three separate training sets for pure Mg, pure Nd, and Mg-Nd.

Energies and forces for the structures in our dataset are computed with DFT as implemented in the \emph{Vienna Ab-initio Simulation Package} (VASP) \cite{kresse_ab_1994, kresse_ab_1995, kresse_efficiency_1996, kresse_ultrasoft_1999}. These calculations used the Perdew-Burke-Ernzerhof exchange-correlation functional within the generalized gradient approximation \cite{perdew_generalized_1996, perdew_generalized_1997}. The projector-augmented wave pseudopotentials treated the 2p$^6$3s$^2$ and 5s$^2$5p$^6$4f$^1$6s$^2$ electrons as valence for Mg and Nd respectively. All calculations utilized a plane-wave cutoff of 500 eV, a reciprocal $\Gamma$-centered k-point mesh with smallest allowed spacing of $\approx$ 0.02 \AA$^{-1}$ for Brillouin zone sampling and a second order Methfessel-Paxton smearing of 0.2 eV. Formation energies of structures were computed relative to hcp-Mg, and dhcp-Nd as:
\begin{equation}
    \label{eq:formation_energy}
    e_{f} = \frac{E(N, M) - N e_{hcp} - M e_{dhcp}}{N + M}
\end{equation}
where $E(N, M)$ is the total energy of the configuration, $N$ and $M$ are the numbers of Mg and Nd atoms, and $e_{hcp}$ and $e_{dhcp}$ are the per-atom energies of hcp-Mg, and dhcp-Nd.

\section{Dataset construction} \label{sect:dataset_construction}

\begin{figure}[!ht]
    \centering
    \subfloat[][\label{fig:summary_data_color_coded}]{\includegraphics[width=\linewidth]{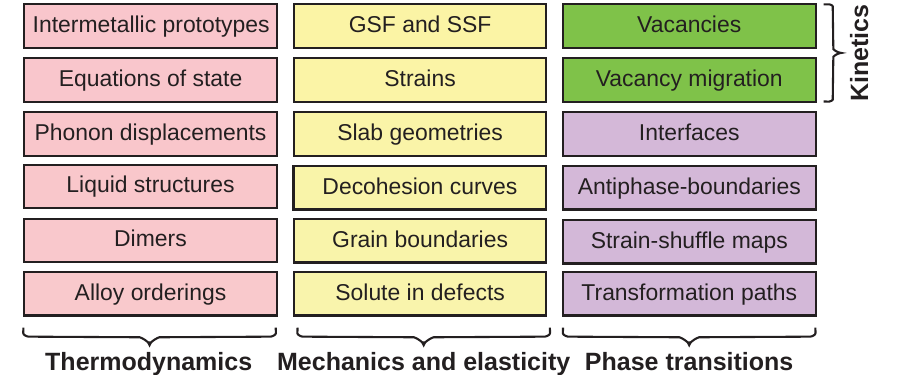}}\\
    \subfloat[][\label{fig:active_learning_loop}]{\includegraphics[width=0.75\columnwidth ]{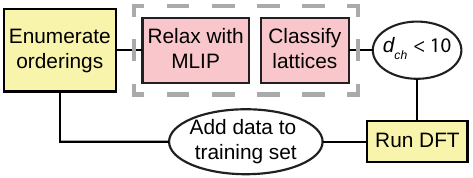}}
    \caption{(a) Structures in the reference database and their relationship to material properties. (b) Active learning workflow for discovering new low-energy orderings.}
\end{figure}

To parameterize an accurate interatomic potential, a training set must encompass the structural and chemical environments relevant to the properties of interest. We constructed our training dataset by systematically generating a wide range of structures relevant to the precipitation process in Mg-Nd alloys. Several studies \cite{kobayashi_neural_2017, maresca_screw_2018, marchand_machine_2020, stricker_machine_2020} outline established strategies for constructing training sets for multicomponent alloy potentials. \Cref{fig:summary_data_color_coded} summarizes the types of structures included in our training dataset and their relationship to specific material properties. In the following, we highlight strategies used to improve modeling accuracy for precipitation in the Mg-Nd system.

The starting point for our training dataset was a catalogue of 267 structural prototypes of intermetallics, containing a range of diverse atomic environments from simple close packed arrangements (e.g., fcc and hcp) to complex topologically close packed phases \cite{kolli_discovering_2020}. This enables us to sample a large degree of structural diversity within relatively small supercells. To span a wide range of densities, we included the volumetric distortions of several low-energy prototypes. The addition of thermally disordered structures sampled from high-temperature ab-initio molecular dynamics simulations provided finite temperature information.

Capturing the properties of pure Mg is important for modeling precipitation in dilute alloys. The training database included symmetrically-distinct deformations of hcp-Mg \cite{thomas_exploration_2017}. The database also contains a variety of structural defects, such as stable stacking faults (SSF), generalized stacking faults (GSF), grain boundaries (GB), vacancies and surfaces. Some of the planar defects were sourced from the Crystallium database \cite{tran_anisotropic_2019} and the work of Stricker et al. \cite{stricker_machine_2020} respectively.

Constructing the reference database for the binary alloy was significantly more complex. Crystal structures of experimentally reported phases were obtained from the ICSD and OQMD databases \cite{zagorac_recent_2019, kirklin_open_2015}. Symmetrically-distinct arrangements of the two elements where enumerated on hcp, bcc, fcc and C15 Laves parent structures using the \verb|CASM| \cite{puchala_casm_2023} code with the algorithm of Hart and Forcade \cite{hart_algorithm_2008, hart_generating_2009}. We generated orderings in supercells containing up to 5, 8, 8 and 2 primitive unit cells for each parent structure. The atomic coordinates and lattice vectors of each resulting structure were then fully relaxed with DFT. The training database also included structures that capture the mechanical response \cite{thomas_exploration_2017} of all stable and metastable phases in the binary alloy. Chemically and structurally disordered configurations were created with Monte Carlo molecular dynamics simulations using the MEAM potential of Kim et al. \cite{kim_modified_2017, thompson_lammps_2022}. The energies and forces for a representative selection of these disordered configurations were then recomputed with DFT for inclusion in the database. Mg-rich compositions are oversampled in our training dataset, as the purpose of the MLIP is to model precipitation in Mg-rich alloys. A detailed description of the configurations and DFT data in the reference database is provided in \cref{supp-tab:Mg_database,supp-tab:Nd_database,supp-tab:MgNd_database}.

An MLIP must accurately reproduce the thermodynamic stability of stable and metastable phases. While specialized methods such as bespoke weighting schemes, loss functions, and cluster selection algorithms address this challenge for on-lattice cluster expansions \cite{puchala_thermodynamics_2013, huang_construction_2017, goiri_recursive_2018}, MLIPs require a different approach. We implemented an iterative active learning scheme to augment the reference database and training sets, as illustrated in \cref{fig:active_learning_loop}. First, we trained a preliminary potential on a subset of the data. We then used this preliminary potential to relax approximately 0.5 million symmetrically distinct orderings in supercells not present in the training set. We assigned the MLIP-relaxed structures to parent lattices (hcp, bcc, fcc, and C15 Laves) based on a similarity metric computed from their ACE descriptors. \Cref{supp-sup:descriptor_classification} describes the algorithm for classifying crystal structures based on ACE descriptors. We compared the relaxed formation energies of these new decorations against the convex hull of MLIP-predicted formation energies for structures in the training dataset, constructing separate convex hulls for orderings on each parent crystal structure. The structures in the training dataset were classified using the structure matching algorithm of Thomas et al. \cite{thomas_comparing_2021}. Finally, we collected any configuration within 10 meV/atom of each convex hull with compositions $x_{\text{Nd}}$ < 0.5, along with all newly predicted ground states at higher Nd content, and recomputed them with DFT. The resulting energies and forces were added to the training set. We then retrained the MLIP and repeated the active learning loop until no new low-energy ground states were discovered. 

\subsection{Enumerating transformation pathways} \label{sect:enumerating_paths}

\begin{figure}
    \centering
    \includegraphics[width=\columnwidth]{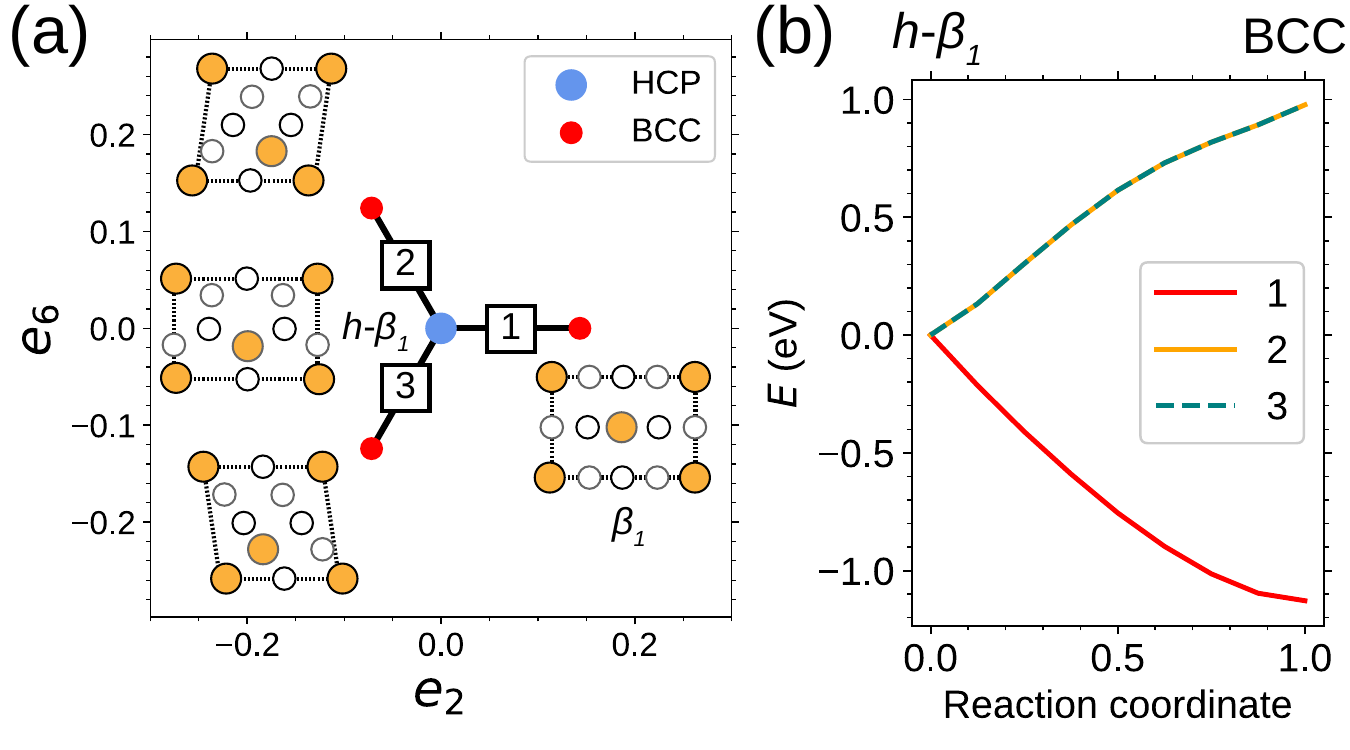}
    \caption{(a) Crystal structures and symmetry-distinct transformation pathways of $h$-$\beta_1$ illustrated in the two-dimensional strain order parameter space ($e_2$, $e_6$) \cite{thomas_exploration_2017,natarajan_2019_understandingdeformation}. (b) DFT energy profiles along the three transformation paths.}
    \label{fig:hbeta1_pathways}
\end{figure}

In addition to capturing the thermodynamic stability of stable and metastable phases, the MLIP must also accurately reproduce the energy landscapes for structural phase transformations relevant to the precipitation process. The structural phase transitions connecting a pair of crystal structures can be enumerated in single-component materials through either domain knowledge or through the automated algorithm of Thomas \emph{et al.}\cite{thomas_comparing_2021}. The lower symmetry of an ordered phase compared to its undecorated parent structure complicates the computation of these energy landscapes in multicomponent alloys. Arranging two or more elements on a parent crystal structure breaks the symmetry of the parent structure, making transformation pathways that are equivalent in a pure element inequivalent in an ordered phase.

As an example, \cref{fig:hbeta1_pathways}a illustrates the three possible variants of bcc that can be accessed through the Burgers transformation starting from $h$-$\beta_1$. Pathway 1 leads to the $\beta_1$ structure. Deforming the $h$-$\beta_1$ structure along pathway 2 results in an atomic arrangement on the bcc lattice distinct from the $\beta_1$ phase. The Nd atoms are arranged in pairs of columns distributed on the $(001)_{bcc}$ atomic planes. The bcc arrangement obtained from a deformation along pathway 3 is identical to that from pathway 2. These differences in the final atomic arrangements are reflected in their respective energy landscapes as shown in \cref{fig:hbeta1_pathways}b. For the transformation of $h$-$\beta_1$ to $\beta_1$ along pathway 1, the initial structure is unstable and transforms to the bcc phase without an energy barrier. In contrast, for pathways 2 and 3, the final bcc decorations are higher in energy than $h$-$\beta_1$, causing the energy to increase along these transformation paths. Parameterizing an MLIP that reproduces the energy landscapes for structural phase transitions requires a training dataset that contains transformations connecting distinct orderings on two or more parent crystal structures. The systematic generation of training and validation sets for an MLIP must account for the underlying symmetry of the transformation pathway and structural decorations.

Any transformation pathway between two crystal structures can be decomposed into a homogeneous lattice distortion and atomic shuffles. An initial structure $\mathcal{P}_{\textrm{initial}}$, is defined by its lattice vectors $\mathbf{L}_{\textrm{initial}} = [\vec{a}, \vec{b}, \vec{c}]$, and the Cartesian coordinates of its $n$ basis atoms, $\mathbf{C} = [\vec{r}_{1}, \vec{r}_{2}, \cdots, \vec{r}_{n}]$. A supercell of the initial structure is generally required to describe the transformation pathway due to differing periodicities between the initial and final structures. This supercell lattice, $\mathbf{S}_{\textrm{initial}} = \mathbf{L}_{\textrm{initial}} \mathbf{T}$, is computed using an integer transformation matrix $\mathbf{T}$. The homogeneous distortion along the transformation pathway is described by a $3\times 3$ deformation tensor $\mathbf{F}$, which maps the initial supercell to the final lattice: $\mathbf{S}_{\textrm{final}} = \mathbf{F} \mathbf{S}_{\textrm{initial}}$. The atomic shuffles are given by a displacement field, $\mathbf{D}$, which is a $3\times d$ matrix containing the displacement vectors for each of the $d$ atoms to attain the final structure $\mathcal{P}_{\textrm{final}}$. The complete transformation pathway can be defined by the tuple $\mathcal{T} = (\mathbf{S}, \mathbf{F}, \mathbf{D})$.

Our algorithm for generating transformation pathways between orderings on two different parent crystal structures consists of three steps. First, a single transformation pathway $\mathcal{T}_{1}$ , is generated between the two undecorated parent structures. This is either computed with the structure mapping algorithm of \cite{thomas_comparing_2021} or from user input. All symmetrically distinct variants of the pathway are enumerated by applying the symmetry operations of the initial parent structure, $\mathcal{P}_{\textrm{initial}}$ to $\mathcal{T}_{1}$. This procedure yields a set of pathways, $\mathcal{S} = \{\mathcal{T}_{1}, \mathcal{T}_{2},\cdots \}$. For example, in the hcp$\rightarrow$bcc transformation, one of the pathways identified by the algorithm of \cite{thomas_comparing_2021} will correspond to the Burgers transformation. The Burgers pathway has three distinct variants that connect the parent hcp structure to three different bcc orientations. These three variants are symmetrically equivalent if the starting point is an undecorated crystal of hcp.

In the second step, we enumerate all possible arrangements of the elements in the alloy on the parent crystal structure $\mathcal{P}_{\textrm{initial}}$. Symmetrically-distinct orderings are enumerated with the algorithm of Hart and Forcade \cite{hart_algorithm_2008, hart_generating_2009}. Each decoration, $\sigma_{i}$ represents a distinct arrangement of the elements on $\mathcal{P}_{\textrm{initial}}$. The set of orderings on the parent structure are collected in a set $\sigma = \{\sigma_{1}, \sigma_{2}, \cdots \}$.

In the final step of the algorithm, the symmetry group of each ordering is used to partition the set of transformation pathways, $\mathcal{S}$, into equivalent groups. The decoration of a crystal structure with several elements can lower the symmetry of the ordered phase relative to the unordered structure. The group of symmetry operations, $\mathcal{G}_{\sigma_{i}}$, that leaves an ordering $\sigma_{i}$ invariant is a sub-group of the symmetry operations that leave the undecorated parent structure unchanged, $\mathcal{G}_{\sigma_{l}} \subseteq \mathcal{G}_{\mathcal{P}_{\textrm{initial}}}$. Applying the symmetry group of the ordered phase to the elements of $\mathcal{S}$ will partition the set into groups of transformation pathways that are related by symmetry. For instance, in the case of the hcp$\rightarrow$bcc transformation of $h$-$\beta_1$ shown in \cref{fig:hbeta1_pathways}, the three transformation pathways, $\mathcal{T}_{1}, \mathcal{T}_{2}, \mathcal{T}_{3}$ are partitioned into two symmetrically inequivalent sets: $\{\mathcal{T}_{1}\}$ and $\{\mathcal{T}_{2}, \mathcal{T}_{3}\}$. Pathways $\mathcal{T}_{2}$ and $\mathcal{T}_{3}$ are related by the symmetry operations of $h$-$\beta_1$, whereas no symmetry operation in $\mathcal{G}_{h-\beta_1}$ maps $\mathcal{T}_{1}$ to either of the other two. This equivalence is reflected in the transformation energy landscapes of \cref{fig:hbeta1_pathways}. The energies for $\mathcal{T}_{2}$ and $\mathcal{T}_{3}$ are identical, while $\mathcal{T}_{1}$ is distinct, showing the initial hcp ordering to be unstable. Constructing a minimal training dataset only requires generating structures along one representative pathway from each distinct set. Our algorithm implements this by selecting one pathway from each set (e.g. $\mathcal{T}_{1}$ and $\mathcal{T}_{2}$ for $h$-$\beta_1$) for structure generation.

\section{Validating interatomic potentials} \label{sect:validation}

The quality of a MLIP is typically quantified by error metrics that measure the difference between its predicted energies, forces, and stresses and the reference values in training and validation datasets. These global error metrics are supplemented by comparisons of material properties, such as elastic moduli and stacking fault energies, and by a qualitative assessment of defect structures like dislocation cores \cite{marchand_machine_2020, marchand_machine_2022}. While global error metrics, such as the root mean squared error (RMSE) or mean absolute error (MAE), provide a general measure of the potential's predictive accuracy, these property-specific calculations are necessary to ensure that the MLIP accurately captures important quantities for large-scale atomistic simulations.

\subsection{Assessing phase stability}
\begin{figure}[htbp]
    \centering
    \includegraphics[width=0.7\linewidth]{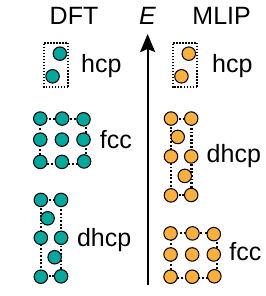}
    \caption{Schematic energy spectrum of elemental polymorphs, depicting a scenario where multiple stacking-order phases such as fcc, hcp and dhcp lie close to the global ground state.}
    \label{fig:kendall_tau}
\end{figure}

Beyond conventional error metrics, an interatomic potential for studying precipitation must correctly reproduce the thermodynamic stability of all relevant phases. For instance, consider the structures and energies of a pure element shown in \cref{fig:kendall_tau}, where the double-hcp (dhcp) structure, with an $ABAC$ stacking of triangular layers, is the lowest energy polymorph. An MLIP that incorrectly predicts another structure, such as fcc, to be more stable than the dhcp ground state would be unreliable for simulating the element's thermodynamic properties. This type of failure can be difficult to detect using only global error metrics like RMSE or MAE, as an incorrect energy ranking between two similar, low-energy polymorphs may only marginally increase the total error. Accurate reproduction of the relative stabilities of low-energy structures is important, as they may be thermally accessible during finite-temperature simulations. In contrast, the exact energy ordering of very high-energy polymorphs is less important, as these structures are unlikely to form under typical simulation conditions.

The weighted Kendall-$\tau$ \cite{vigna_weighted_2015} ($\tau_w$) quantifies the similarities and differences between the energy spectra of a set of structural polymorphs computed with DFT and the MLIP:
\begin{equation}
    \label{eq:weighted_kendall_tau}
    \tau_w = \frac{1}{N_{\tau}} \sum_{i, j > i} w(i, j) sgn(e^{DFT}_{i} - e^{DFT}_{j}) sgn(e^{MLIP}_{i} - e^{MLIP}_{j})
\end{equation}
$e^{DFT}_{i}$ and $e^{MLIP}_{i}$ are the normalized formation energies of a structural polymorph $i$ of an element relative to a reference state, $w(i, j)$ is a weighting function, $N_{\tau}$ is a normalization constant ($N_{\tau}= \sum_{i, j > i} w(i, j)$), and $sgn(x)$ is defined as:
\begin{equation}
    \label{eq:sgn_definition}
    \quad sgn(x) =
    \begin{cases}
        +1 & x > 0 \\
        0  & x = 0 \\
        -1 & x < 0
    \end{cases}
\end{equation}
The weighted Kendall-$\tau$ takes a value of +1 if the ordering of structural polymorphs between DFT and the MLIP is identical, and a value of -1 if the ordering of the structural polymorphs predicted by the MLIP is exactly opposite to the ordering of energies computed with DFT. Using a uniform weight of 1 for all pairs of structures, transforms the $\tau_w$ to the well known standard unweighted Kendall-$\tau$, denoted $\tau$ \cite{vigna_weighted_2015}. The weighted Kendall-$\tau$ is particularly useful for analyzing large datasets with a wide energy spread, whereas the unweighted Kendall-$\tau$ is better suited for small datasets with a moderate energy spread.

The unweighted Kendall-$\tau$ considers all possible pairs of structural polymorphs and counts how often the sign of the energy difference predicted by the MLIP agrees with DFT. If the sign is the same, the pair is \textit{concordant}, and the coefficient registers a point of similarity (+1). If the sign is different, the pair is \textit{discordant}, and it is registered as a point of dissimilarity (-1). The prefactor in \cref{eq:weighted_kendall_tau} is then the total number of pairs and normalizes the coefficient. The $\tau$ coefficient is similar to the concordance index used in \cite{liu_assessing_2024}.

The weighting function, $w(i,j)$, can be tuned to prioritize the importance of low-energy polymorphs over high-energy structural polymorphs. Two commonly used weighting functions are either additive, $w(i,j)=w(i)+w(j)$, or multiplicative, $w(i,j) = w(i)\times w(j)$. The weights for individual structural polymorphs, $w(i)$, can be any function that decays with the energy of the polymorph relative to the ground state crystal structure. Adjusting the weighting functions $w(i)$ and $w(i, j)$ enables one to resolve the energy ranking of specific subsets of polymorphs. As long as the coefficient is normalized by $N_\tau$, the $\tau_w$ lies between -1 and 1. Values closer to 1 indicate closer reproduction of the energy ordering of highly weighted structural polymorphs.

The Kendall-$\tau$ described by \cref{eq:weighted_kendall_tau} can be broadly applied to estimate the quality of the MLIP on any energy spectrum where the relative ordering of energies is important to reproduce. For example, MLIPs must reproduce not just the magnitude of surface energies of various surface orientations, but also capture the relative order of the surface energies correctly. Similarly, unstable stacking fault energies must also be reproduced to capture the properties of dislocations and cracks. The Kendall-$\tau$ is readily applied to this problem and can provide a quantitative metric that, in conjunction with error metrics such as the RMSE, could guide the parameterization of MLIPs.

\begin{figure}[htbp]
    \centering
    \includegraphics[width=\linewidth]{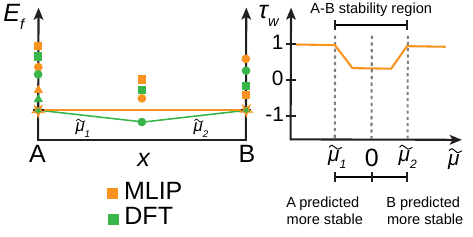}
    \caption{Schematic of formation energies for a binary A-B alloy computed via DFT and MLIP. While the energy ordering of configurations is correctly predicted at each composition, the AB compound fails to lie on the convex hull in the MLIP prediction. This discrepancy manifests as a decrease in $\tau_w$ within the chemical potential range corresponding to AB stability.}
    \label{fig:gc_kendall_tau}
\end{figure}

Utilizing the Kendall-$\tau$ to compare phase stability predictions in a multicomponent alloy is more challenging. A naive approach would be to evaluate the $\tau_w$ coefficient at individual compositions. Although this approach provides an indication of the accuracy of the potential, it neglects the relative stabilities between phases at different compositions. For example, \cref{fig:gc_kendall_tau} illustrates a comparison between the true and MLIP-predicted convex hulls for a hypothetical alloy. Despite relative energies of orderings being reproduced exactly for all compositions, the equiatomic ordering is predicted to be unstable because the formation energies of all orderings are more positive, as predicted by the MLIP, compared with the ground truth formation energies. The Kendall-$\tau$ needs to be modified to compare the relative orderings of energies across multiple compositions.

The stability of phases with varying compositions can be compared within the semi-grand canonical ensemble. The semi-grand canonical potential energy, $\omega(\tilde{\mu}) = e_f(x_{B}) - \tilde{\mu} x_{B}$, can be used to compare the stabilities of various degrees of order across crystal structures and compositions. At a fixed exchange chemical potential, $\tilde{\mu} = \mu_{B}-\mu_{A}$, the phase with the lowest value of $\omega$ will be the stable state. The relative values of the semi-grand canonical potential then indicate the degree of metastability of individual phases.

A metric of the predictive accuracy of the MLIP with respect to phase stability can be obtained by computing the weighted Kendall-$\tau$ value as a function of the exchange chemical potentials. At each value of the exchange chemical potential, the semi-grand canonical potential energy is computed for every phase, and the Kendall-$\tau$ of \cref{eq:weighted_kendall_tau} is calculated by using the semi-grand canonical potential energy in place of the formation energies. This is then repeated for all relevant values of the exchange chemical potential. The output of this procedure is a $\tau_w$--$\tilde{\mu}$ trace that captures the quantitative accuracy of the MLIP in reproducing both the prediction of phase stability and the prediction of the chemical potential for coexistence between stable phases. Reproducing coexistence chemical potentials can be important for applications such as electrochemical materials, corrosion, and hydrogen storage, where experimentally controlled parameters can be directly related to the chemical potentials.

Several choices are possible for the weight function when computing the grand canonical Kendal-$\tau$. In this study we utilize a weight $w(\tilde{\mu}, i)$ given by:
\begin{equation}\label{eq:weight_sgc}
    w(\tilde{\mu}, i) = \frac{n_{\tilde{\mu}}}{(\omega^{DFT}(\tilde{\mu}, i) - \omega^{DFT}_{min}(\tilde{\mu}) + \epsilon)^2}
\end{equation}
where $\omega^{DFT}(\tilde{\mu}, i)$ is the semi-grand potential of structure $i$, $\epsilon$ is a softening factor set to 0.01 eV, and $n_{\tilde{\mu}}$ is a normalizing constant.

\begin{figure}[htbp]
    \centering
    \includegraphics[width=\linewidth]{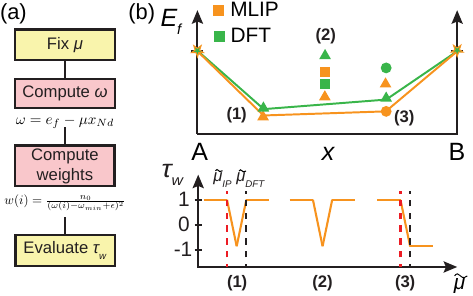}
    \caption{(a) Workflow for computing the semi-grand canonical $\tau_w$. (b) Schematic of formation energies illustrating three common MLIP fitting errors and their corresponding $\tau_w$--$\tilde{\mu}$ traces.}
    \label{fig:gc_kendall_tau_summary}
\end{figure}

\Cref{fig:gc_kendall_tau_summary}a summarizes the algorithm to compute the $\tau_w$ with semi-grand canonical potential energies. First, the range of relevant exchange chemical potentials is estimated based on the formation energies computed with DFT. The interval of exchange chemical potentials is discretized and at each $\tilde{\mu}$ grid-point the semi-grand canonical potentials of the configurations are calculated from DFT and the MLIP. Next, weights are computed from the $\omega(\tilde{\mu})$ obtained with DFT. Finally, the $\tau_w(\tilde{\mu})$ coefficient is evaluated. The trace of this quantity can be plotted, and the similarities and differences relative to a value of 1 can be used to infer the quality of the MLIP. For example, \cref{fig:gc_kendall_tau} illustrates the $\tau_w$--$\tilde{\mu}$ trace for the MLIP predictions. At low values of $\tilde{\mu}$ the $\tau_w$ is equal to 1 as the MLIP reproduces the stable structure of element A. As the exchange chemical potential increases, the MLIP does not predict the ground state with a composition of 0.5 to be stable. Consequently, as illustrated in \cref{fig:gc_kendall_tau}, the Kendall-$\tau$ degrades to a value significantly below 1. At larger chemical potentials, the MLIP and DFT predict the same stable phase and the Kendall-$\tau$ approaches 1 again.

The $\tau_w$--$\tilde{\mu}$ trace can be more complex than the simple scenario illustrated in \cref{fig:gc_kendall_tau}.
\Cref{fig:gc_kendall_tau_summary}b shows three other possible scenarios. In scenario 1, the temporary degradation in Kendall-$\tau$ is due to a mismatch in the coexistence chemical potentials of the ground state between DFT and the MLIP. The MLIP predicts a slightly more negative coexistence chemical potential than DFT. As shown in the trace of the $\tau_w$, the coefficient degrades between $\tilde{\mu}_{MLIP}$ and $\tilde{\mu}_{DFT}$. The $\tau_w$ recovers a value of 1 at chemical potentials above $\tilde{\mu}_{DFT}$. \Cref{fig:gc_kendall_tau_summary}b also illustrates a similar degradation at exchange chemical potential values that are away from coexistence chemical potentials. In case 2, the drop in accuracy is due to the incorrect prediction of phase stability away from the ground states. While such mismatches can result in quantitative discrepancies between the predictions of the electronic structure calculations and the MLIP, the MLIP is usually still useful in predicting the thermodynamics of the alloy. In the final scenario illustrated in \cref{fig:gc_kendall_tau_summary}b, the MLIP predicts a ground state that is different from DFT. Though the MLIP and DFT predict a ground state at B-rich compositions, they differ in the precise structure. The $\tau_w$ at high values of the exchange chemical potential consequently degrades to a small value. The degradation of $\tau_w$ over an extended chemical potential range indicates a discrepancy in ground states. The MLIP will then need additional fine-tuning to reproduce the low-temperature thermodynamics of the material.

\subsection{Assessing structural phase transitions}
Beyond phase stability, we must also quantify the MLIP's accuracy in reproducing transformation energy landscapes connecting pairs of crystal structures. A simple error metric such as the RMSE computed over structures corresponding to structural transformations provides a useful estimate of MLIP quality. However, ensuring qualitative agreement between MLIP predictions and DFT calculations is equally important.

\begin{figure}[!ht]
    \centering
    \includegraphics[width=0.95\columnwidth]{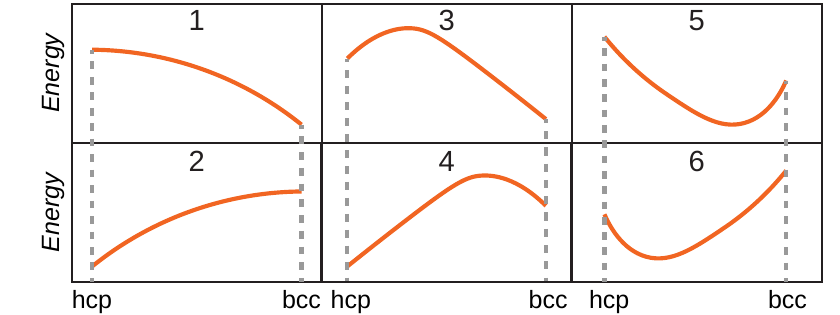}
    \caption{Schematic of the six energy profile classes for the Burgers transformation. These classes span physical scenarios ranging from barrierless transitions to pathways exhibiting barriers and stable intermediate states.}
    \label{fig:pathway_classes}
\end{figure}

\Cref{fig:pathway_classes} illustrates six possible energy profiles for structural phase transformations relating an ordering on the hcp parent crystal structure to its bcc counterpart. Pathways 1 and 2 show landscapes where the ordering is unstable in one parent structure and undergoes a barrierless homogeneous transformation to an ordering on the other parent structure. Pathways 3 and 4 illustrate a different scenario with an energy barrier separating the higher-energy phase from the lower-energy structure. The last two pathway classes illustrate scenarios where the ordering is unstable in both parent structures and transforms to an intermediate state between them.

The MLIP must capture qualitative trends in these energy profiles to reliably simulate precipitation. Categorizing pathways by their qualitative behavior rather than precise profile shape allows us to assess mismatches between DFT and MLIP using a confusion matrix. We first assign each pathway in the dataset to one of the classes in \cref{fig:pathway_classes} based on its DFT-computed energy landscape. We then use the MLIP to predict energies along the transformation path and classify the pathway accordingly. A confusion matrix quantifies the MLIP's accuracy in reproducing each class.

\section{Results} \label{sect:results}

We used a large dataset of electronic structure calculations to parameterize the binary interatomic potential for Mg-Nd. We validated the interatomic potential with conventional error measures and the error metrics introduced in this study.

The dataset for pure Mg contains the crystal structures of common intermetallic compounds, various configurations of bulk hcp-Mg (strains, GSFs, SSFs), volumetric distortions of high-symmetry and low-energy prototypes, grain boundaries, surface slabs, vacancy point defects, and liquid configurations. The training set for Nd is less diverse, containing the crystal structures of common intermetallic compounds, volumetric distortions of high-symmetry and low-energy prototypes, and liquid configurations. The Mg-Nd training set includes the crystal structures of the experimental phases, solute point defects in various hcp-Mg configurations (bulk hcp, stacking faults, and grain boundaries), Nd pair clusters, vacancy-Nd clusters, vacancy point defects in the precipitate phases, symmetry-distinct orderings on simple crystal structures, liquid-like configurations, and elastic distortions of all stable phases and some metastable phases. We also included structures along the Burgers transformation connecting hcp and bcc that can be acccommdated within a supercell of hcp that is twice the primitive cell size and structures related to the $\beta_1$ phase transformations. The datasets for Mg, Nd and Mg-Nd contain 7055, 4002 and 7985 configurations respectively. A detailed description of the dataset is included in \cref{supp-sup:training_datasets}.

\subsection{Unary potential}

\begin{table}[!ht]
    \centering
    \begin{tabular}{|c|c|c|c|}
        \hline
        Property                             & Exp.                                    & DFT                             & MLIP  \\ \hline
        \multicolumn{4}{|c|}{Lattice parameters}                                                                                 \\
        $a$ (\r{A})                          & 3.21 (298K) \cite{walker_lattice_1959}  & 3.21                            & 3.20  \\
        $c/a$                                & 1.624 (298K) \cite{walker_lattice_1959} & 1.614                           & 1.617 \\
        \multicolumn{4}{|c|}{Mechanical properties (GPa)}                                                                        \\
        $C_{11}$                             & 63.48 \cite{slutsky_elastic_1957}       & 62.5                            & 62.4  \\
        $C_{12}$                             & 25.94 \cite{slutsky_elastic_1957}       & 22.3                            & 23.4  \\
        $C_{13}$                             & 21.7 \cite{slutsky_elastic_1957}        & 21                              & 21    \\
        $C_{33}$                             & 66.45 \cite{slutsky_elastic_1957}       & 63.8                            & 64.8  \\
        $C_{44}$                             & 18.42 \cite{slutsky_elastic_1957}       & 17.6                            & 17.4  \\
        $C_{66}$                             & 18.75 \cite{slutsky_elastic_1957}       & 20.1                            & 19.5  \\
        $B$                                  & 36.7 \cite{nishimura_volume_nodate}     & 36.2                            & 36.4  \\
        \multicolumn{4}{|c|}{Surface/interface properties (mJ/m$^2$)}                                                            \\
        $\gamma_{SSF}$ Basal I1              & 33 \cite{ahmad_designing_2019}          & 34 \cite{stricker_machine_2020} & 38    \\
        $\gamma_{SSF}$ Prism                 & -                                       & 225                             & 187   \\
        $\gamma_{SSF}$ Pyramid I             & -                                       & 160                             & 149   \\
        $\gamma_{SSF}$ Pyramid II            & -                                       & 155                             & 165   \\
        $E_{surf}$ Basal (0001)              & -                                       & 549                             & 573   \\
        $E_{surf}$ Prism (10$\bar{1}$0)      & -                                       & 607                             & 615   \\
        $E_{surf}$ Pyramid I (10$\bar{1}$1)  & -                                       & 639                             & 665   \\
        $E_{surf}$ Pyramid II (11$\bar{2}$1) & -                                       & 771                             & 767   \\
        $E_{surf}$ $\Sigma$ GB (0001)        & -                                       & 168                             & 177   \\
        \multicolumn{4}{|c|}{Vacancy properties}                                                                                 \\
        $E_{vac}$ (eV)                       & 0.79 \cite{tzanetakis_formation_1976}   & 0.813                           & 0.818 \\
        $E_{act}$ basal hop (meV)            & -                                       & 390                             & 401   \\
        $E_{act}$ pyramidal hop (meV)        & -                                       & 403                             & 420   \\ \hline
    \end{tabular}
    \caption{Comparison of selected Mg properties from MLIP, DFT, and experiment.}
    \label{tab:properties_Mg}
\end{table}

\begin{table}[!ht]
    \centering
    \begin{tabular}{| c | c c c | c c c | c |}
        \hline
        \multirow{2}{*}{ \centering Surfaces } & \multicolumn{3}{|c|}{Highest (mJ/m$^2$)} & \multicolumn{3}{|c|}{Lowest (mJ/m$^2$)} & \multirow{2}{*}{ \centering $\tau$}                              \\ \cline{2-7}
                                               & Plane                                    & DFT                                     & MLIP                                & Plane  & DFT & MLIP &      \\ \hline
        hcp slabs                              & (20$\bar{2}$1)                           & 784                                     & 793                                 & (0001) & 549 & 573  & 0.91 \\
        fcc $\Sigma$ GB                        & (110)                                    & 475                                     & 471                                 & (111)  & 129 & 140  & 0.90 \\ \hline
    \end{tabular}
    \caption{Comparison of energy rankings for various slab and grain boundary geometries. The Kendall-$\tau$ coefficient is unweighted.}
    \label{tab:surfaces_Mg}
\end{table}

\begin{figure}[!ht]
    \centering
    \subfloat[][\label{fig:e3_shuffle}]{\includegraphics[width=\columnwidth ]{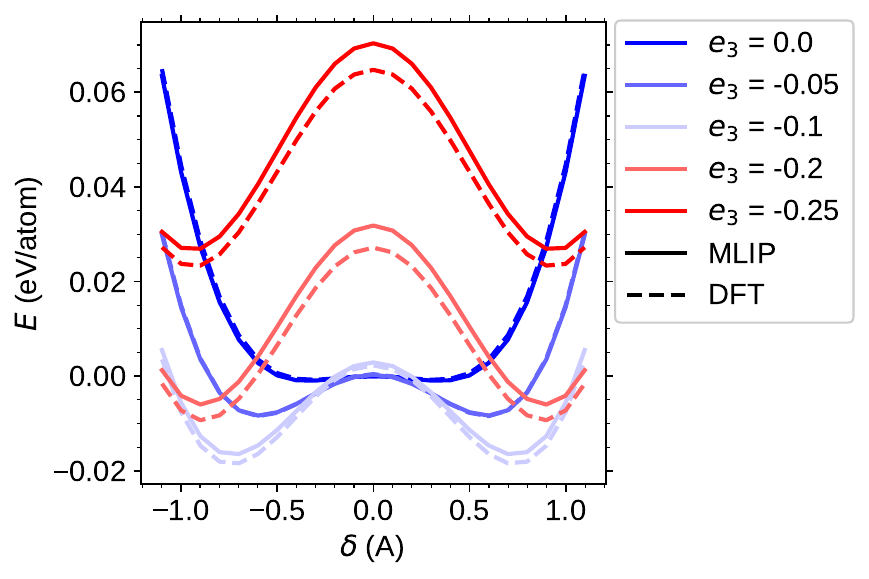}}\\
    \subfloat[][\label{fig:e1_shuffle}]{\includegraphics[width=\columnwidth]{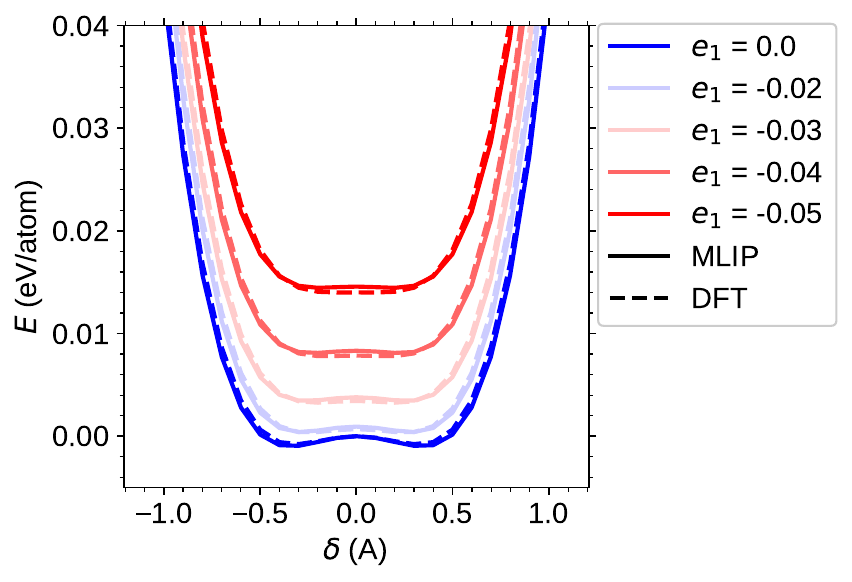}}
    \caption{Energy variation with the Burgers shuffle at various strain states relative to bcc. Panels show the effect of varying (a) tetragonal distortion $e_{3}$ and (b) volumetric strain $e_{1}$.}
    \label{fig:strain_shuffle_Mg}
\end{figure}

The Mg potential achieves training MAEs of 1.9 meV/atom and 1.3 meV/\AA. Validation MAEs are 5.6 meV/atom for energies and 6.3 meV/\AA{} for forces. \Cref{tab:properties_Mg} compares the predicted hcp-Mg properties with DFT and experimental values. The potential reproduces lattice parameters, elastic constants, and the bulk modulus. It also captures migration barriers despite the absence of this class of data in the training set. Furthermore, \cref{tab:surfaces_Mg} shows that the MLIP accurately predicts planar defect properties, including surface, stacking fault, and grain boundary energies. The potential preserves the energetic ordering and magnitudes of surface energies, achieving a Kendall-$\tau \geq 0.9$. We used finite-temperature molecular dynamics to verify the potential's stability at elevated temperatures, as illustrated in \cref{supp-fig:npt_pure_Mg}.

The MLIP also reproduces the energy landscape of the structural phase transition relating hcp-Mg to bcc-Mg through the Burgers transformation. \Cref{fig:strain_shuffle_Mg} illustrates the energy variation from the shuffle of alternate $(110)$ bcc-Mg planes at different strain states. These shuffles are necessary to attain the alternating stacking of triangular layers in the hcp structure. \Cref{fig:e3_shuffle} compares MLIP and DFT shuffle energies along the $e_3$ strain order parameter, which tracks the degree of Burgers transformation. Previous studies\cite{thomas_exploration_2017,natarajan_2019_understandingdeformation} rigorously define the strain order parameters $e_{1}$ and $e_{3}$ and their relationship to the Burgers transformation. For the ideal bcc structure at zero shuffle magnitude and $e_{1}=0$, both DFT and the MLIP indicate that the $(110)$ shuffle is only marginally more stable. As $e_3$ becomes negative, the bcc $(110)$ planes distort toward a triangular arrangement. With increasing distortion, the ideal bcc structure becomes less stable, and local minima emerge at non-zero shuffle amplitudes. The hcp structure, the lowest energy state, is attained at $e_3 \approx -0.1$. \Cref{fig:e1_shuffle} illustrates the response to volumetric compression, $e_1$, relative to the bcc reference. As the bcc structure is compressed, DFT predicts the shuffle instability diminishes. The MLIP qualitatively reproduces this trend.

\begin{figure}[!ht]
    \centering
    \subfloat[][\label{fig:error_tau_MgNd}]{\includegraphics[width=0.9\columnwidth]{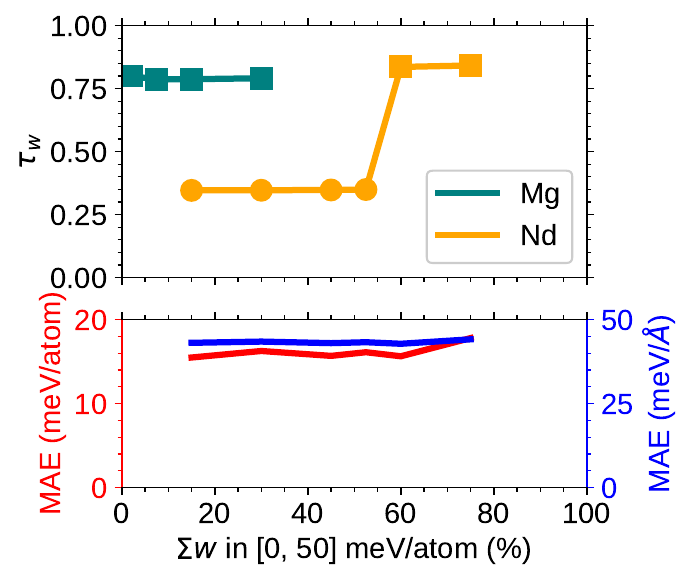}}\\
    \subfloat[][\label{fig:EOS_Nd}]{\includegraphics[width=0.75\columnwidth]{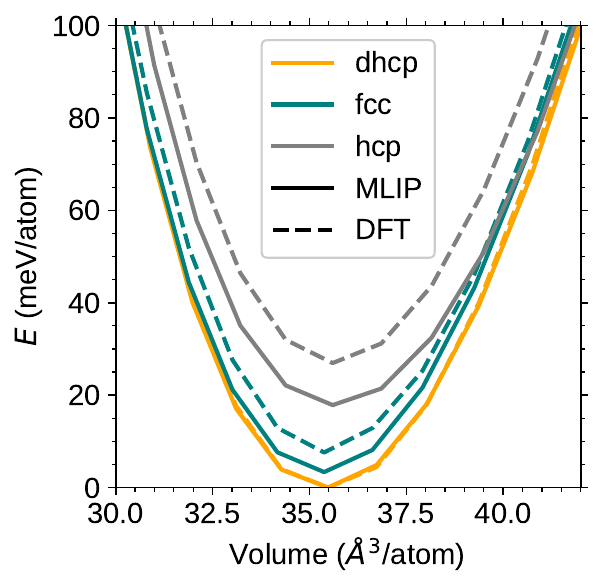}}
    \caption{(a) The upper panel shows $\tau_w$ as a function of the total weight assigned to data points in the 0--50 meV/atom range for Mg and Nd. Circles and squares denote MLIPs predicting incorrect and correct ground states, respectively. The lower panel illustrates the lack of correlation between MAE and $\tau_w$ for Nd MLIPs trained with varying low-energy weights. (b) Energy-volume curves for Nd dhcp, fcc, and hcp. DFT and MLIP energies are referenced to the dhcp ground state.}
\end{figure}

Parameterizing the Nd potential was more challenging as multiple crystal structures are close in energy to the 0K ground state. DFT predicts the 0K ground state is dhcp, with hcp and fcc structures within 20 meV/atom. A naive parameterization of the interatomic potential for Nd failed to reproduce the dhcp ground state. \Cref{fig:error_tau_MgNd} plots the $\tau_w$ metric for Mg and Nd MLIPs as a function of the weight fraction assigned to structures within 50 meV/atom of the minimum energy structure. The coefficient is calculated solely over the data of structural prototypes of intermetallics using a multiplicative weighting function. A Mg potential parameterized on a comparable dataset (\cref{supp-tab:Mg_tau_set}) maintains $\tau_w \approx 1$ regardless of the weight assigned to low-energy polymorphs. In contrast, the Nd $\tau_w$ drops to $\approx 0.4$ when the cumulative weight of low-energy structures falls below 50\%, as the MLIP incorrectly predicts the ground state to be fcc. Increasing the weight of low-energy structures improves the predictive quality of the MLIP. As shown by the case of Nd, standard MAEs are largely invariant to varying weights and fail to capture the discrepancies with DFT-predicted ground states. An analysis like that of \cref{fig:error_tau_MgNd} provides a quantitative guide towards adjusting weighing schemes such that MLIPs reproduce low-temperature phase stability.

The final Nd potential achieved training MAEs of 9.2 meV/atom and 9.4 meV/\AA, and validation MAEs of 42.5 meV/atom and 31.5 meV/\AA\ on held-out data. The volumetric distortions for dhcp, fcc, and hcp (\cref{fig:EOS_Nd}) indicate the potential reproduces relative phase stabilities across a range of volumes. An NPT MD heating-cooling cycle confirmed the potential's stability, \cref{supp-fig:npt_pure_Nd}. Finally, \cref{tab:properties_Nd} shows the MLIP reproduces the lattice parameters and bulk modulus of the dhcp ground state.

\begin{table}[!ht]
    \centering
    \begin{tabular}{|c|c|c|c|}
        \hline
        Property  & Exp.                               & DFT   & MLIP  \\ \hline
        a (\r{A}) & 3.65 \cite{villars_nd_nodate}      & 3.71  & 3.70  \\
        c/a       & 3.226 \cite{villars_nd_nodate}     & 3.216 & 3.220 \\
        $B$ (GPa) & 31.8 \cite{noauthor_handbook_1978} & 33.4  & 33.8  \\ \hline
    \end{tabular}
    \caption{Comparison of lattice parameters and mechanical properties of dhcp-Nd from experiment, DFT, and MLIP. }
    \label{tab:properties_Nd}
\end{table}

\subsection{Binary potential}
\label{sect:binary_results}
\begin{figure}[htbp]
    \centering
    \includegraphics[width=0.75\columnwidth ]{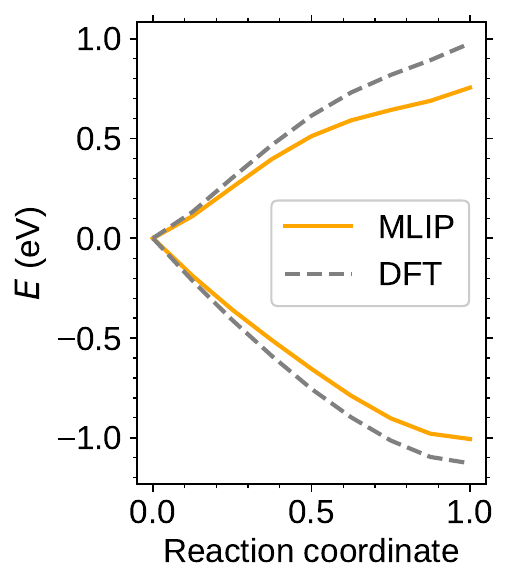}
    \caption{Comparison of Burgers transformation pathways originating from $h$-$\beta_1$.}
    \label{fig:burgers_path_beta1_comparison}
\end{figure}

The Mg-Nd potential achieved training MAEs of 6.4 meV/atom and 22.6 meV/\r{A}. Validation errors were 3.6 meV/atom and 16.1 meV/\r{A}. A single round of the active learning loop was sufficient to converge the hcp, bcc and C15 Laves phase convex hulls with respect to orderings not contained in the initial training set. \Cref{tab:active_learning} breaks down the $\approx$0.5 million configurations across the 4 available parent lattices. \Cref{tab:properties_MgNd_poor} summarizes the structural and mechanical properties of the relevant precipitate phases. The MLIP-predicted lattice parameters match DFT predictions. The bulk moduli and elastic constants for these phases agree with DFT. The potential also reproduces properties of stable phases that do not form during precipitation in magnesium alloys containing dilute Nd concentrations. For example, \cref{tab:properties_MgNd_rich} compares the MLIP-predicted lattice parameters and bulk moduli of Mg$_{2}$Nd and MgNd (B2) phases with DFT values. The interatomic potential reproduces the transformation energy landscape for the Burgers transformation of $h$-$\beta_1$ to $\beta_1$, as shown in \cref{fig:burgers_path_beta1_comparison}.

\begin{table}[!ht]
    \centering
    \begin{tabular}{|c|c|c|c|c|c|c|c|c|c|}
        \hline
        \multicolumn{2}{|c|}{Enumeration} & \multicolumn{4}{|c|}{After relaxation} & \multicolumn{4}{|c|}{$d_{ch}$ < 10 meV/atom}                                                   \\ \hline
        Parent                            & Orderings                              & hcp                                          & bcc    & fcc   & C15 & hcp  & bcc  & fcc  & C15 \\ \hline
        hcp                               & 332272                                 & 60037                                        & 201010 & 7635  & 8   & 1157 & 1159 & 1954 & 5   \\
        bcc                               & 66905                                  & 165                                          & 41452  & 2758  & 0   & 11   & 800  & 599  & 0   \\
        fcc                               & 66905                                  & 49                                           & 38782  & 11167 & 0   & 4    & 690  & 570  & 0   \\
        C15                               & 31268                                  & 1630                                         & 7752   & 45    & 203 & 36   & 12   & 2    & 14  \\ \hline
    \end{tabular}
    \caption{Breakdown of symmetry-distinct orderings enumerated and relaxed during the active learning loop. A subset of configurations did not correspond to any of the four parent lattices.}
    \label{tab:active_learning}
\end{table}

\begin{table}[!ht]
    \centering
    \begin{tabular}{|c |c c|c c|c c|c c|}
        \hline
        \multirow{2}{*}{Property} & \multicolumn{2}{|c|}{$\beta'$} & \multicolumn{2}{|c|}{$\beta''$} & \multicolumn{2}{|c|}{$\beta_1$} & \multicolumn{2}{|c|}{$\beta_e$}                                                                         \\ \cline{2-9}
                                  & DFT                            & MLIP                            & DFT                             & MLIP                            & DFT             & MLIP            & DFT             & MLIP            \\ \hline
        \multicolumn{9}{|c|}{Lattice parameters (\AA)}                                                                                                                                                                                           \\
        $a$                       & 7.07                           & 7.19                            & 6.81                            & 6.81                            & 5.24            & 5.26            & 11.63           & 11.65           \\
        $b$                       & 10.87                          & 10.75                           & $\equiv a$                      & $\equiv a$                      & $\equiv a$      & $\equiv a$      & $\equiv a$      & $\equiv a$      \\
        $c$                       & 5.12                           & 5.11                            & 5.25                            & 5.25                            & $\equiv a$      & $\equiv a$      & $\equiv a$      & $\equiv a$      \\
        \multicolumn{9}{|c|}{Mechanical properties (GPa)}                                                                                                                                                                                        \\
        $B$                       & 37.8                           & 38.1                            & 38.3                            & 38.8                            & 40.1            & 38.7            & 39.4            & 39.6            \\
        $C_{11}$                  & 54.4                           & 53.7                            & 64.0                            & 63.6                            & 59.6            & 57.2            & 72.3            & 67.6            \\
        $C_{12}$                  & 35.9                           & 39.4                            & 31.2                            & 31.9                            & 28.8            & 31.4            & 21.3            & 23.0            \\
        $C_{13}$                  & 21.2                           & 22.7                            & 18.1                            & 22.1                            & $\equiv C_{12}$ & $\equiv C_{12}$ & 21.9            & 26.2            \\
        $C_{33}$                  & 70.1                           & 70.5                            & 77.1                            & 73.4                            & $\equiv C_{11}$ & $\equiv C_{11}$ & 71.7            & 64.5            \\
        $C_{44}$                  & 14.6                           & 11.6                            & 10.8                            & 15.5                            & 39.0            & 37.1            & 27.5            & 22.9            \\
        $C_{55}$                  & 30.7                           & 28.0                            & $\equiv C_{44}$                 & $\equiv C_{44}$                 & $\equiv C_{44}$ & $\equiv C_{44}$ & $\equiv C_{44}$ & $\equiv C_{44}$ \\
        $C_{66}$                  & 27.6                           & 24.4                            & 16.4                            & 15.8                            & $\equiv C_{44}$ & $\equiv C_{44}$ & 16.3            & 12.26           \\ \hline
    \end{tabular}
    \caption{Comparison of structural and mechanical properties for phases with $x_{\text{Nd}} \leq 0.25$ from DFT, and MLIP.}
    \label{tab:properties_MgNd_poor}
\end{table}

\begin{table}[!ht]
    \centering
    \begin{tabular}{|c |c c|c c|}
        \hline
        \multirow{2}{*}{Property} & \multicolumn{2}{|c|}{Mg$_2$Nd} & \multicolumn{2}{|c|}{MgNd}               \\ \cline{2-5}
                                  & DFT                            & MLIP                       & DFT  & MLIP \\ \hline
        $a$ (\r{A})               & 3.88                           & 3.89                       & 6.14 & 6.15 \\
        $B$ (GPa)                 & 38.6                           & 41.3                       & 40.0 & 41.4 \\ \hline
    \end{tabular}
    \caption{Comparison of structural and mechanical properties for phases with $x_{\text{Nd}} \geq 0.25$ from DFT, and MLIP.}
    \label{tab:properties_MgNd_rich}
\end{table}

\begin{figure*}[!ht]
    \centering
    \subfloat[][\label{fig:global_hull}]{\includegraphics[width=\columnwidth]{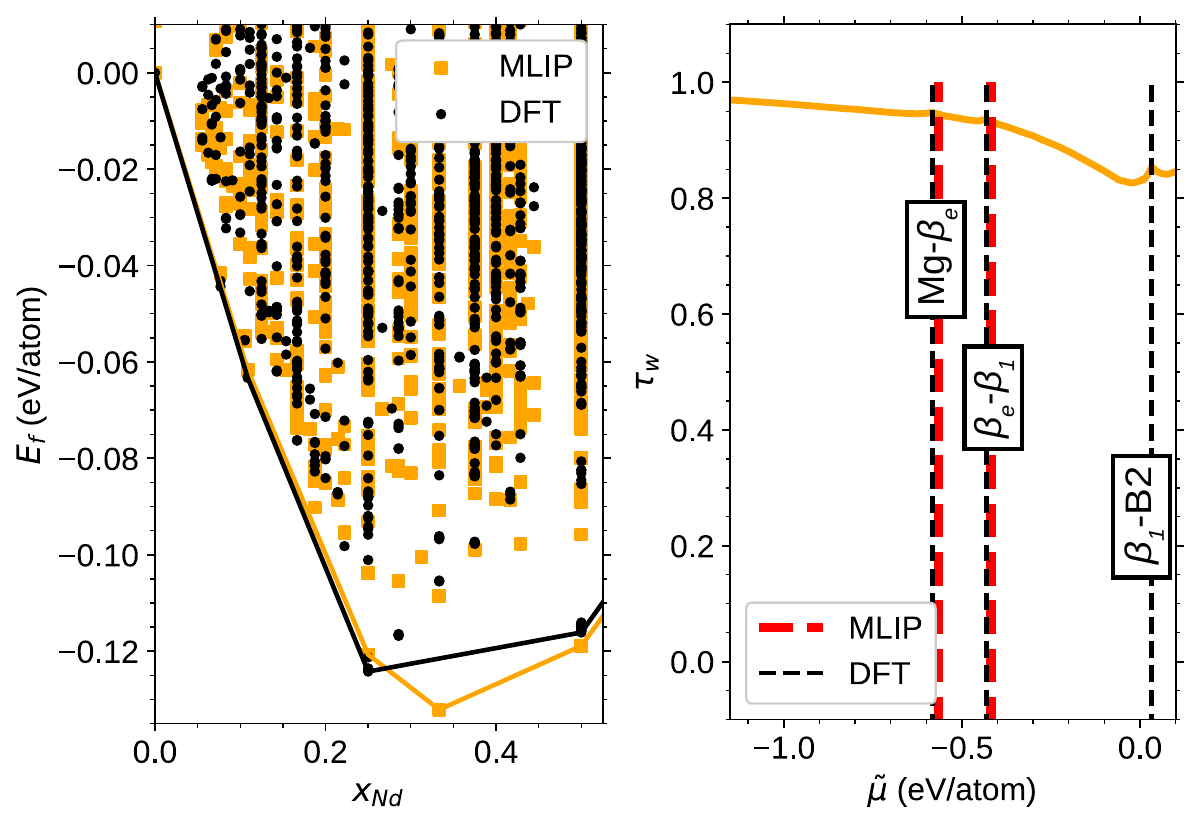}}
    \subfloat[][\label{fig:bcc_hull}]{\includegraphics[width=\columnwidth]{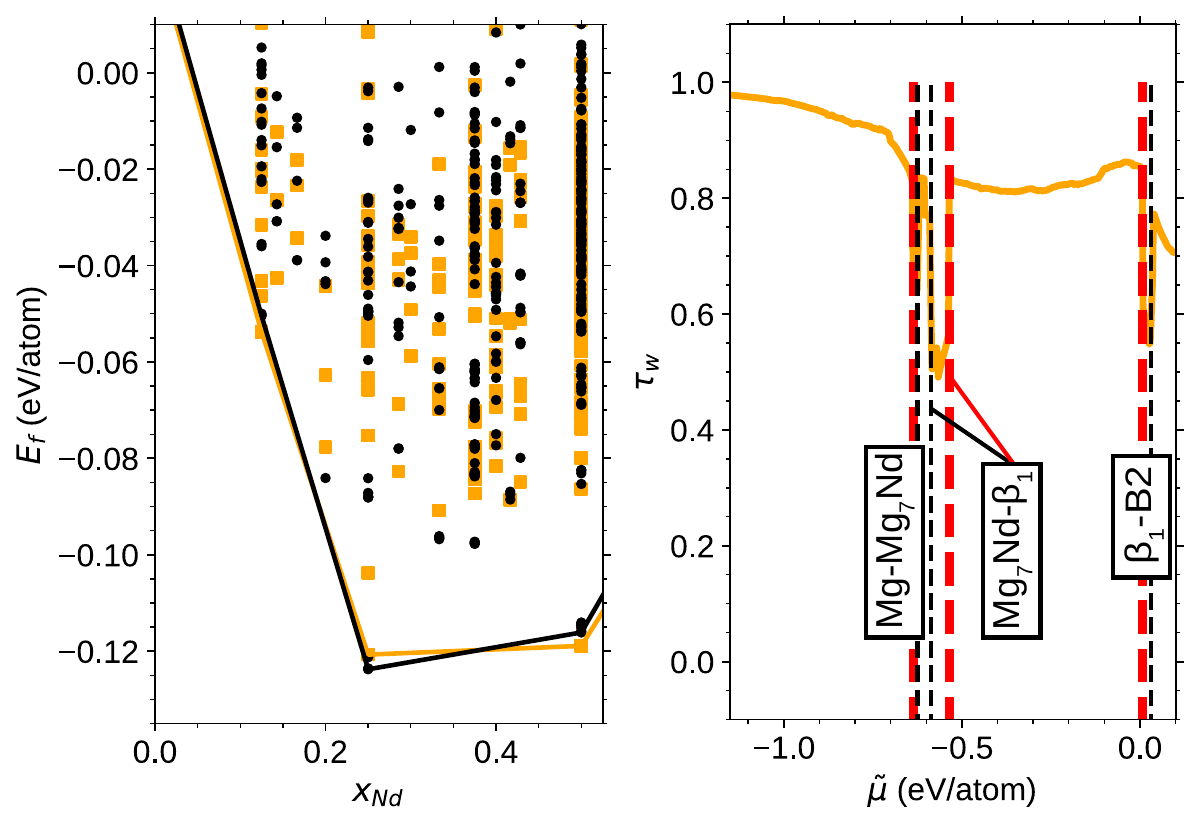}} \\
    \subfloat[][\label{fig:hcp_hull}]{\includegraphics[width=\columnwidth, valign=c]{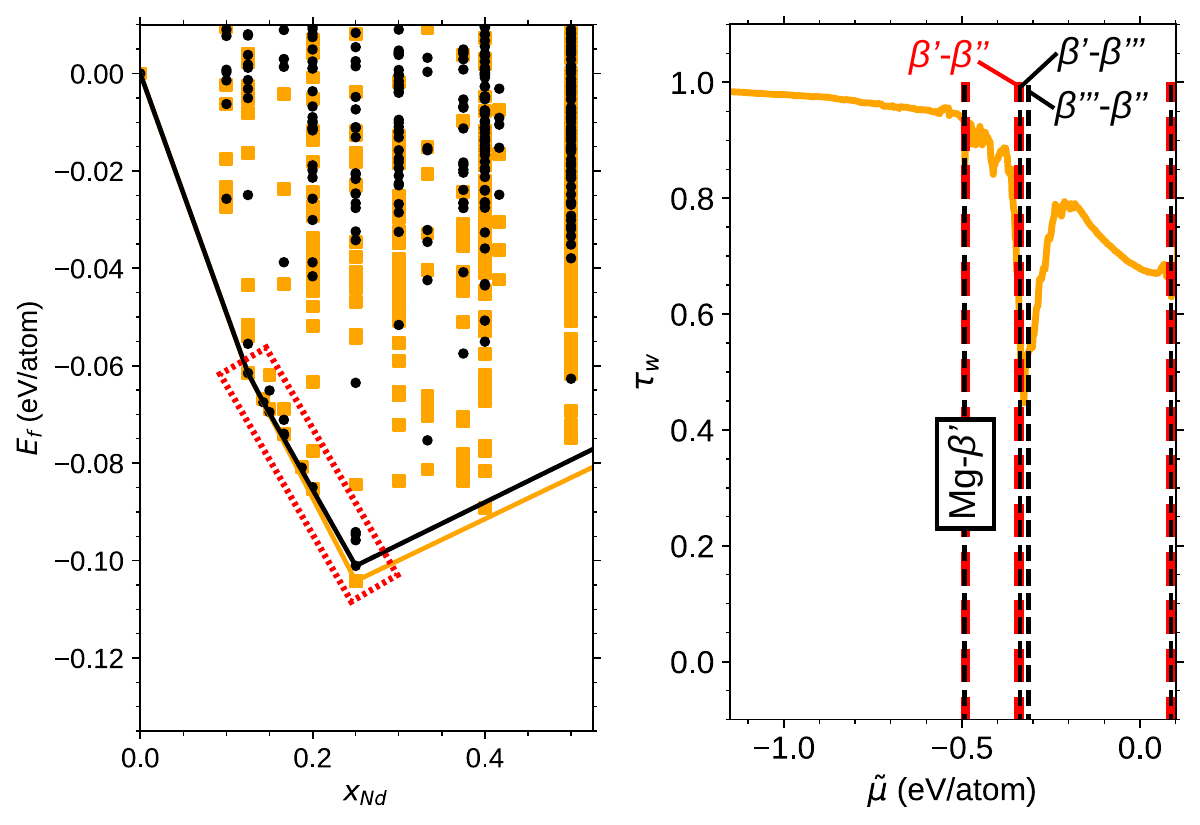}}
    \subfloat[][\label{fig:betappp_hull}]{\includegraphics[width=0.5\columnwidth, valign=c]{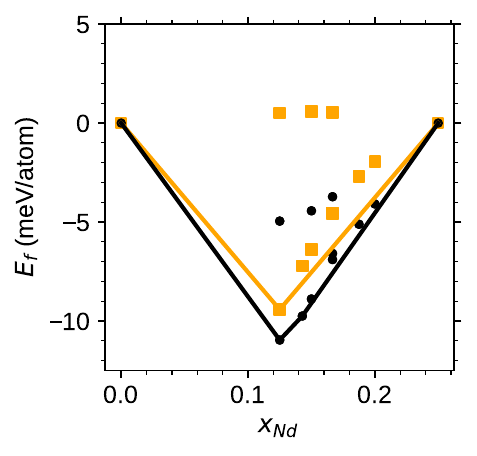}}
    \caption{Convex hulls and $\tau_w$ metrics for (a) the global system, (b) the bcc phase, and (c) the hcp phase. The weighting scheme is additive in (a) and multiplicative in (b, c). Vertical dashed lines indicate coexistence chemical potentials predicted by the MLIP (red) and DFT (grey). (d) Energies of of $\beta'''$ orderings (dotted box in (c)). Energies are referenced to pure Mg and $\beta''$ ($x_{\text{Nd}} = 0.25$).}
\end{figure*}

We assessed phase stability by examining zero-Kelvin convex hulls for Nd compositions $0 \leq x_{\text{Nd}} \leq 0.5$ and analyzing the weighted Kendall-$\tau$ trace as a function of the neodymium exchange chemical potential ($\tilde{\mu}=\mu_{\text{Nd}}-\mu_{\text{Mg}}$). \Cref{fig:global_hull,fig:bcc_hull,fig:hcp_hull} compare MLIP formation energies with DFT values. The $\tau_w$--$\tilde{\mu}$ trace computed with additive weights over the formation energies of all structures in \cref{fig:global_hull} remains close to 1 up to $\tilde{\mu} \approx -0.4$ eV/atom, the coexistence chemical potential between the $\beta_1$ and $\beta_{e}$ phases. This $\tau_{w}$ value indicates that the potential accurately captures stable and metastable phases up to $x_{\text{Nd}} = 0.25$. Near this coexistence chemical potential, the metric drops due to a spurious ground state at $x_{\text{Nd}} = 1/3$. This incorrectly predicted ground state is an ordering of Mg and Nd over the $\omega$ parent structure. DFT places this phase $\approx 25$ meV/atom above the convex hull. This discrepancy likely stems from parameterization choices, such as basis set size. While this error precludes using the potential for $x_{\text{Nd}} > 0.25$, it does not affect the study of precipitation in Mg-rich alloys that contain less than 1\% of neodymium.

The MLIP also reproduces metastable convex hulls for chemical decorations of bcc and hcp parent crystal structures. In \cref{fig:bcc_hull}, the multiplicative $\tau_w$ remains near 1, with deviations arising from mismatches in predicted coexistence chemical potentials. The metastable hcp convex hull in \cref{fig:hcp_hull} exhibits a more complex $\tau_w$--$\tilde{\mu}$ trace. Also in this case the weighting function is multiplicative. The metric remains near 1 for exchange chemical potentials up to $\approx -0.5$ eV/atom, corresponding to the hcp-$\beta^{\prime}$ coexistence. At higher exchange chemical potentials, the $\tau_w$ declines. As shown in \cref{fig:betappp_hull}, the potential fails to predict the $\beta^{\prime\prime\prime}$ orderings on the convex hull. Capturing these long-period orderings is challenging due to small energy differences ($\approx 2$ meV/atom), the influence of long-range strain interactions, and inherent DFT errors. However, the overall discrepancy is small, and the potential remains sufficiently accurate to model the precipitation process. Finally, for chemical potentials $\tilde{\mu} > -0.35$ eV/atom, the $\tau_w$ degrades due to incorrect phase stability predictions at $x_{\text{Nd}} > 0.25$.

\begin{figure}
    \centering
    \includegraphics[width=0.6\columnwidth]{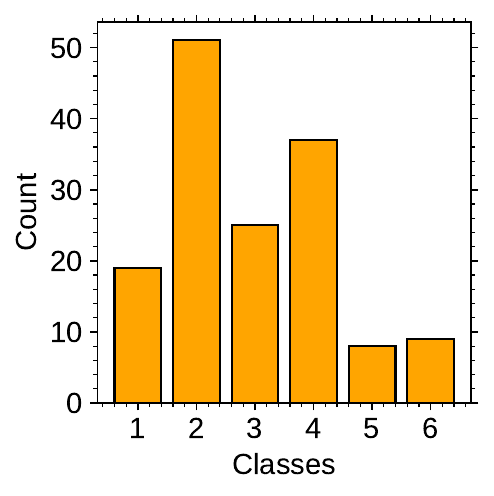}
    \caption{Number of pathways assigned to each class defined in \cref{fig:pathway_classes}. Classification is based on the DFT energy profiles.}
    \label{fig:transformation_pathways_data}
\end{figure}

Having achieved sufficient accuracy to model the low-temperature thermodynamics of Mg-rich Mg-Nd alloys with the MLIP, we assessed its reproduction of the hcp-to-bcc Burgers transformation using symmetrically distinct transformation pathways enumerated with the algorithm outlined in \cref{sect:enumerating_paths}. Our algorithm enumerated 149 symmetrically distinct pathways. The dataset includes all pathways obtained from symmetrically distinct chemical decorations in supercells up to twice the size of the primitive hcp cell, orderings with Nd compositions of 1/8 and 1/4 in supercells containing 6 and 8 atoms, and specific orderings in supercells containing 16, 24, and 96 atoms. The larger supercells capture structural phase transitions between precursor orderings on hcp and $\beta_1$. As shown in \cref{fig:transformation_pathways_data}, our systematic enumeration produces all six classes of transformation energy landscapes defined in \cref{fig:pathway_classes}.

\begin{figure}[!ht]
    \centering
    \includegraphics[width=0.7\columnwidth]{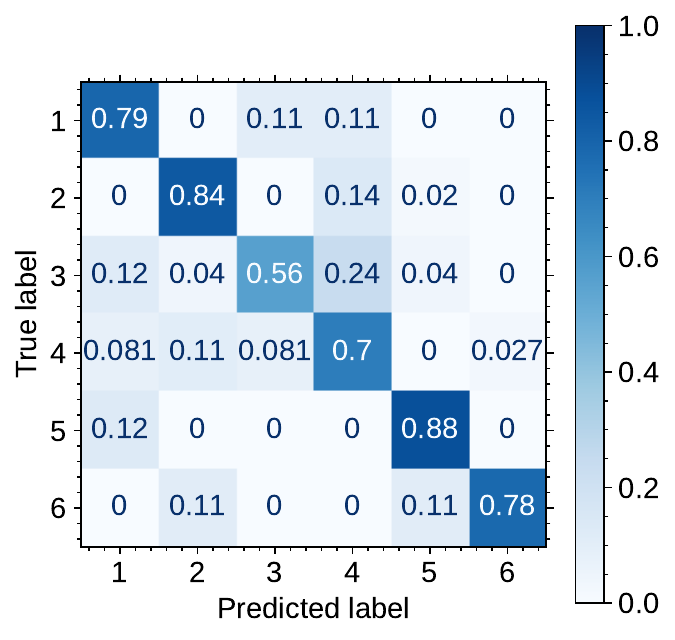}
    \caption{Confusion matrix comparing MLIP and DFT predictions for the qualitative shapes of transformation pathways. Pathway classes correspond to those defined in \cref{fig:pathway_classes}.}
    \label{fig:conf_mat}
\end{figure}

\Cref{fig:conf_mat} shows the confusion matrix comparing DFT and MLIP classifications. Energies of structures along transformation pathways connecting hcp and bcc orderings spanned 10--50 meV/atom. Misclassifications occur most frequently in pathways with barriers (classes 3--4). In both classes, the MLIP erroneously predicts $\approx 10\%$ of pathways to lack a barrier. End-state stability inversion is another common error, causing 24\% of class 3 pathways to be mislabeled as class 4. Further analysis revealed that pathways with inverted end-state energies have an average endpoint energy difference of $\approx 4.5$ meV/atom, well within the potential's validation error.

\begin{figure*}[!ht]
    \includegraphics[width=0.8\linewidth]{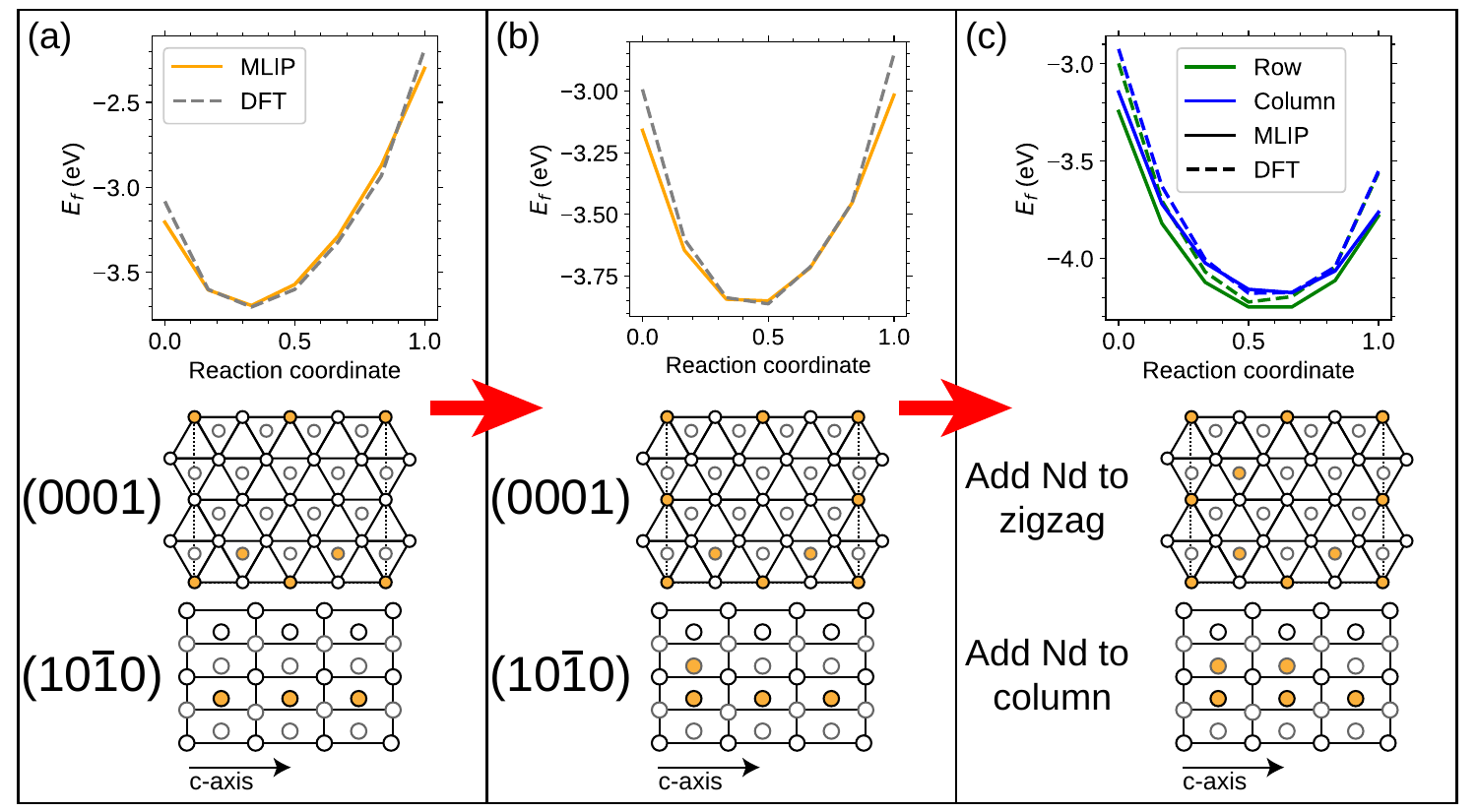}
    \caption{Schematic of the initial steps of $h$-$\beta_1$ nucleation starting from $\beta'$. Energy profiles represent Burgers pathways from the depicted hcp ordering to the lowest-energy bcc state. Crystal structures shown in (a) (no Nd added) and (b) (one Nd addition), the basal and prismatic projections correspond to the same structure. In (c) (two Nd additions), the basal projection shows Nd addition along a zigzag path, while the prismatic projection illustrates addition along a column.}
    \label{fig:row_column_growth_summary}
\end{figure*}
Modeling precipitate growth requires accurate relative energies for point and extended defects within precipitates. For instance, the $\beta^\prime$ phase transforms into $h$-$\beta_1$ by replacing some Mg atoms with Nd as shown in \cref{fig:burgers_transformation}. \Cref{fig:row_column_growth_summary} depicts two substitution mechanisms that progressively approach the $h$-$\beta_1$ ordering. \Cref{fig:row_column_growth_summary}a plots the energy profile for the structural transformation from $\beta^{\prime}$ to its bcc counterpart, showing close agreement between MLIP and DFT. Substituting a Nd atom at a site that would transform the $\beta^{\prime}$ ordering toward $h$-$\beta_1$ shifts the energy minimum toward bcc, as illustrated in \cref{fig:row_column_growth_summary}b. We also computed transformation energies for orderings obtained by adding a second Nd atom along either the hcp $c$-axis or the $[10\overline{1}0]$ direction. As shown in \cref{fig:row_column_growth_summary}c, both DFT and the MLIP indicate that addition along the $[10\overline{1}0]$ direction is energetically favorable.

\begin{figure*}[!ht]
    \centering
    \subfloat[][\label{fig:NdNd_int}]{\includegraphics[width=0.7\columnwidth]{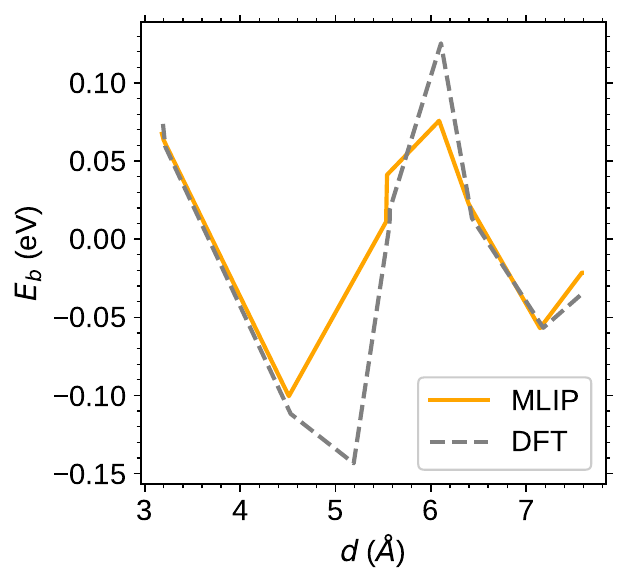}}
    \subfloat[][\label{fig:SSF_int}]{\includegraphics[width=0.7\columnwidth]{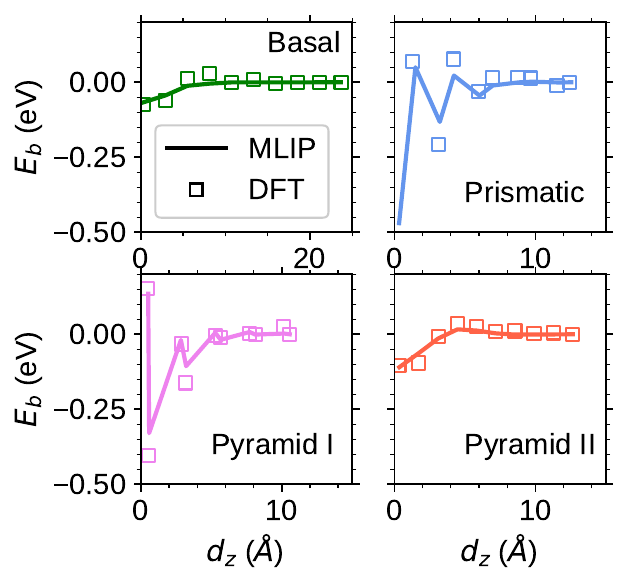}}
    \subfloat[][\label{fig:GB_int}]{\includegraphics[width=0.7\columnwidth]{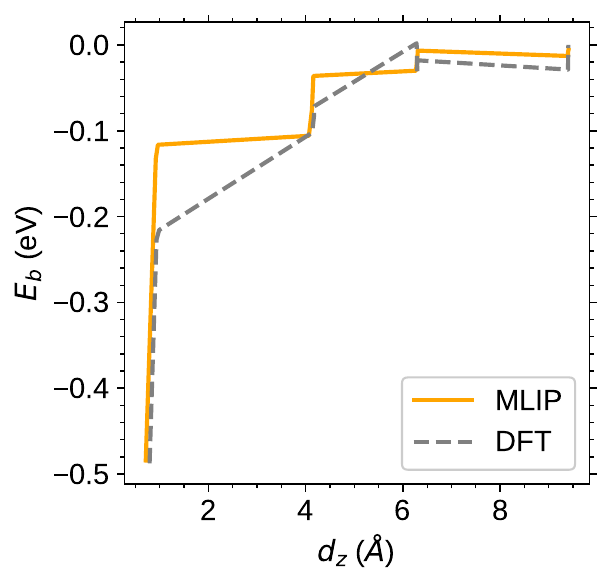}}
    \caption{Comparison of DFT and MLIP energies for (a) Nd-Nd binding, (b) Nd-SSF binding, and (c) Nd-$\Sigma$ grain boundary interaction.}
\end{figure*}

In addition to capturing defect energies within precipitates, the MLIP must also reproduce defect energetics within the disordered matrix, including point defects such as dilute Nd in hcp-Mg and Nd segregation to planar defects like stacking faults and grain boundaries. DFT predicts the Nd dissolution energy in a $4\times 4\times 3$ hcp supercell to be -34 meV relative to pure reference states, while the MLIP predicts -41 meV. The MLIP also accurately describes Nd-Nd and Nd-defect interactions. \Cref{fig:NdNd_int} shows that Nd-Nd binding energies within hcp-Mg agree with DFT. Similarly, binding energies to stable stacking faults (\cref{fig:SSF_int}) and the $\Sigma$ grain boundary (\cref{fig:GB_int}) match DFT predictions.

\begin{figure*}[!ht]
    \centering
    \subfloat[][\label{vacancy_structures}]{\includegraphics[width=0.9\columnwidth]{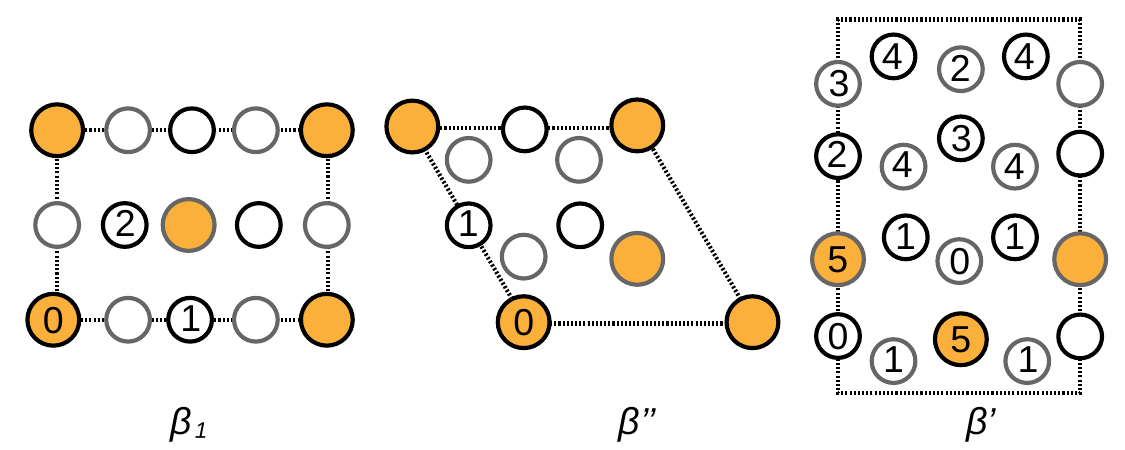}}
    \subfloat[][\label{beta1_betapp}]{
        \begin{tabular}[b]{|c|c|c c|}
            \hline
            Phase                      & Vacancy & DFT (eV)  & MLIP (eV) \\ \hline
            \multirow{3}{*}{$\beta_1$} & 0       & 0.63 & 0.47 \\
                                       & 1       & 1.84 & 1.61 \\
                                       & 2       & 0.92 & 0.72 \\ \hline
            \multirow{2}{*}{$\beta''$} & 0       & 1.20 & 1.22 \\
                                       & 1       & 1.69 & 1.58 \\ \hline
        \end{tabular}
    }
    \subfloat[][\label{betap}]{
        \begin{tabular}[b]{|c|c|c c|}
            \hline
            Phase                     & Vacancy & DFT (eV)  & MLIP (eV) \\ \hline
            \multirow{6}{*}{$\beta'$} & 0       & 1.53 & 1.29 \\
                                      & 1       & 1.12 & 1.04 \\
                                      & 2       & 1.17 & 1.05 \\
                                      & 3       & 0.51 & 0.41 \\
                                      & 4       & 0.51 & 0.41 \\
                                      & 5       & 1.66 & 1.57 \\ \hline
        \end{tabular}
    }
    \caption{(a) Vacancy sites in the $\beta_1$, $\beta''$, and $\beta'$ structures. (b)--(c) Comparison of vacancy formation energies computed via DFT and MLIP.}
    \label{fig:vacancy_comparison}
\end{figure*}

\begin{figure}[!ht]
    \centering
    \includegraphics[width=0.7\columnwidth]{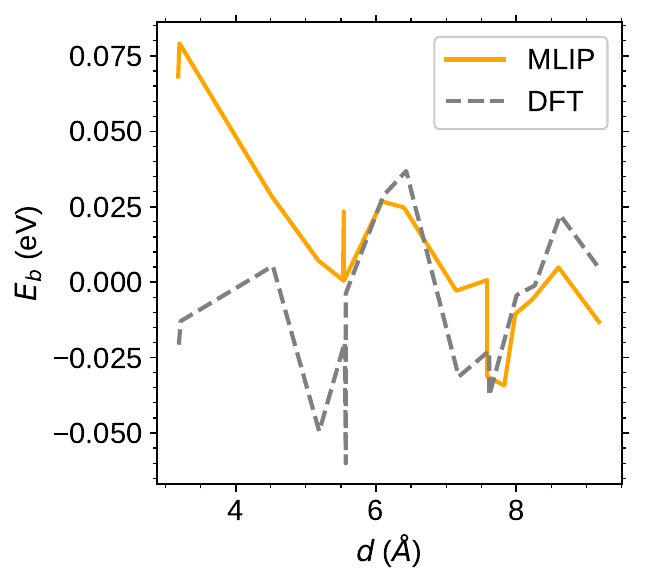}
    \caption{Comparison of Nd-vacancy binding energies in hcp-Mg calculated via MLIP and DFT.}
    \label{fig:Nd_vac_binding}
\end{figure}

The MLIP also accurately captures vacancy defect energetics, reproducing vacancy formation energies in the relevant ground states (\cref{fig:vacancy_comparison}). Capturing the Nd-vacancy binding energy in the hcp-Mg matrix proved more difficult. As shown in \cref{fig:Nd_vac_binding}, DFT predicts a slight attraction at short distances, whereas the MLIP predicts a slight repulsion. This discrepancy persisted despite increased weighting during training. Consequently, caution is warranted when using this potential to compute kinetic parameters for dilute Nd-containing alloys.

\begin{figure*}[!ht]
    \centering
    \subfloat[][\label{fig:precipitate_growth_schematic}]{\includegraphics[width=\columnwidth]{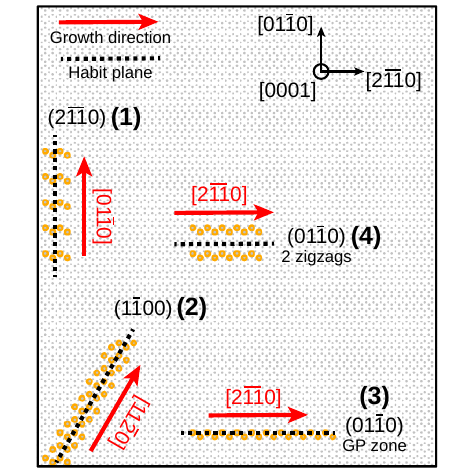}}
    \subfloat[][\label{fig:precipitate_assembly}]{\includegraphics[width=\columnwidth]{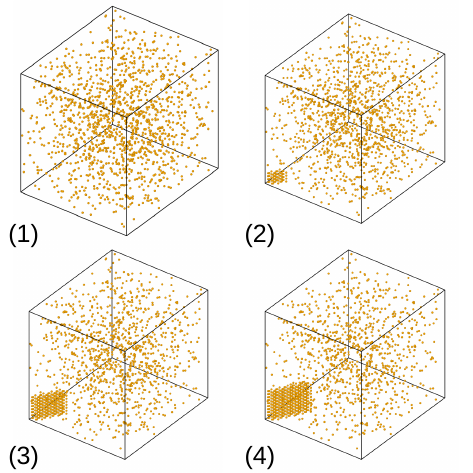}}\\
    \subfloat[][\label{fig:precipitation_DFT_check}]{\includegraphics[width=0.35\linewidth]{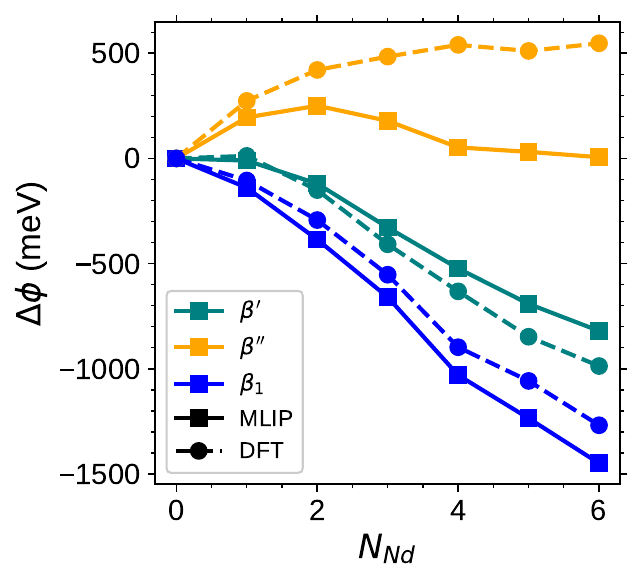}}
    \subfloat[][\label{fig:precipitate_formation_energy change}]{\includegraphics[width=0.32\linewidth]{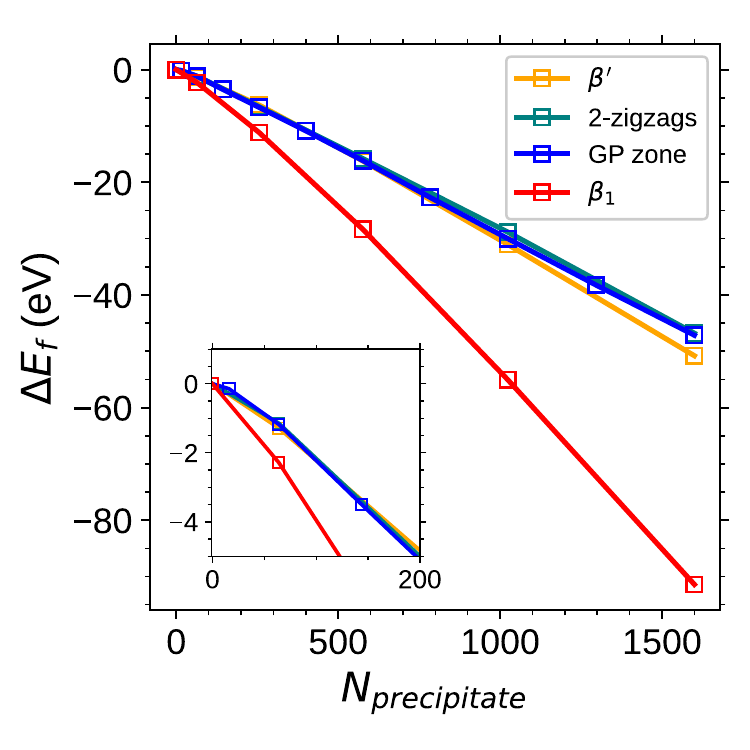}}
    \subfloat[][\label{fig:stein_precipitate}]{\includegraphics[width=0.32\linewidth]{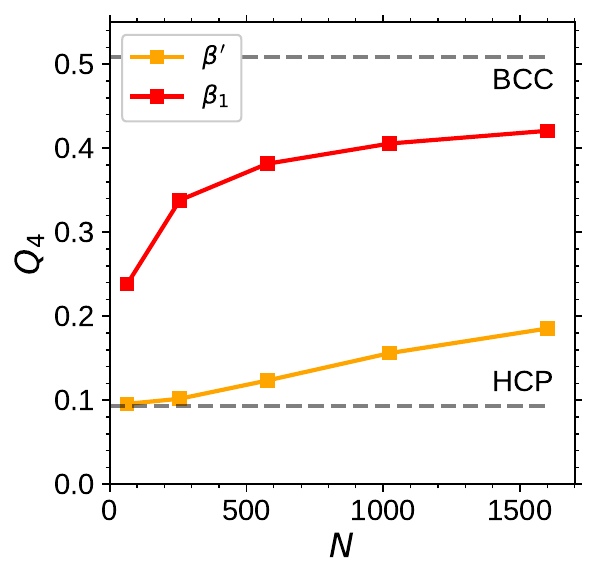}}
    \caption{(a) Schematic of the four precipitate types and their orientations within a disordered matrix. (b) Snapshots of $\beta_1$ formation within a disordered matrix phase. Only Nd atoms are shown. (c) Comparison of MLIP and DFT nucleation energies for small precipitates. (d) Nucleation energies for precipitate formation from the disordered matrix. (e) Average $Q_4$ order parameter as a function of precipitate size.}
\end{figure*}

\section{Discussion} \label{sect:discussion}
Precipitation hardening is widely used in engineering alloys to improve material strength. This study presents methods for parameterizing reliable MLIPs to simulate precipitation and structural phase transitions in metallic alloys. We introduce error metrics and data generation schemes for MLIPs that model solid-state phase transformations during precipitation. We demonstrate these methods by parameterizing an MLIP for the Mg-Nd binary alloy that captures stable and metastable thermodynamics, elastic moduli, point defect energetics, and vacancy formation energies. The weighted Kendall-$\tau$ ($\tau_w$) and its semi-grand canonical generalization, illustrated in \cref{fig:kendall_tau,fig:gc_kendall_tau_summary,fig:gc_kendall_tau}, provide interpretable metrics for scoring an MLIP's prediction of low-temperature thermodynamics in multi-component systems. We can compute $\tau_{w}$ using several weighting schemes, though the relative merits of different schemes require further exploration. The algorithm for enumerating symmetry-distinct transformation pathways complements the phase stability analysis by mapping the potential energy surface along pathways connecting chemical orderings on different parent crystal structures. Although we demonstrated this algorithm for the Mg-Nd system using the Burgers transformation linking hcp and bcc orderings, it applies to any well-defined pathway connecting different crystal structures. This methodology enables error quantification through a confusion matrix as shown in \cref{fig:conf_mat}.

Systematic enumeration of crystal structures, combined with careful fine-tuning of the MLIP based on the error metrics developed here, produces an interatomic potential that generalizes well to data not included during training. We tested the MLIP on several interface and kinetic properties. These included planar fault energies such as anti-phase boundary energies in $\beta^{\prime}$, generalized stacking fault energies in $\beta^{\prime}$, $\beta^{\prime\prime}$ and $\beta_1$, and the energies of several interfaces between hcp-Mg, $\beta^{\prime}$ and $\beta_1$. We analyzed vacancy migration barriers for symmetrically-distinct vacancy hops within $\beta^{\prime}$ and for a vacancy adjacent to a Nd atom in otherwise pure hcp-Mg. \Cref{supp-fig:gamma_surfaces,supp-fig:solute_Mg_hops,supp-fig:Nd_vac_exchange,supp-fig:kra,supp-fig:vacancy_hops_betap,supp-tab:interface_energies} show comparisons between DFT and the MLIP. The MLIP matches our DFT results and literature values \cite{guo_anti-phase_2021, dewitt_misfit-driven_2017, choudhuri_interfacial_2017, agarwal_exact_2017} across all these properties, indicating that the potential is well parameterized and captures features not explicitly included in training. The precise origin of this transferability remains unclear and warrants further investigation.

The MLIP also reproduces precipitate energies across varying sizes when benchmarked against DFT. We computed the energies of small nuclei within a 384-atom hcp supercell, corresponding to a $3\times2\times4$ supercell of the $\beta^{\prime}$ crystal structure. Simulations included the formation of $\beta^{\prime}$ within hcp-Mg, as well as $\beta^{\prime\prime}$ and $\beta_1$ embedded within $\beta^{\prime}$. Precipitates of varying sizes were generated by systematically adding Nd atoms one at a time to the lowest-energy site while accounting for atomic relaxations. We utilized the MLIP to identify the site with the lowest Nd substitution energy, after which the energy was computed with DFT. Only the sublattice relevant to the precipitating phase was filled, with precipitate size restricted to at most 1/8 of the supercell.

\Cref{fig:precipitation_DFT_check} compares nucleation energies predicted by the MLIP against DFT for the formation of three precipitates. We approximate the nucleation energy at 0K as:
\begin{equation}
    \Delta \phi = E_f(N_{Nd}) - E_f(0) - N_{Nd} \tilde{\mu}
\end{equation}
where $E_f(N_{Nd})$ is the formation energy of the supercell after adding $N_{Nd}$ solute atoms and $\tilde{\mu}$ is the difference in chemical potential between Nd and Mg. We set the chemical potential to the dissolution energy of a single Nd atom within a hcp matrix of Mg.

The MLIP predictions are in excellent agreement with the DFT calculations shown in \cref{fig:precipitation_DFT_check}. At the chemical potential and the small simulation cell sizes used in \cref{fig:precipitation_DFT_check}, the formation of $\beta_1$ within $\beta^{\prime}$ is thermodynamically favored over that of $\beta^{\prime\prime}$. This agrees with experimental results from several studies \cite{solomon_early_2017, liu_structure_2017, meier_stimulating_2025}. The MLIP predictions match the DFT results even though the comparison is made in the larger supercell. The energy differences between DFT and MLIP values are $< 2$ meV/atom, suggesting that the accuracy of this MLIP for modeling precipitation is better than what the validation errors would indicate.

Having established that the MLIP captures precipitate formation energies, we used the atomistic model to perform a preliminary study of precipitation in a dilute Mg-Nd alloy. \Cref{fig:precipitate_growth_schematic} summarizes the precipitate geometries considered. Previous experiments \cite{nie_precipitation_2012, solomon_early_2017, natarajan_early_2016, liu_structure_2017} and phase field simulations \cite{liu_simulation_2014-1, dewitt_misfit-driven_2017} indicate that precipitates in Mg-Nd alloys form as flat plates with habit planes of $\{1\bar{2}10\}_{hcp}$ for $\beta'$ and $\{1\bar{1}00\}_{hcp}$ for $\beta_1$. GP zones typically form through Nd atoms arranged in a zigzag pattern with a habit plane of $\{01\bar{1}0\}_{hcp}$ \cite{natarajan_early_2016}. We also studied the formation of precipitates with two rows of Nd atoms arranged as zigzags with a habit plane of $\{01\bar{1}0\}_{hcp}$ and spacing matching that of $\beta^{\prime}$. This morphology has not been observed experimentally. In \cref{fig:precipitate_growth_schematic}, these geometries are labeled precipitates 1 through 4, corresponding to $\beta'$, $\beta_1$, Guinier-Preston zones, and the double zigzag structure, respectively.

In principle, the energies of precipitates of all possible shapes at a fixed size should be computed to assess nucleation energies. We use insights from experiments and previous simulations to select specific precipitate aspect ratios. Following the phase field simulations in \cite{dewitt_misfit-driven_2017} for $\beta'$ and \cite{liu_simulation_2014} for $\beta_1$, we set the ratio between the height (along $[0001]_{hcp}$) and the length (along the growth direction indicated by the red arrow in \cref{fig:precipitate_growth_schematic}) to approximately 1. Experimental data at 15 minutes of aging \cite{dewitt_misfit-driven_2017} suggests a $\beta'$ plate thickness (normal to the habit plane) of $\approx$10\AA, corresponding to 4 Nd atoms in the zigzag. Due to the lack of experimental data, we imposed the same thickness on the $\beta_1$ plate. We based the GP zone and 2-zigzag precipitate geometries on crystallographic information described in \cite{nie_precipitation_2012, natarajan_early_2016}. The modeled precipitation mechanism involves simultaneous size increase along $[0001]_{hcp}$ and the growth direction (red arrow in \cref{fig:precipitate_growth_schematic}) while keeping the plate thickness fixed.

We assessed precipitate formation energetics by computing the energy difference required to form a coherent precipitate from a 0.5 at.\% Nd solid solution. The initial state consisted of a large orthorhombic simulation box, $173\times199\times186$ \AA, containing randomly distributed Nd atoms at the target composition. Precipitates were created by extracting solute atoms from the disordered matrix and rearranging them to match the shape, arrangement, and orientation of one of the precipitates shown in \cref{fig:precipitate_growth_schematic}. The precipitate-containing cells were then relaxed to track the variation in system energy as a function of the number of Nd atoms in the precipitate. All precipitates were assumed to be stoichiometric and coherent. \Cref{fig:precipitate_assembly} illustrates the assembly process for the $\beta_1$ precipitate.

\Cref{fig:precipitate_formation_energy change} shows the change in energy of the simulation cell as precipitate size increases. Nucleation of $\beta_1$ produces the largest energy reduction, as it is more stable than $\beta^{\prime}$ or other orderings on hcp. Growth of $\beta'$ along the $[01\bar{1}0]_{hcp}$ direction is favored over the GP zone and 2-zigzag precipitate, consistent with experimental observations. The energy reduction for the formation of $\beta'$, the GP zone, and the 2-zigzag structure are similar at small sizes, suggesting competition among these ordered precipitates during early aging. As precipitates grow, however, the energy of the experimentally observed coherent $\beta^{\prime}$ precipitate decreases more rapidly, making it more favorable than other ordered phases on hcp.

The predictions of \cref{fig:precipitate_formation_energy change} show no nucleation barrier. This likely results from the large thermodynamic driving force for precipitation at 0K. The absence of finite-temperature contributions significantly destabilizes the disordered phase, causing ordered precipitates to form without any barrier. For a sufficiently dilute alloy where solute interactions are negligible, the chemical potential of the random solid solution can be approximated by the dissolution energy of the solute. Within this approximation, the chemical potential of the disordered Mg-Nd matrix, $\tilde{\mu}_{SS}$, is -0.063 eV/atom. This value is much higher than the coexistence chemical potential between hcp-Mg and $\beta'$ or hcp-Mg and $\beta_1$ which is predicted to be $\approx$ -0.49 eV/atom. As a result the driving force for nucleation is very large and completely overcomes any interfacial or strain energy penalties. Further study is needed to determine the effects of temperature on the presence (or absence) of a nucleation barrier in this alloy system.

Steinhardt bond order parameters \cite{steinhardt_bond-orientational_1983} can quantify the extent of structural phase transformation when precipitates such as $\beta_1$ form within a matrix phase of a different crystal structure. \Cref{fig:stein_precipitate} shows the $Q_4$ order parameter, averaged over atoms in the precipitate, as a function of the size of $\beta'$ and $\beta_1$. Horizontal dashed lines indicate ideal order parameter values for perfect hcp and bcc.

For $\beta^{\prime}$, the $Q_4$ value at small precipitate sizes is close to the ideal hcp value but progressively deviates from hcp as precipitate size increases. This deviation arises from the large stress-free transformation strains accompanying $\beta'$ formation, which have been predicted to be as high as $\approx12\%$ along the $[2\bar{1}\bar{1}0]_{hcp}$ direction \cite{natarajan_early_2016,natarajan_2017_unifieddescription}. As $\beta^{\prime}$ precipitates grow, the energy gained by relaxing atoms away from perfect hcp positions offsets the energy penalties from straining the surrounding Mg matrix.

The $\beta_{1}$ precipitates in \cref{fig:stein_precipitate} exhibit a different trend. Even at small sizes, atoms within the $\beta_{1}$ precipitate relax away from their ideal hcp positions, resulting in structural order parameters that deviate from hcp more significantly than those of $\beta^{\prime}$. As the precipitate grows, the average order parameter for $\beta_1$ gradually shifts toward bcc, indicating a continuous phase transition that transforms the underlying hcp crystal structure of the $h$-$\beta_1$ ordering into a bcc structure resembling $\beta_1$. Achieving perfect bcc-like coordination requires larger precipitate sizes, as the energy gain from the structural phase transition must offset penalties from interface mismatch and the strain energy needed to maintain precipitate coherency.

\section{Conclusions}\label{sect:conclusions}
In this work, we present a strategy for parameterizing MLIPs to study precipitation and structural phase transitions in metallic alloys. The training and validation database was large and diverse, encompassing the crystalline environments required to capture the features of the Mg-Nd system and to model nucleation. We supplemented conventional fitting procedures with tailored error metrics and data-generation schemes, including the weighted Kendall-$\tau$ coefficient and an algorithm to enumerate symmetry-distinct transformation pathways. The $\tau_w$ provides a measure of the MLIP's ability to predict the relative stability of different polymorphs and to optimize weighting schemes. Its semi-grand canonical formulation quantifies the MLIP's reliability in describing phase stability across the full composition range. Through this careful parameterization and assessment, the Mg-Nd MLIP generalized well to crystal structures absent from the training set, particularly interfaces and vacancy hop configurations. A preliminary study of $\beta'$ and $\beta_1$ precipitation showed that $\beta_1$ nucleation produces the largest energy reduction, with $\beta^{\prime}$ formation also lowering the total energy but to a lesser extent. These results agree well with previous experimental studies of precipitation in this alloy system. The methodologies developed here provide a general framework for studying precipitation in other alloy systems and are readily applicable to multi-component Al alloys \cite{sigli_recent_2018, bourgeois_transforming_2020, ding_orderdisorder_2025}, Ti alloys \cite{li_origin_2020, fu_atomic-scale_2022}, and high-entropy alloys \cite{muller_first-principles_2024}.

\section{Acknowledgments}
This research was supported by the Swiss National Science Foundation (grant number 215178). The authors also acknowledge access to Eiger at the Swiss National Supercomputing Centre with project ID mr30 under the NCCR MARVEL's share, a National Centre of Competence in Research funded by the Swiss National Science Foundation (grant number 205602). The authors acknowledge the use of \verb|pymatgen| \cite{ong_python_2013}, \verb|ase| \cite{hjorth_larsen_atomic_2017}, \verb|spglib| \cite{togo_spglib_2024}, and the implementation of the weighted Kendall-$\tau$ in \verb|scipy.stats.weightedtau| \cite{vigna_weighted_2015}.

\bibliographystyle{unsrt}
\bibliography{references}

\begin{thebibliography}{10}

\bibitem{starikov_atomic_2025}
Sergei Starikov, Yury Lysogorskiy, Minaam Qamar, Anton Bochkarev, Matous
  Mrovec, and Ralf Drautz.
\newblock Atomic cluster expansion for the aluminum-magnesium-hydrogen system.
\newblock {\em Physical Review Materials}, 9(10):103606, October 2025.

\bibitem{marchand_foundation_2025}
Daniel Marchand.
\newblock Foundation models for metallurgy?
\newblock {\em MRS Bulletin}, 50(7):805--818, July 2025.

\bibitem{srinivasan_atomic_2025}
Prashanth Srinivasan, Siddharth Puri, Karen Pacho~Dominguez, Andrew~P.
  Horsfield, Mark~R. Gilbert, and Duc Nguyen-Manh.
\newblock Atomic cluster expansion interatomic potentials for lithium:
  {Investigating} the solid and liquid phases.
\newblock {\em Physical Review B}, 112(5):054108, August 2025.

\bibitem{liu_discrepancies_2023}
Yunsheng Liu, Xingfeng He, and Yifei Mo.
\newblock Discrepancies and error evaluation metrics for machine learning
  interatomic potentials.
\newblock {\em npj Computational Materials}, 9(1):174, September 2023.

\bibitem{liu_assessing_2024}
Yunsheng Liu and Yifei Mo.
\newblock Assessing the accuracy of machine learning interatomic potentials in
  predicting the elemental orderings: {A} case study of {Li}-{Al} alloys.
\newblock {\em Acta Materialia}, 268:119742, April 2024.

\bibitem{nie_precipitation_2012}
Jian-Feng Nie.
\newblock Precipitation and {Hardening} in {Magnesium} {Alloys}.
\newblock {\em Metallurgical and Materials Transactions A}, 43(11):3891--3939,
  November 2012.

\bibitem{xu_shear_2014}
Z.~Xu, M.~Weyland, and J.F. Nie.
\newblock Shear transformation of coupled beta1/$\beta$' precipitates in
  {Mg}–{RE} alloys: {A} quantitative study by aberration corrected {STEM}.
\newblock {\em Acta Materialia}, 81:58--70, December 2014.

\bibitem{liu_structure_2017}
H.~Liu, Y.M. Zhu, N.C. Wilson, and J.F. Nie.
\newblock On the structure and role of $\beta$ {F}' in beta1 precipitation in
  {Mg}–{Nd} alloys.
\newblock {\em Acta Materialia}, 133:408--426, July 2017.

\bibitem{xie_diffusional-displacive_2021}
Hongbo Xie, Xiaobo Zhao, Jingchun Jiang, Junyuan Bai, Shanshan Li, Hucheng Pan,
  Xueyong Pang, Hongxiao Li, Yuping Ren, and Gaowu Qin.
\newblock Diffusional-displacive transformation mechanism for the beta1
  precipitate in a model {Mg}-rare-earth alloy.
\newblock {\em Materials Characterization}, 174:111018, April 2021.

\bibitem{natarajan_early_2016}
Anirudh~Raju Natarajan, Ellen~L.S. Solomon, Brian Puchala, Emmanuelle~A.
  Marquis, and Anton Van Der~Ven.
\newblock On the early stages of precipitation in dilute {Mg}–{Nd} alloys.
\newblock {\em Acta Materialia}, 108:367--379, April 2016.

\bibitem{natarajan_2017_unifieddescription}
Anirudh~Raju Natarajan and Anton {Van der Ven}.
\newblock A unified description of ordering in {{HCP Mg-RE}} alloys.
\newblock {\em Acta Materialia}, 124:620--632, 2017.

\bibitem{natarajan_2019_understandingdeformation}
Anirudh~Raju Natarajan and Anton {Van der Ven}.
\newblock Toward an {{Understanding}} of {{Deformation Mechanisms}} in
  {{Metallic Lithium}} and {{Sodium}} from {{First-Principles}}.
\newblock {\em Chemistry of Materials}, 31(19):8222--8229, 2019.

\bibitem{supp_info}
Lorenzo Piersante and Anirudh~Raju Natarajan.
\newblock See supplementary material at url-will-be-inserted-by-publisher for
  summary of crystallography, details concerning the ace mlip, in-depth
  description of the database and training dataset, and supplementary results
  on unary and binary potentials.

\bibitem{drautz_atomic_2019}
Ralf Drautz.
\newblock Atomic cluster expansion for accurate and transferable interatomic
  potentials.
\newblock {\em Physical Review B}, 99(1):014104, January 2019.

\bibitem{lysogorskiy_performant_2021}
Yury Lysogorskiy, Cas Van~Der Oord, Anton Bochkarev, Sarath Menon, Matteo
  Rinaldi, Thomas Hammerschmidt, Matous Mrovec, Aidan Thompson, Gábor Csányi,
  Christoph Ortner, and Ralf Drautz.
\newblock Performant implementation of the atomic cluster expansion ({PACE})
  and application to copper and silicon.
\newblock {\em npj Computational Materials}, 7(1):97, June 2021.

\bibitem{bochkarev_efficient_2022}
Anton Bochkarev, Yury Lysogorskiy, Sarath Menon, Minaam Qamar, Matous Mrovec,
  and Ralf Drautz.
\newblock Efficient parametrization of the atomic cluster expansion.
\newblock {\em Physical Review Materials}, 6(1):013804, January 2022.

\bibitem{ibrahim_atomic_2023}
Eslam Ibrahim, Yury Lysogorskiy, Matous Mrovec, and Ralf Drautz.
\newblock Atomic {Cluster} {Expansion} for a {General}-{Purpose} {Interatomic}
  {Potential} of {Magnesium}, May 2023.
\newblock arXiv:2305.03577 [cond-mat].

\bibitem{mahalanobis_reprint_2018}
P.~C. Mahalanobis.
\newblock Reprint of: {Mahalanobis}, {P}.{C}. (1936) "{On} the {Generalised}
  {Distance} in {Statistics}.".
\newblock {\em Sankhya A}, 80, January 2018.

\bibitem{kresse_ab_1994}
G.~Kresse and J.~Hafner.
\newblock \textit{{Ab} initio} molecular-dynamics simulation of the
  liquid-metal–amorphous-semiconductor transition in germanium.
\newblock {\em Physical Review B}, 49(20):14251--14269, May 1994.

\bibitem{kresse_ab_1995}
G~Kresse.
\newblock Ab initio molecular dynamics for liquid metals.
\newblock 1995.

\bibitem{kresse_efficiency_1996}
G.~Kresse and J.~Furthmüller.
\newblock Efficiency of ab-initio total energy calculations for metals and
  semiconductors using a plane-wave basis set.
\newblock {\em Computational Materials Science}, 6(1):15--50, July 1996.

\bibitem{kresse_ultrasoft_1999}
G.~Kresse and D.~Joubert.
\newblock From ultrasoft pseudopotentials to the projector augmented-wave
  method.
\newblock {\em Physical Review B}, 59(3):1758--1775, January 1999.

\bibitem{perdew_generalized_1996}
John~P. Perdew, Kieron Burke, and Matthias Ernzerhof.
\newblock Generalized {Gradient} {Approximation} {Made} {Simple}.
\newblock {\em Physical Review Letters}, 77(18):3865--3868, October 1996.

\bibitem{perdew_generalized_1997}
John~P. Perdew, Kieron Burke, and Matthias Ernzerhof.
\newblock Generalized {Gradient} {Approximation} {Made} {Simple} [{Phys}.
  {Rev}. {Lett}. 77, 3865 (1996)].
\newblock {\em Physical Review Letters}, 78(7):1396--1396, February 1997.

\bibitem{kobayashi_neural_2017}
Ryo Kobayashi, Daniele Giofré, Till Junge, Michele Ceriotti, and William~A.
  Curtin.
\newblock Neural network potential for {Al}-{Mg}-{Si} alloys.
\newblock {\em Physical Review Materials}, 1(5):053604, October 2017.

\bibitem{maresca_screw_2018}
Francesco Maresca, Daniele Dragoni, Gábor Csányi, Nicola Marzari, and
  William~A. Curtin.
\newblock Screw dislocation structure and mobility in body centered cubic {Fe}
  predicted by a {Gaussian} {Approximation} {Potential}.
\newblock {\em npj Computational Materials}, 4(1):69, December 2018.

\bibitem{marchand_machine_2020}
Daniel Marchand, Abhinav Jain, Albert Glensk, and W.~A. Curtin.
\newblock Machine learning for metallurgy {I}. {A} neural-network potential for
  {Al}-{Cu}.
\newblock {\em Physical Review Materials}, 4(10):103601, October 2020.

\bibitem{stricker_machine_2020}
Markus Stricker, Binglun Yin, Eleanor Mak, and W.~A. Curtin.
\newblock Machine learning for metallurgy {II}. {A} neural-network potential
  for magnesium.
\newblock {\em Physical Review Materials}, 4(10):103602, October 2020.

\bibitem{kolli_discovering_2020}
Sanjeev~Krishna Kolli, Anirudh~Raju Natarajan, John~C. Thomas, Tresa~M.
  Pollock, and Anton Van Der~Ven.
\newblock Discovering hierarchies among intermetallic crystal structures.
\newblock {\em Physical Review Materials}, 4(11):113604, November 2020.

\bibitem{thomas_exploration_2017}
John~C. Thomas and Anton Van Der~Ven.
\newblock The exploration of nonlinear elasticity and its efficient
  parameterization for crystalline materials.
\newblock {\em Journal of the Mechanics and Physics of Solids}, 107:76--95,
  October 2017.

\bibitem{tran_anisotropic_2019}
Richard Tran, Xiang-Guo Li, Joseph~H. Montoya, Donald Winston, Kristin~Aslaug
  Persson, and Shyue~Ping Ong.
\newblock Anisotropic work function of elemental crystals.
\newblock {\em Surface Science}, 687:48--55, September 2019.

\bibitem{zagorac_recent_2019}
D.~Zagorac, H.~Müller, S.~Ruehl, J.~Zagorac, and S.~Rehme.
\newblock Recent developments in the {Inorganic} {Crystal} {Structure}
  {Database}: theoretical crystal structure data and related features.
\newblock {\em Journal of Applied Crystallography}, 52(5):918--925, October
  2019.

\bibitem{kirklin_open_2015}
Scott Kirklin, James~E Saal, Bryce Meredig, Alex Thompson, Jeff~W Doak,
  Muratahan Aykol, Stephan Rühl, and Chris Wolverton.
\newblock The {Open} {Quantum} {Materials} {Database} ({OQMD}): assessing the
  accuracy of {DFT} formation energies.
\newblock {\em npj Computational Materials}, 1(1):15010, December 2015.

\bibitem{puchala_casm_2023}
Brian Puchala, John~C. Thomas, Anirudh~Raju Natarajan, Jon~Gabriel Goiri,
  Sesha~Sai Behara, Jonas~L. Kaufman, and Anton Van Der~Ven.
\newblock {CASM} — {A} software package for first-principles based study of
  multicomponent crystalline solids.
\newblock {\em Computational Materials Science}, 217:111897, January 2023.

\bibitem{hart_algorithm_2008}
Gus L.~W. Hart and Rodney~W. Forcade.
\newblock Algorithm for generating derivative structures.
\newblock {\em Physical Review B}, 77(22):224115, June 2008.

\bibitem{hart_generating_2009}
Gus L.~W. Hart and Rodney~W. Forcade.
\newblock Generating derivative structures from multilattices: {Algorithm} and
  application to hcp alloys.
\newblock {\em Physical Review B}, 80(1):014120, July 2009.

\bibitem{kim_modified_2017}
Ki-Hyun Kim and Byeong-Joo Lee.
\newblock Modified embedded-atom method interatomic potentials for {Mg}-{Nd}
  and {Mg}-{Pb} binary systems.
\newblock {\em Calphad}, 57:55--61, June 2017.

\bibitem{thompson_lammps_2022}
Aidan~P. Thompson, H.~Metin Aktulga, Richard Berger, Dan~S. Bolintineanu,
  W.~Michael Brown, Paul~S. Crozier, Pieter~J. In~'T~Veld, Axel Kohlmeyer,
  Stan~G. Moore, Trung~Dac Nguyen, Ray Shan, Mark~J. Stevens, Julien Tranchida,
  Christian Trott, and Steven~J. Plimpton.
\newblock {LAMMPS} - a flexible simulation tool for particle-based materials
  modeling at the atomic, meso, and continuum scales.
\newblock {\em Computer Physics Communications}, 271:108171, February 2022.

\bibitem{puchala_thermodynamics_2013}
B.~Puchala and A.~Van Der~Ven.
\newblock Thermodynamics of the {Zr}-{O} system from first-principles
  calculations.
\newblock {\em Physical Review B}, 88(9):094108, September 2013.

\bibitem{huang_construction_2017}
Wenxuan Huang, Alexander Urban, Ziqin Rong, Zhiwei Ding, Chuan Luo, and
  Gerbrand Ceder.
\newblock Construction of ground-state preserving sparse lattice models for
  predictive materials simulations.
\newblock {\em npj Computational Materials}, 3(1):30, August 2017.

\bibitem{goiri_recursive_2018}
Jon~Gabriel Goiri and Anton Van Der~Ven.
\newblock Recursive alloy {Hamiltonian} construction and its application to the
  {Ni}-{Al}-{Cr} system.
\newblock {\em Acta Materialia}, 159:257--265, October 2018.

\bibitem{thomas_comparing_2021}
John~C. Thomas, Anirudh~Raju Natarajan, and Anton Van Der~Ven.
\newblock Comparing crystal structures with symmetry and geometry.
\newblock {\em npj Computational Materials}, 7(1):164, October 2021.

\bibitem{marchand_machine_2022}
Daniel Marchand and W.~A. Curtin.
\newblock Machine learning for metallurgy {IV}: {A} neural network potential
  for {Al}-{Cu}-{Mg} and {Al}-{Cu}-{Mg}-{Zn}.
\newblock {\em Physical Review Materials}, 6(5):053803, May 2022.

\bibitem{vigna_weighted_2015}
Sebastiano Vigna.
\newblock A {Weighted} {Correlation} {Index} for {Rankings} with {Ties}.
\newblock In {\em Proceedings of the 24th {International} {Conference} on
  {World} {Wide} {Web}}, pages 1166--1176, Florence Italy, May 2015.
  International World Wide Web Conferences Steering Committee.

\bibitem{walker_lattice_1959}
G.B Walker and M~Marezio.
\newblock Lattice parameters and zone overlap in solid solutions of lead in
  magnesium.
\newblock {\em Acta Metallurgica}, 7(12):769--773, December 1959.

\bibitem{slutsky_elastic_1957}
L.~J. Slutsky and C.~W. Garland.
\newblock Elastic {Constants} of {Magnesium} from 4.2°{K} to 300°{K}.
\newblock {\em Physical Review}, 107(4):972--976, August 1957.

\bibitem{nishimura_volume_nodate}
M~Nishimura, K~Kinoshita, Y~Akahama, and H~Kawamura.
\newblock Volume compression of {Mg} and {Al} to multimegabar pressure.

\bibitem{ahmad_designing_2019}
Rasool Ahmad, Binglun Yin, Zhaoxuan Wu, and W.A. Curtin.
\newblock Designing high ductility in magnesium alloys.
\newblock {\em Acta Materialia}, 172:161--184, June 2019.

\bibitem{tzanetakis_formation_1976}
P.~Tzanetakis, J.~Hillairet, and G.~Revel.
\newblock The {Formation} {Energy} of {Vacancies} in {Aluminium} and
  {Magnesium}.
\newblock {\em physica status solidi (b)}, 75(2):433--439, 1976.

\bibitem{villars_nd_nodate}
Nd {Crystal} {Structure}: {Datasheet} from “{PAULING} {FILE} {Multinaries}
  {Edition} – 2022” in {SpringerMaterials}.

\bibitem{noauthor_handbook_1978}
Handbook on the physics and chemistry of rare earths, 1978.
\newblock ISSN: 0168-1273 Place: Amsterdam.

\bibitem{guo_anti-phase_2021}
Yanlin Guo, Bin Liu, Wei Xie, Qun Luo, and Qian Li.
\newblock Anti-phase boundary energy of $\beta$ series precipitates in
  {Mg}-{Y}-{Nd} system.
\newblock {\em Scripta Materialia}, 193:127--131, March 2021.

\bibitem{dewitt_misfit-driven_2017}
Stephen DeWitt, Ellen~L.S. Solomon, Anirudh~Raju Natarajan, Vicente
  Araullo-Peters, Shiva Rudraraju, Larry~K. Aagesen, Brian Puchala,
  Emmanuelle~A. Marquis, Anton Van Der~Ven, Katsuyo Thornton, and John~E.
  Allison.
\newblock Misfit-driven $\beta$''' precipitate composition and morphology in
  {Mg}-{Nd} alloys.
\newblock {\em Acta Materialia}, 136:378--389, September 2017.

\bibitem{choudhuri_interfacial_2017}
D.~Choudhuri, R.~Banerjee, and S.~G. Srinivasan.
\newblock Interfacial structures and energetics of the strengthening
  precipitate phase in creep-resistant {Mg}-{Nd}-based alloys.
\newblock {\em Scientific Reports}, 7(1):40540, January 2017.

\bibitem{agarwal_exact_2017}
Ravi Agarwal and Dallas~R. Trinkle.
\newblock Exact {Model} of {Vacancy}-{Mediated} {Solute} {Transport} in
  {Magnesium}.
\newblock {\em Physical Review Letters}, 118(10):105901, March 2017.

\bibitem{solomon_early_2017}
Ellen~L.S. Solomon, Vicente Araullo-Peters, John~E. Allison, and Emmanuelle~A.
  Marquis.
\newblock Early precipitate morphologies in {Mg}-{Nd}-({Zr}) alloys.
\newblock {\em Scripta Materialia}, 128:14--17, February 2017.

\bibitem{meier_stimulating_2025}
Janet~M. Meier, Jiashi Miao, Lisa DeBeer-Schmitt, Jan Ilavsky, and Alan~A. Luo.
\newblock Stimulating $\beta$-{Series} {Precipitation} in {Mg}–{Nd} {Alloys}
  {Via} {Microalloying}: {A} {Comparison} of {Electron} {Microscopy} and
  {Small}-{Angle} {Scattering} {Techniques}.
\newblock {\em Metallurgical and Materials Transactions A}, 56(3):914--927,
  March 2025.

\bibitem{liu_simulation_2014-1}
H.~Liu, Y.~Gao, Y.M. Zhu, Y.~Wang, and J.F. Nie.
\newblock A simulation study of $\beta$ 1 precipitation on dislocations in an
  {Mg}–rare earth alloy.
\newblock {\em Acta Materialia}, 77:133--150, September 2014.

\bibitem{liu_simulation_2014}
H.~Liu, Y.~Gao, Y.M. Zhu, Y.~Wang, and J.F. Nie.
\newblock A simulation study of $\beta$ 1 precipitation on dislocations in an
  {Mg}–rare earth alloy.
\newblock {\em Acta Materialia}, 77:133--150, September 2014.

\bibitem{steinhardt_bond-orientational_1983}
Paul~J. Steinhardt, David~R. Nelson, and Marco Ronchetti.
\newblock Bond-orientational order in liquids and glasses.
\newblock {\em Physical Review B}, 28(2):784--805, July 1983.

\bibitem{sigli_recent_2018}
Christophe Sigli, Frédéric De~Geuser, Alexis Deschamps, Joël Lépinoux, and
  Michel Perez.
\newblock Recent advances in the metallurgy of aluminum alloys. {Part} {II}:
  {Age} hardening.
\newblock {\em Comptes Rendus. Physique}, 19(8):688--709, November 2018.

\bibitem{bourgeois_transforming_2020}
Laure Bourgeois, Yong Zhang, Zezhong Zhang, Yiqiang Chen, and Nikhil~V.
  Medhekar.
\newblock Transforming solid-state precipitates via excess vacancies.
\newblock {\em Nature Communications}, 11(1):1248, March 2020.

\bibitem{ding_orderdisorder_2025}
Lipeng Ding, Flemming J.~H. Ehlers, Rong Hu, Zezhong Zhang, Hiromi Nagaum,
  Christopher Hutchinson, Qing Liu, and Zhihong Jia.
\newblock On the order–disorder transformation within a main hardening
  precipitate in {Al}–{Mg}–{Si} alloys.
\newblock {\em Philosophical Magazine}, 105(3):169--188, February 2025.

\bibitem{li_origin_2020}
Mingjia Li and Xiaohua Min.
\newblock Origin of $\omega$-phase formation in metastable $\beta$-type
  {Ti}-{Mo} alloys: cluster structure and stacking fault.
\newblock {\em Scientific Reports}, 10(1):8664, May 2020.

\bibitem{fu_atomic-scale_2022}
Xiaoqian Fu, Xu-Dong Wang, Beikai Zhao, Qinghua Zhang, Suyang Sun, Jiang-Jing
  Wang, Wei Zhang, Lin Gu, Yangsheng Zhang, Wen-Zheng Zhang, Wen Wen, Ze~Zhang,
  Long-qing Chen, Qian Yu, and En~Ma.
\newblock Atomic-scale observation of non-classical nucleation-mediated phase
  transformation in a titanium alloy.
\newblock {\em Nature Materials}, 21(3):290--296, March 2022.

\bibitem{muller_first-principles_2024}
Yann~L. Müller and Anirudh~Raju Natarajan.
\newblock First-principles thermodynamics of precipitation in
  aluminum-containing refractory alloys.
\newblock {\em Acta Materialia}, 274:119995, August 2024.

\bibitem{ong_python_2013}
Shyue~Ping Ong, William~Davidson Richards, Anubhav Jain, Geoffroy Hautier,
  Michael Kocher, Shreyas Cholia, Dan Gunter, Vincent~L. Chevrier, Kristin~A.
  Persson, and Gerbrand Ceder.
\newblock Python {Materials} {Genomics} (pymatgen): {A} robust, open-source
  python library for materials analysis.
\newblock {\em Computational Materials Science}, 68:314--319, February 2013.

\bibitem{hjorth_larsen_atomic_2017}
Ask Hjorth~Larsen, Jens Jørgen~Mortensen, Jakob Blomqvist, Ivano~E Castelli,
  Rune Christensen, Marcin Dułak, Jesper Friis, Michael~N Groves, Bjørk
  Hammer, Cory Hargus, Eric~D Hermes, Paul~C Jennings, Peter Bjerre~Jensen,
  James Kermode, John~R Kitchin, Esben Leonhard~Kolsbjerg, Joseph Kubal,
  Kristen Kaasbjerg, Steen Lysgaard, Jón Bergmann~Maronsson, Tristan Maxson,
  Thomas Olsen, Lars Pastewka, Andrew Peterson, Carsten Rostgaard, Jakob
  Schiøtz, Ole Schütt, Mikkel Strange, Kristian~S Thygesen, Tejs Vegge, Lasse
  Vilhelmsen, Michael Walter, Zhenhua Zeng, and Karsten~W Jacobsen.
\newblock The atomic simulation environment—a {Python} library for working
  with atoms.
\newblock {\em Journal of Physics: Condensed Matter}, 29(27):273002, July 2017.

\bibitem{togo_spglib_2024}
Atsushi Togo, Kohei Shinohara, and Isao Tanaka.
\newblock Spglib: a software library for crystal symmetry search.
\newblock {\em Science and Technology of Advanced Materials: Methods},
  4(1):2384822, December 2024.

\end{thebibliography}


\begin{thebibliography}{10}

\bibitem{delfino_phase_1990}
S.~Delfino, A.~Saccone, and R.~Ferro.
\newblock Phase relationships in the neodymium-magnesium alloy system.
\newblock {\em Metallurgical Transactions A}, 21(8):2109--2114, August 1990.

\bibitem{kirklin_open_2015}
Scott Kirklin, James~E Saal, Bryce Meredig, Alex Thompson, Jeff~W Doak,
  Muratahan Aykol, Stephan Rühl, and Chris Wolverton.
\newblock The {Open} {Quantum} {Materials} {Database} ({OQMD}): assessing the
  accuracy of {DFT} formation energies.
\newblock {\em npj Computational Materials}, 1(1):15010, December 2015.

\bibitem{zagorac_recent_2019}
D.~Zagorac, H.~Müller, S.~Ruehl, J.~Zagorac, and S.~Rehme.
\newblock Recent developments in the {Inorganic} {Crystal} {Structure}
  {Database}: theoretical crystal structure data and related features.
\newblock {\em Journal of Applied Crystallography}, 52(5):918--925, October
  2019.

\bibitem{natarajan_early_2016}
Anirudh~Raju Natarajan, Ellen~L.S. Solomon, Brian Puchala, Emmanuelle~A.
  Marquis, and Anton Van Der~Ven.
\newblock On the early stages of precipitation in dilute {Mg}–{Nd} alloys.
\newblock {\em Acta Materialia}, 108:367--379, April 2016.

\bibitem{kim_modified_2017}
Ki-Hyun Kim and Byeong-Joo Lee.
\newblock Modified embedded-atom method interatomic potentials for {Mg}-{Nd}
  and {Mg}-{Pb} binary systems.
\newblock {\em Calphad}, 57:55--61, June 2017.

\bibitem{bochkarev_efficient_2022}
Anton Bochkarev, Yury Lysogorskiy, Sarath Menon, Minaam Qamar, Matous Mrovec,
  and Ralf Drautz.
\newblock Efficient parametrization of the atomic cluster expansion.
\newblock {\em Physical Review Materials}, 6(1):013804, January 2022.

\bibitem{kolli_discovering_2020}
Sanjeev~Krishna Kolli, Anirudh~Raju Natarajan, John~C. Thomas, Tresa~M.
  Pollock, and Anton Van Der~Ven.
\newblock Discovering hierarchies among intermetallic crystal structures.
\newblock {\em Physical Review Materials}, 4(11):113604, November 2020.

\bibitem{thomas_exploration_2017}
John~C. Thomas and Anton Van Der~Ven.
\newblock The exploration of nonlinear elasticity and its efficient
  parameterization for crystalline materials.
\newblock {\em Journal of the Mechanics and Physics of Solids}, 107:76--95,
  October 2017.

\bibitem{tran_anisotropic_2019}
Richard Tran, Xiang-Guo Li, Joseph~H. Montoya, Donald Winston, Kristin~Aslaug
  Persson, and Shyue~Ping Ong.
\newblock Anisotropic work function of elemental crystals.
\newblock {\em Surface Science}, 687:48--55, September 2019.

\bibitem{togo_implementation_2023}
Atsushi Togo, Laurent Chaput, Terumasa Tadano, and Isao Tanaka.
\newblock Implementation strategies in phonopy and phono3py.
\newblock {\em Journal of Physics: Condensed Matter}, 35(35):353001, September
  2023.

\bibitem{puchala_casm_2023}
Brian Puchala, John~C. Thomas, Anirudh~Raju Natarajan, Jon~Gabriel Goiri,
  Sesha~Sai Behara, Jonas~L. Kaufman, and Anton Van Der~Ven.
\newblock {CASM} — {A} software package for first-principles based study of
  multicomponent crystalline solids.
\newblock {\em Computational Materials Science}, 217:111897, January 2023.

\bibitem{thomas_comparing_2021}
John~C. Thomas, Anirudh~Raju Natarajan, and Anton Van Der~Ven.
\newblock Comparing crystal structures with symmetry and geometry.
\newblock {\em npj Computational Materials}, 7(1):164, October 2021.

\bibitem{kresse_ab_1994}
G.~Kresse and J.~Hafner.
\newblock \textit{{Ab} initio} molecular-dynamics simulation of the
  liquid-metal–amorphous-semiconductor transition in germanium.
\newblock {\em Physical Review B}, 49(20):14251--14269, May 1994.

\bibitem{kresse_ab_1995}
G~Kresse.
\newblock Ab initio molecular dynamics for liquid metals.
\newblock 1995.

\bibitem{kresse_efficiency_1996}
G.~Kresse and J.~Furthmüller.
\newblock Efficiency of ab-initio total energy calculations for metals and
  semiconductors using a plane-wave basis set.
\newblock {\em Computational Materials Science}, 6(1):15--50, July 1996.

\bibitem{thompson_lammps_2022}
Aidan~P. Thompson, H.~Metin Aktulga, Richard Berger, Dan~S. Bolintineanu,
  W.~Michael Brown, Paul~S. Crozier, Pieter~J. In~'T~Veld, Axel Kohlmeyer,
  Stan~G. Moore, Trung~Dac Nguyen, Ray Shan, Mark~J. Stevens, Julien Tranchida,
  Christian Trott, and Steven~J. Plimpton.
\newblock {LAMMPS} - a flexible simulation tool for particle-based materials
  modeling at the atomic, meso, and continuum scales.
\newblock {\em Computer Physics Communications}, 271:108171, February 2022.

\bibitem{jonsson_nudged_1998}
Hannes Jónsson, Greg Mills, and Karsten~W. Jacobsen.
\newblock Nudged elastic band method for finding minimum energy paths of
  transitions.
\newblock In {\em Classical and {Quantum} {Dynamics} in {Condensed} {Phase}
  {Simulations}}, pages 385--404, LERICI, Villa Marigola, June 1998. WORLD
  SCIENTIFIC.

\bibitem{dewitt_misfit-driven_2017}
Stephen DeWitt, Ellen~L.S. Solomon, Anirudh~Raju Natarajan, Vicente
  Araullo-Peters, Shiva Rudraraju, Larry~K. Aagesen, Brian Puchala,
  Emmanuelle~A. Marquis, Anton Van Der~Ven, Katsuyo Thornton, and John~E.
  Allison.
\newblock Misfit-driven $\beta$''' precipitate composition and morphology in
  {Mg}-{Nd} alloys.
\newblock {\em Acta Materialia}, 136:378--389, September 2017.

\bibitem{choudhuri_interfacial_2017}
D.~Choudhuri, R.~Banerjee, and S.~G. Srinivasan.
\newblock Interfacial structures and energetics of the strengthening
  precipitate phase in creep-resistant {Mg}-{Nd}-based alloys.
\newblock {\em Scientific Reports}, 7(1):40540, January 2017.

\bibitem{guo_anti-phase_2021}
Yanlin Guo, Bin Liu, Wei Xie, Qun Luo, and Qian Li.
\newblock Anti-phase boundary energy of $\beta$ series precipitates in
  {Mg}-{Y}-{Nd} system.
\newblock {\em Scripta Materialia}, 193:127--131, March 2021.

\bibitem{nie_precipitation_2012}
Jian-Feng Nie.
\newblock Precipitation and {Hardening} in {Magnesium} {Alloys}.
\newblock {\em Metallurgical and Materials Transactions A}, 43(11):3891--3939,
  November 2012.

\bibitem{liu_simulation_2013}
H.~Liu, Y.~Gao, J.Z. Liu, Y.M. Zhu, Y.~Wang, and J.F. Nie.
\newblock A simulation study of the shape of $\beta$' precipitates in
  {Mg}–{Y} and {Mg}–{Gd} alloys.
\newblock {\em Acta Materialia}, 61(2):453--466, January 2013.

\bibitem{liu_structure_2017}
H.~Liu, Y.M. Zhu, N.C. Wilson, and J.F. Nie.
\newblock On the structure and role of $\beta$ {F}' in beta1 precipitation in
  {Mg}–{Nd} alloys.
\newblock {\em Acta Materialia}, 133:408--426, July 2017.

\bibitem{agarwal_exact_2017}
Ravi Agarwal and Dallas~R. Trinkle.
\newblock Exact {Model} of {Vacancy}-{Mediated} {Solute} {Transport} in
  {Magnesium}.
\newblock {\em Physical Review Letters}, 118(10):105901, March 2017.

\end{thebibliography}

\end{document}

\title{Supplementary information: Machine learning interatomic potentials for solid-state precipitation}
\author{Lorenzo Piersante}
\affiliation{Laboratory of materials design and simulation (MADES), Institute of Materials, \'{E}cole Polytechnique F\'{e}d\'{e}rale de Lausanne}
\author{Anirudh Raju Natarajan}
\email{anirudh.natarajan@epfl.ch}
\affiliation{Laboratory of materials design and simulation (MADES), Institute of Materials, \'{E}cole Polytechnique F\'{e}d\'{e}rale de Lausanne}
\affiliation{National Centre for Computational Design and Discovery of Novel Materials (MARVEL), \'{E}cole Polytechnique F\'{e}d\'{e}rale de Lausanne}

\maketitle

\section{Crystallography of the Mg-Nd binary system}
\begin{figure*}
    \centering
    \includegraphics[width=\linewidth]{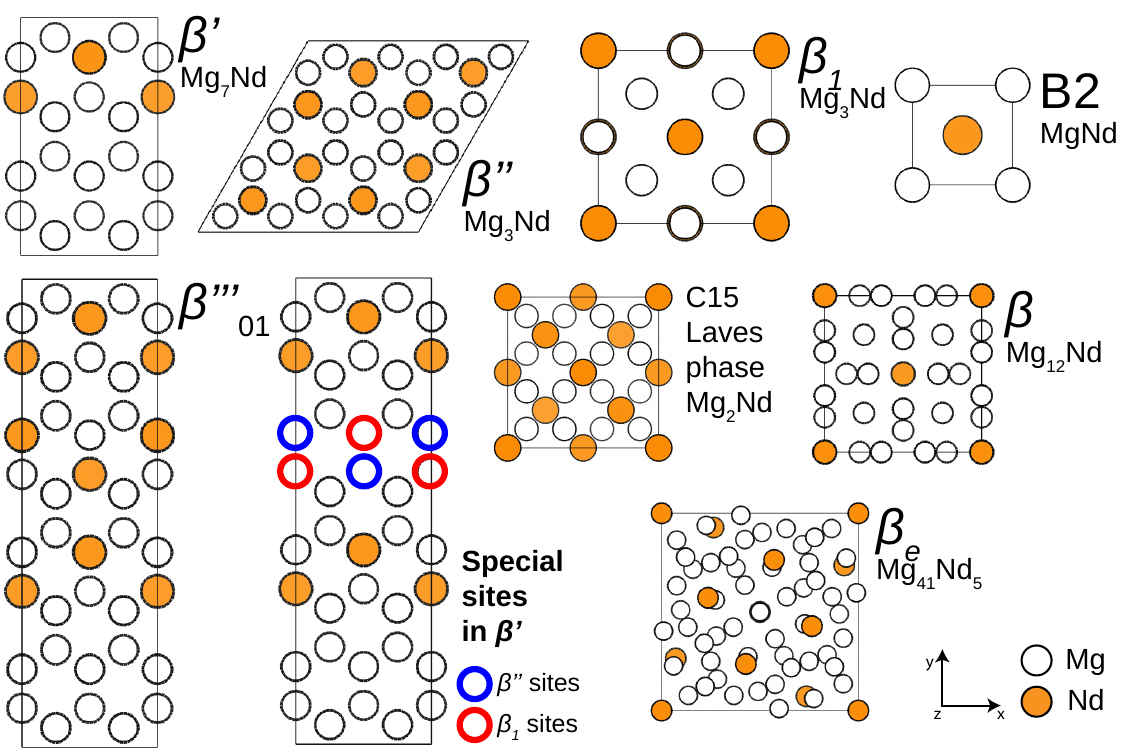}
    \caption{Schematic crystal structures of the stable and metastable phases in the Mg-Nd system. All crystal structures are projected along the $c$-axis.}
    \label{fig:crystallography_mgnd}
\end{figure*}

\Cref{fig:crystallography_mgnd} illustrates the crystal structures of all known metastable and stable phases in the Mg-Nd system \cite{delfino_phase_1990, kirklin_open_2015, zagorac_recent_2019, natarajan_early_2016}. Precipitation in a dilute Mg-Nd alloy ($x_{\text{Nd}} \leq 0.5$ at.\%) proceeds as follows \cite{natarajan_early_2016}:
\begin{gather}
    \text{solid solution} \rightarrow \text{GP zones} \rightarrow \beta''' \rightarrow \beta_1 \rightarrow \beta \rightarrow \beta_e \\
    hcp \rightarrow hcp \rightarrow hcp \rightarrow bcc \rightarrow I4/mmm \rightarrow I4/m
\end{gather}
In the early stages of aging, Guinier-Preston (GP) zones form first. Next, the $\beta^{\prime\prime\prime}$ precipitate emerges via an order-disorder transformation. The $\beta^{\prime\prime\prime}$ phase comprises an infinite family of structures generated by stacking zigzag and hexagonal chains of Nd atoms along the $[10\overline{1}0]$ direction of hcp \cite{natarajan_early_2016}. \Cref{fig:crystallography_mgnd} shows one such arrangement, labeled $\beta'''_{01}$. As heat treatment continues, the $\beta_1$ phase forms, followed by the metastable $\beta$ and equilibrium $\beta_e$ phases at longer aging times. Precipitation of these later phases reduces alloy strength from its peak value. At higher Nd concentrations, two additional equilibrium phases occur: the high-temperature C15 Laves phase (Mg$_{2}$Nd) and an equiatomic B2 (MgNd) intermetallic compound.

\section{MEAM benchmarks} \label{sup:meam_results}

\begin{figure}[!ht]
    \centering
    \includegraphics[width=0.75\columnwidth]{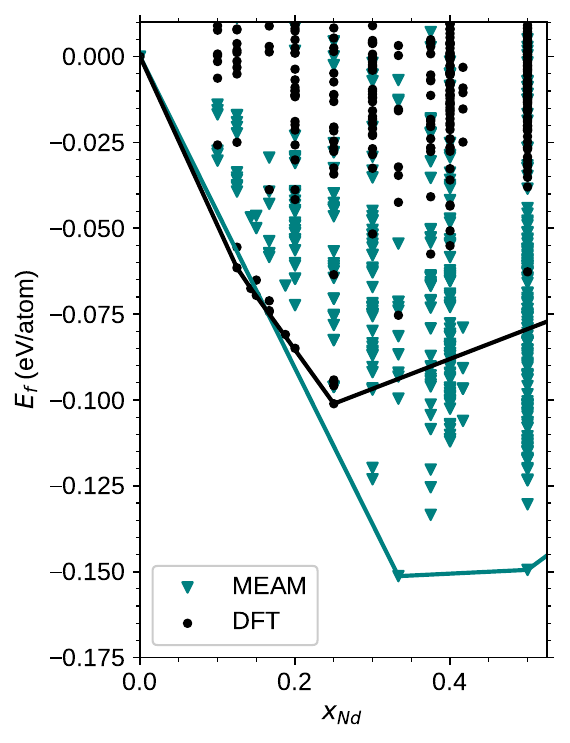}
    \caption{Comparison of the HCP convex hull obtained with the MEAM potential and DFT calculations. \label{fig:meam_hull}}
\end{figure}

\Cref{fig:meam_hull} shows the metastable hcp convex hull obtained with the MEAM potential of Kim et al. \cite{kim_modified_2017}. All symmetry-distinct orderings identified as hcp from their relaxed DFT structures were relaxed with the MEAM potential, and their formation energies are plotted in \cref{fig:meam_hull}. A large discrepancy between DFT and MEAM predictions is evident. Two spurious ground states appear at compositions $x_{\text{Nd}}$ > 0.25 with formation energies so negative that the experimentally observed $\beta^{\prime\prime\prime}$ phases no longer lie on the hcp convex hull. As a result, the MEAM potential is not suitable for studying precipitation in dilute Mg-Nd alloys.

\section{Atomic cluster expansion potential} \label{sup:ace}
The ACE MLIP fitted in this work uses spherical Bessel functions as radial basis functions, with a radial parameter of 5.25 and a cosine cutoff \cite{bochkarev_efficient_2022}. At short interatomic distances, the potential interpolates to a ZBL repulsion. An inner cutoff function mediates this transition by smoothly switching off the radial basis functions within the region $[r_{in} - \Delta_{in}, r_{in}]$. At short distances, when $r < r_{in} - \Delta_{in}$, the radial basis functions vanish and only the ZBL repulsion contributes \cite{bochkarev_efficient_2022}. \Cref{tab:core_repulsion} summarizes the values of $r_{in}$ and $\Delta_{in}$ used in this work.

\begin{table}[!ht]
    \centering
    \begin{tabular}{|c|c c|}
        \hline
        Block & $r_{in}$ (\r{A}) & $\Delta_{in}$ (\r{A}) \\ \hline
        Mg    & 1.2              & 0.2                   \\
        Nd    & 2.2              & 0.4                   \\
        Mg-Nd & 2.0              & 0.2                   \\ \hline
    \end{tabular}
    \caption{\centering Inner cutoff parameters of the ACE potential.}
    \label{tab:core_repulsion}
\end{table}

\section{Reference database}
\begin{table}[!ht]
    \centering
    \begin{tabular}{|c|c|c|}
        \hline
        Type                                                  & Structures & Datapoints \\ \hline
        Intermetallics \cite{kolli_discovering_2020}          & 379        & 1644       \\
        Dimers                                                & 39         & 39         \\
        hcp strains \cite{thomas_exploration_2017}            & 228        & 515        \\
        hcp cohesive energy                                   & 36         & 36         \\
        hcp GSFs                                              & 81         & 81         \\
        hcp SSFs                                              & 3          & 3          \\
        hcp bulk/dot/rod geometries                           & 14         & 14         \\
        VD of 39 prototypes                                  & 907        & 2471       \\
        Vacancy structures                                    & 1510       & 31328      \\
        hcp/fcc surfaces \cite{tran_anisotropic_2019}         & 27         & 83         \\
        hcp surface decohesion                                & 184        & 184        \\
        hcp/fcc/bcc $\Sigma$ GBs \cite{tran_anisotropic_2019} & 27         & 75         \\
        Phonon displacements \cite{togo_implementation_2023}  & 8          & 8          \\
        AIMD snapshots                                        & 1820       & 1820       \\
        Transformation pathways                               & 1167       & 1167       \\
        bcc strain-shuffle maps                               & 966        & 966        \\ \hline
        TOTAL                                                 & 7396       & 40434      \\ \hline
    \end{tabular}
    \caption{Number of structures and datapoints in the reference database for pure Mg. Abbreviations: generalized stacking fault (GSF), stable stacking fault (SSF), volumetric distortion (VD), grain boundary (GB), ab-initio molecular dynamics (AIMD).}
    \label{tab:Mg_database}
\end{table}

\begin{table}[!ht]
    \centering
    \begin{tabular}{|c|c|c|}
        \hline
        Type                                         & Structures & Datapoints \\ \hline
        Intermetallics \cite{kolli_discovering_2020} & 349        & 1169       \\
        Dimers                                       & 39         & 39         \\
        VD of 34 prototypes                         & 714        & 3256       \\
        AIMD snapshots                               & 2104       & 2104       \\ \hline
        TOTAL                                        & 3206       & 6568       \\ \hline
    \end{tabular}
    \caption{Number of structures and datapoints in the reference database for pure Nd. Abbreviations: volumetric distortion (VD), ab-initio molecular dynamics (AIMD).}
    \label{tab:Nd_database}
\end{table}

\begin{table}[!ht]
    \centering
    \begin{tabular}{|c|c|c|}
        \hline
        Type                                                                          & Structures & Datapoints \\ \hline
        Experimental phases \cite{kirklin_open_2015, zagorac_recent_2019}             & 14         & 33         \\
        Symmetry-distinct orderings \cite{puchala_casm_2023, thomas_comparing_2021}   & 1883       & 7451       \\
        Active learning orderings                                                     & 297        & 1261       \\
        Dimers                                                                        & 39         & 39         \\
        Burgers pathways                                                              & 248        & 435        \\
        $h$-$\beta_1$ strain-shuffle maps                                             & 4605       & 90247      \\
        Nd dissolution in hcp                                                         & 2          & 101        \\
        Nd-Nd binding in hcp                                                          & 12         & 215        \\
        Nd in hcp $\Sigma$ GB                                                         & 12         & 38         \\
        Nd in hcp SSFs                                                                & 40         & 674        \\
        VD of convex hull states                                                     & 273        & 1476       \\
        $\beta'$/$\beta''$/$\beta_1$/$\beta_e$ strains \cite{thomas_exploration_2017} & 510        & 3979       \\
        Nd-vacancy interaction in hcp                                                 & 16         & 184        \\
        $\beta'$/$\beta''$/$\beta'''$/$\beta_1$/NdMg$_2$ vacancies                    & 69         & 470        \\
        $\beta'$/$\beta''$/$\beta_1$ GSFE profiles                                    & 72         & 72         \\
        $\beta'$/$\beta''$ APBs                                                       & 5          & 27         \\
        Random phonon displacements                                                   & 1300       & 1300       \\
        Liquid structures                                                             & 1594       & 1594       \\ \hline
        TOTAL                                                                         & 10991      & 109596     \\ \hline
    \end{tabular}
    \caption{Number of structures and datapoints in the reference database for binary Mg-Nd. Abbreviations: grain boundary (GB), stable stacking fault (SSF), volumetric distortion (VD), generalized stacking fault energy (GSFE), anti-phase boundary (APB).}
    \label{tab:MgNd_database}
\end{table}

\Cref{tab:Mg_database,tab:Nd_database,tab:MgNd_database} lists all structures contained in our DFT database. Calculations along each relaxation trajectory were retained, resulting in more DFT data points than reference configurations.

The database for pure Mg includes structures along the well-known Bain ($bcc \leftrightarrow fcc$), Burgers ($bcc \leftrightarrow hcp$), and trigonal ($bcc \rightarrow SC \rightarrow fcc$) transformation pathways. We also included structures along strain distortions connecting different orientational variants of bcc relative to hcp. Ab initio molecular dynamics (AIMD) simulations were performed with VASP \cite{kresse_ab_1994, kresse_ab_1995, kresse_efficiency_1996} in the NVT ensemble at various temperatures and densities for pure Mg and Nd. Symmetrically-distinct chemical decorations were enumerated in hcp, bcc, fcc, and C15 Laves parent crystals with the \verb|CASM| code \cite{puchala_casm_2023}. Liquid structures in binary Mg-Nd were sampled at compositions $x_{\text{Nd}}$ = 1/8, 1/4, 1/3, and 1/2 with Monte Carlo/molecular dynamics in the NPT ensemble as implemented in LAMMPS \cite{thompson_lammps_2022} using the MEAM potential of Kim et al. \cite{kim_modified_2017}. Crystal structures with atomic displacements were created either by applying random displacements to atomic coordinates or by using the enumeration routines in Phonopy \cite{togo_implementation_2023}. In pure Mg, atomic displacements were enumerated in hcp, bcc, fcc, dhcp, and $\omega$ structures. In Mg-Nd, we created rattled structures for all stable and metastable convex hull states up to $x_{\text{Nd}}$ = 1/2. Lastly, strain-shuffle maps denote structures that systematically sample the Burgers transformation across strain and displacement.

\section{Active learning with Mahalanobis distances}\label{sup:active_learning_mahalanobis}

The mean and covariance matrix used to compute the Mahalanobis distance in \cref{main-fig:dataset_selection} are updated with the following two equations:
\begin{equation}
    \vec{\mu}_{N+1} = \frac{N}{N + 1} {\vec{\mu}}_{N} + \frac{1}{N + 1} \vec{Z}
    \label{sup_eq:mean_update}
\end{equation}
\begin{equation}
    \Sigma_{N+1} = \frac{N-1}{N} \Sigma_{N} + \frac{({\vec{\mu}}_{N} - \vec{Z}) ({\vec{\mu}}_{N} - \vec{Z})^T}{N + 1}
    \label{sup_eq:covariance_update}
\end{equation}
where $N$ is the number of atomic sites in the initial dataset, $\vec{\mu}_N$ and $\Sigma_{N}$ are the mean and covariance matrix of the initial dataset, and $\vec{Z}$ is the descriptor of the newly selected site. In the workflow, we select the structure containing the site with the largest Mahalanobis distance. We then apply \cref{sup_eq:mean_update} and \cref{sup_eq:covariance_update} iteratively to update the mean and covariance matrix over all sites in the newly selected structure.

\section{Descriptor-based classification} \label{sup:descriptor_classification}
The active learning loop described in \cref{main-fig:active_learning_loop} requires assigning nearly 0.5 million structures to four different parent crystal structures. Direct application of the structure-matching algorithm of Thomas et al. \cite{thomas_comparing_2021} is infeasible at this scale, so we instead use a method based on geometric descriptors. A simple Euclidean distance between the average descriptor of a relaxed structure and that of the reference parent structure is also difficult to apply because assignment thresholds are sensitive to small lattice and atomic distortions.

To address this, we compute the Mahalanobis distance of each relaxed structure relative to a small set of structures labeled using the accurate structure-matching method. Because crystal structures on a given convex hull (e.g., fcc) deviate slightly from the ideal parent structure, the Mahalanobis distance relative to such a dataset naturally accounts for the spread in descriptor values. We classified an initial set of symmetry-distinct orderings (\cref{tab:MgNd_database}) using the structure-matching algorithm in \verb|CASM| \cite{puchala_casm_2023, thomas_comparing_2021}, obtaining four separate convex hulls for hcp, bcc, fcc, and C15 Laves parent lattices. These labeled datasets then served as references for computing the Mahalanobis distance of structures generated during active learning. Algorithm \ref{algo:descriptor_based_classification} outlines the classification procedure. We used an ACE basis set of 15/5-0/4 throughout the classification routine.

\begin{algorithm}[!ht]
    \caption{\label{algo:descriptor_based_classification} Descriptor-based classification}
    \begin{algorithmic}[1]
        \State mask the atomic species of the new ordering;
        \State calculate the average descriptor;
        \State evaluate the Mahalanobis distance, $d_M$, of the average descriptor relative to each reference dataset (hcp, bcc, fcc, C15);\\
        \If{$d_M^{hcp}$, $d_M^{bcc}$, $d_M^{fcc}$, $d_M^{C15}$ > tolerance}{
            \State assign the ordering to no lattice;
        }
        \Else{
            \State assign the ordering to the nearest lattice;
        }
    \end{algorithmic}
\end{algorithm}

\section{Training the MLIP} \label{sup:training_datasets}

\begin{figure}[!ht]
    \centering
    \includegraphics[width=0.8\columnwidth]{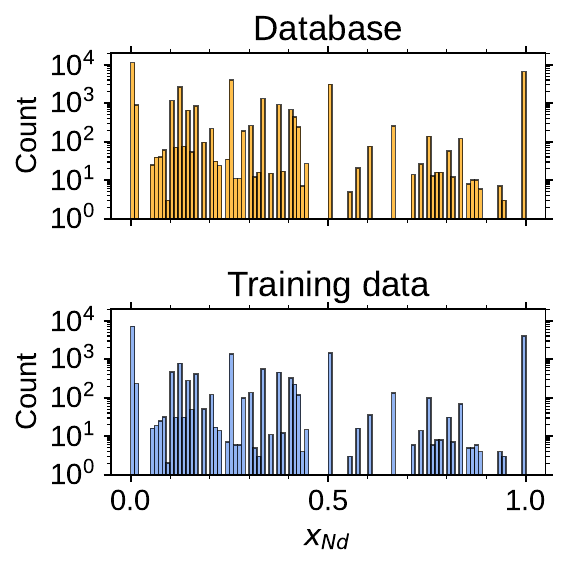}
    \caption{Histograms of crystal structures in the reference database (top) and training datasets (bottom) as a function of composition. Data points from Mg vacancy relaxation trajectories and Mg-Nd $h$-$\beta_1$ strain-shuffle maps are excluded for clarity.}
    \label{fig:data_summary}
\end{figure}

\Cref{fig:data_summary} shows the distribution of data points across composition for the database of all calculations and the training datasets used in constructing the Mg-Nd MLIP. Construction of the training datasets combined domain knowledge with the Mahalanobis distance sampling workflow shown in \cref{main-fig:dataset_selection}. Three training datasets were developed to parameterize the Mg, Nd, and Mg-Nd MLIPs. Each training dataset began with a set of initial structures chosen based on domain knowledge. Additional data points were then selected using the procedure in \cref{main-fig:dataset_selection} to enhance structural and chemical diversity. For the Mg-Nd dataset, we also included structures containing atoms closer than 2.5 Å to capture the core repulsion. The abbreviations employed in the following paragraphs are listed in the captions of \cref{tab:Mg_database,tab:Nd_database,tab:MgNd_database}.

The pure Mg training set began with 4405 data points comprising prototypical intermetallic structures, dimers, deformations of the bulk hcp structure (strains, GSFs, SSFs, bulk/rod/corner geometries, and $\Sigma$ GBs), hcp slab geometries, transformation pathways, volumetric distortions of hcp and bcc, strain-shuffle distortions of bcc related to the Burgers transformation, and vacancies in several prototypical structures (hcp, bcc, fcc, and C15). Using this initial dataset as a reference, we sampled 2650 additional data points from the remaining pool of structures. The final training set for the pure Mg potential contained 7055 data points.

The pure Nd training set was initialized with 2097 data points comprising prototypical intermetallic structures, dimers, and volumetric distortions of all structural prototypes with energies below that of hcp-Nd. Using this initial dataset as a reference, we sampled an additional 708 data points from the volumetric distortions of all other structural prototypes. The updated dataset then served as the reference for sampling 1200 AIMD snapshots. The final pure Nd training set contained 4005 data points.

The initial Mg-Nd training set contained 5563 data points comprising experimental phases, dimers, Nd dissolution configurations, Nd-Nd pairs at varying distances embedded within a hcp-Mg matrix, symmetry-distinct orderings enumerated with \verb|CASM|, symmetry-distinct orderings identified via active learning, volumetric distortions of phases on the convex hull, Nd-SSF interactions, vacancies in $\beta'$/$\beta''$/$\beta_1$/Mg$_2$Nd, Nd-vacancy interactions, symmetry-distinct Burgers pathways starting from hcp orderings in supercells containing up to 4 atoms, and symmetry-distinct Burgers pathways of $\beta'$ with occupation of some $h$-$\beta_1$ sites. Using this initial data as a reference set, we sampled 75 liquid structures at four compositions ($x_{\text{Nd}}$ = 1/8, 1/4, 1/3, and 1/2), 30 rattled structures for each of 13 structural prototypes corresponding to all global and metastable ground states up to $x_{\text{Nd}}$ = 1/2, 500 strained configurations of $\beta'$/$\beta''$/$\beta_1$/$\beta_e$, 400 structures from the Burgers transformation of $h$-$\beta_1$, 30 Nd-$\Sigma$ GB structures, and 625 data points from the remaining volumetric distortions, active learning, Nd-SSF, Nd-vacancy, Nd dissolution, and Nd-Nd binding data. The final binary Mg-Nd training set contained 7985 data points.

\section{MLIP benchmarks}

\subsection{Unary : Mg, Nd}
\begin{figure}[!ht]
    \centering
    \subfloat[][\label{fig:npt_pure_Mg} HCP]{\includegraphics[width=0.75\columnwidth]{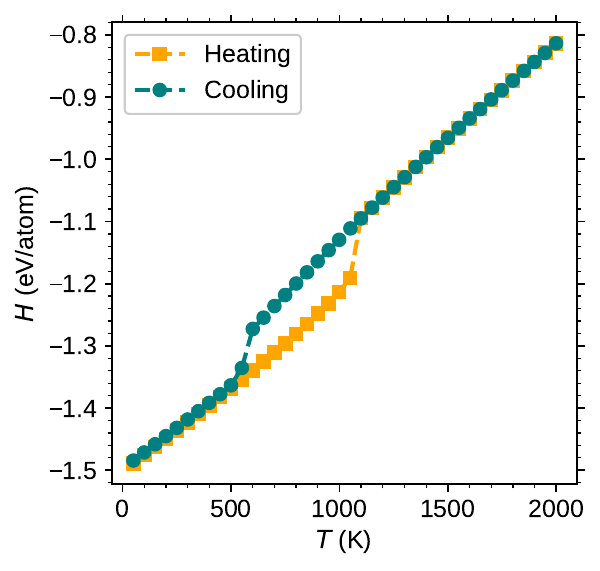}}\\
    \subfloat[][\label{fig:npt_pure_Nd} DHCP]{\includegraphics[width=0.75\columnwidth]{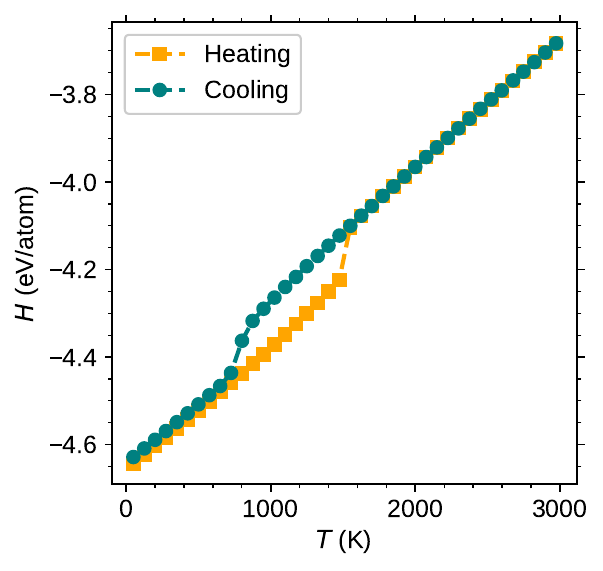}}\\
    \caption{Ensemble-averaged enthalpy as a function of temperature computed via NPT molecular dynamics. Trajectories were initiated at low temperature for (a) hcp-Mg and (b) dhcp-Nd.}
\end{figure}

\Cref{fig:npt_pure_Mg} shows the change in enthalpy during a heating-cooling cycle carried out with molecular dynamics in the NPT ensemble at zero pressure, starting from hcp-Mg. The final structure after cooling contained some defects and had a slightly higher energy. \Cref{fig:npt_pure_Nd} shows the corresponding heating-cooling cycle for dhcp-Nd. The final structure after cooling was predominantly fcc-Nd and had a higher energy than the dhcp-Nd starting point.

\begin{figure}[!ht]
    \centering
    \includegraphics[width=0.7\columnwidth]{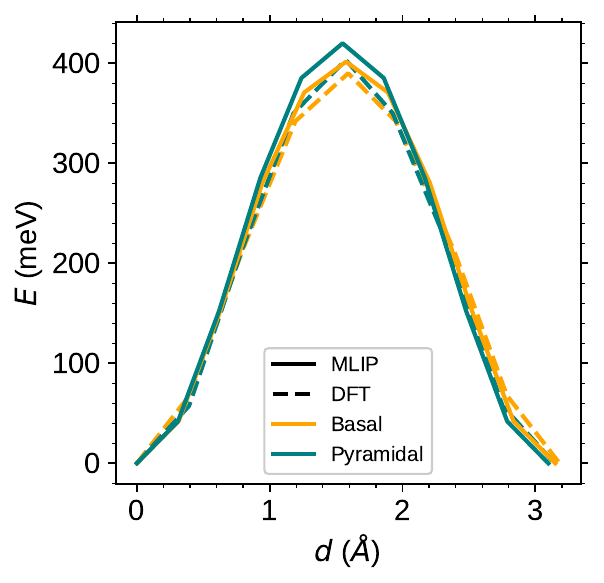}
    \caption{Comparison of vacancy hops in hcp-Mg computed via the NEB method using DFT and MLIP.}
    \label{fig:pure_Mg_vacancy_hop}
\end{figure}

\Cref{fig:pure_Mg_vacancy_hop} compares DFT and MLIP energies of vacancy hops in hcp-Mg along the pyramidal and basal planes computed with the nudged elastic band (NEB) method \cite{jonsson_nudged_1998}. The simulations were performed in a $4\times4\times3$ supercell.

\Cref{tab:Mg_tau_set} summarizes the training dataset used to fit the Mg potentials used in the investigation of the influence of the weighting scheme. The initial dataset contained the prototypes of intermetallics and the dimers. This dataset was the starting point to sample volumetric distortions of several prototypes and AIMD snapshots. The final dataset had 4483 datapoints and closely resembled the dataset used to train the Nd potentials.

\begin{table}[!ht]
    \centering
    \begin{tabular}{| c | c |}
         \hline
         Type &  Datapoints \\ \hline
          Common intermetallics & 1644 \\
          Dimers & 39 \\
          Sampled VD data & 1600 \\
          Sampled AIMD snapshots & 1200 \\ \hline
    \end{tabular}
    \caption{Training dataset to fit the Mg potentials as a function of the low-energy weights. Abbreviations: volumetric distortion (VD), ab-initio molecular dynamics (AIMD). \label{tab:Mg_tau_set}}
\end{table}

\subsection{Binary: Mg-Nd}

\begin{figure*}[!ht]
    \centering
    \subfloat[][\label{fig:gsf_paths}]{\includegraphics[width=0.6\linewidth]{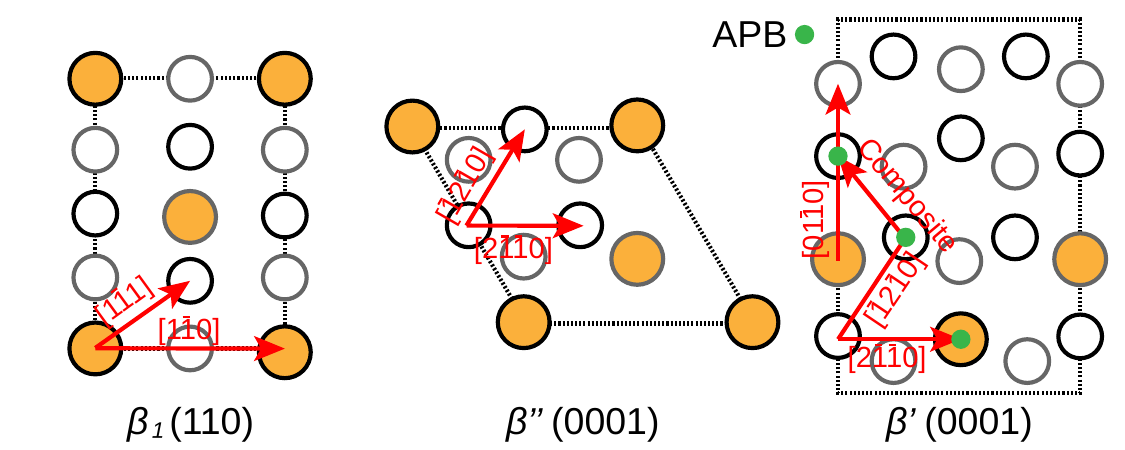}} \\
    \subfloat[][\label{fig:beta1_profile} $\beta_1$]{\includegraphics[width=0.3\linewidth]{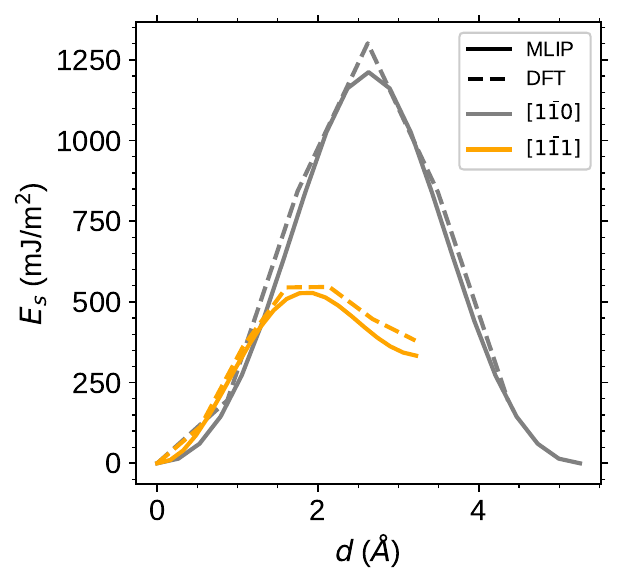}}
    \subfloat[][\label{fig:betapp_profiles} $\beta''$]{\includegraphics[width=0.3\linewidth]{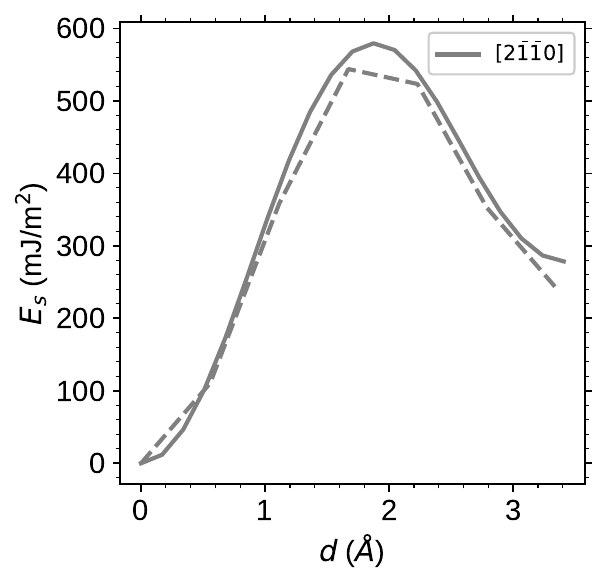}}
    \subfloat[][\label{fig:betap_profiles} $\beta'$]{\includegraphics[width=0.3\linewidth]{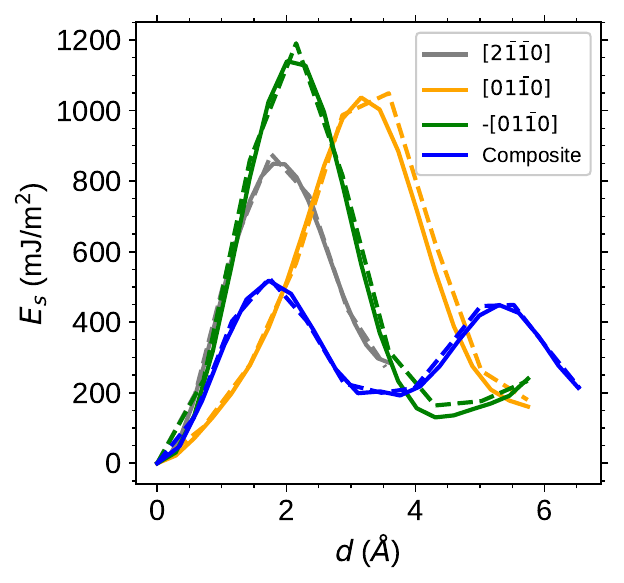}}
    \caption{(a) Schematic of displacement vectors used in the GSFE profile calculation. The green dots in the $\beta'$ schematic mark the locations where APB were calculated. (b-d) Comparison of GSFE profiles computed via DFT and MLIP along specific directions in $\beta_1$, $\beta''$, and $\beta'$.}
    \label{fig:gamma_surfaces}
\end{figure*}

\subsubsection{Interface properties}

\begin{table}[!hb]
    \centering
    \vspace{5mm}
    \begin{tabular}{| c  c | c c | c c | c c | c c | c |}
        \hline
        \multicolumn{2}{|c|}{Phases} & \multicolumn{2}{|c|}{Planes} & \multicolumn{2}{|c|}{Latt. 1} & \multicolumn{2}{|c|}{Average} & \multicolumn{2}{|c|}{Latt. 2} & \multirow{2}{*}{Lit.}                                                                \\
        1                            & 2                            & 1                             & 2                             & DFT                           & IP                    & DFT & IP  & DFT & IP  &                                      \\ \hline
        \multirow{3}{*}{$\beta'$}    & \multirow{3}{*}{hcp}         & $(100)$                       & $(2\bar{1}\bar{1}0)$          & 26                            & 44                    & 17  & 42  & 18  & 38  & 46 \cite{dewitt_misfit-driven_2017}  \\
                                     &                              & $(010)$                       & $(01\bar{1}0)$                & 69                            & 55                    & 75  & 61  & 5   & 47  & 51 \cite{dewitt_misfit-driven_2017}  \\
                                     &                              & $(001)$                       & $(0001)$                      & 49                            & 45                    & 31  & 39  & 23  & 23  & 28 \cite{dewitt_misfit-driven_2017}  \\ \hline
        \multirow{3}{*}{$\beta'$}    & \multirow{3}{*}{$\beta_1$}   & $(100)$                       & $(001)$                       & 36                            & 22                    & 36  & 21  & 37  & 18  & -                                    \\
                                     &                              & $(010)$                       & $(\bar{1}10)$                 & 24                            & 23                    & 21  & 22  & 18  & 21  & -                                    \\
                                     &                              & $(001)$                       & $(110)$                       & 6                             & 5                     & 8   & 5   & 11  & 5   & -                                    \\ \hline
        $\beta_1$                    & hcp                          & $(\bar{1}12)$                 & $(01\bar{1}0)$                & 93                            & 83                    & 88  & 81  & 85  & 79  & 98 \cite{choudhuri_interfacial_2017} \\ \hline
        $\beta_1$                    & $\beta'$                     & $(\bar{1}12)$                 & $(010)$                       & 142                           & 128                   & 212 & 176 & 148 & 107 & -                                    \\ \hline
    \end{tabular}
    \caption{Interfacial energies for various precipitate/matrix combinations. Orientation relationships are detailed in the text. Latt. and Lit. abbreviate lattice parameters and literature values, respectively.}
    \label{tab:interface_energies}
\end{table}

\Cref{fig:beta1_profile}, \cref{fig:betapp_profiles}, \cref{fig:betap_profiles}, and \cref{tab:interface_energies} compare generalized stacking fault energy (GSFE) profiles and interface energies between DFT and the MLIP. \Cref{fig:gsf_paths} shows the basis vectors (red arrows) used to map the GSFE profiles. We compare the MLIP to DFT only along these directions. The green dots in the $\beta'$ schematic mark the locations where we computed anti-phase boundary (APB) energies. We also compare interface energies between DFT and the MLIP at different levels of lattice strain. We evaluated interface energies when two phases were forced to coexist at the lattice parameters of each phase and at their average. Cells were relaxed in the direction perpendicular to the interface. We computed interface energies with the MLIP, followed by static DFT calculations at the MLIP-predicted lattice parameters for validation.

The energy profiles in \cref{fig:beta1_profile,fig:betapp_profiles,fig:betap_profiles} demonstrate close agreement between the MLIP and DFT. The APB produced by a shift along the $[2\bar{1}\bar{1}0]_{hcp}$ vector in $\beta'$ has an MLIP-predicted energy of 173 mJ/m$^2$ against a DFT value of 180 mJ/m$^2$. This result also agrees with the value reported in \cite{guo_anti-phase_2021}. The other two APBs have MLIP-computed energies of 129 and 55 mJ/m$^2$, matching the DFT values of 117 and 66 mJ/m$^2$.

\Cref{tab:interface_energies} compares several interface energies. We analyzed various interfaces between hcp, $\beta'$, and $\beta_1$. We considered two experimentally observed interfaces \cite{nie_precipitation_2012, liu_simulation_2013, liu_structure_2017}: the $\beta'$-hcp interface with orientation relationship $[001]_{\beta'} \parallel [0001]_{hcp}$ $(010)_{\beta'}\parallel (01\bar{1}0)_{hcp}$, and the $\beta_1$-hcp interface with orientation relationship $[110]_{\beta_1} \parallel [0001]_{hcp}$ $(1\bar{1}2)_{\beta_1} \parallel (01\bar{1}0)_{hcp}$. For the $\beta'$-hcp interface, we calculated interface energies on the $(100)_{\beta'}$, $(010)_{\beta'}$, and $(001)_{\beta'}$ planes. We further considered two kinds of $\beta_1$-$\beta'$ interfaces. The first had an orientation relationship $[001]_{\beta'} \parallel [110]_{\beta_1}$ $(010)_{\beta'} \parallel (\bar{1}10)_{\beta_1}$, the second $[110]_{\beta_1} \parallel [001]_{\beta'}$ $(1\bar{1}2)_{\beta_1} \parallel (010)_{\beta'}$. For the former interface, we calculated interface energies on the $(100)_{\beta'}$, $(010)_{\beta'}$, and $(001)_{\beta'}$ planes. Our calculations show agreement between DFT, the MLIP, and literature values.

\begin{figure}[!ht]
    \centering
    \includegraphics[width=0.75\columnwidth]{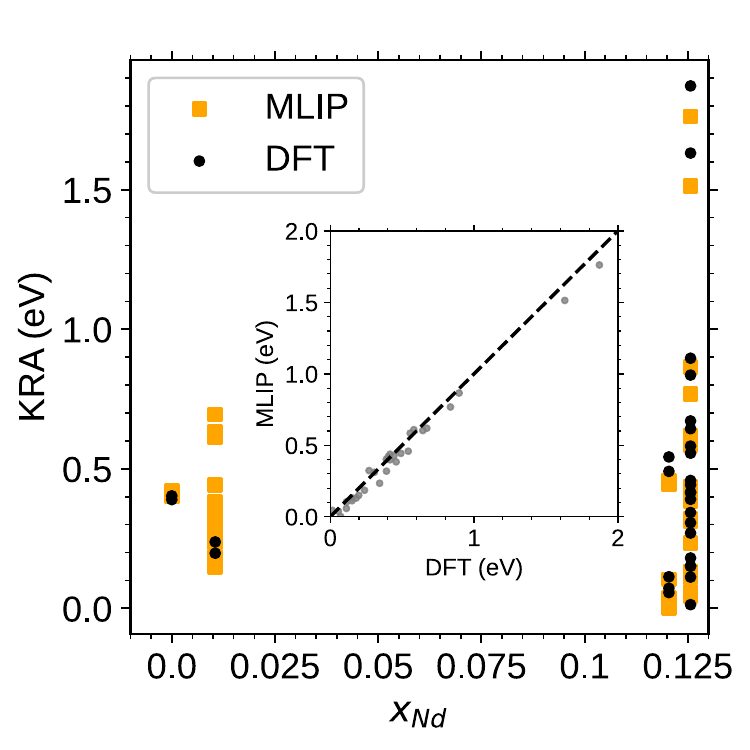}
    \caption{KRA energy barriers for Nd compositions up to $x_{\text{Nd}} = 0.125$. Scatter points clustered around $x_{\text{Nd}} = 0$ and $0.125$ represent hops in hcp-Mg and $\beta'$, respectively. Inset: Parity plot comparing KRA barriers computed with DFT and MLIP.}
    \label{fig:kra}
\end{figure}

\setcounter{figure}{8}
\begin{figure*}[!ht]
    \centering
    \subfloat[\label{fig:Mg_vacancy_exchange}]{\includegraphics[width=0.7\linewidth]{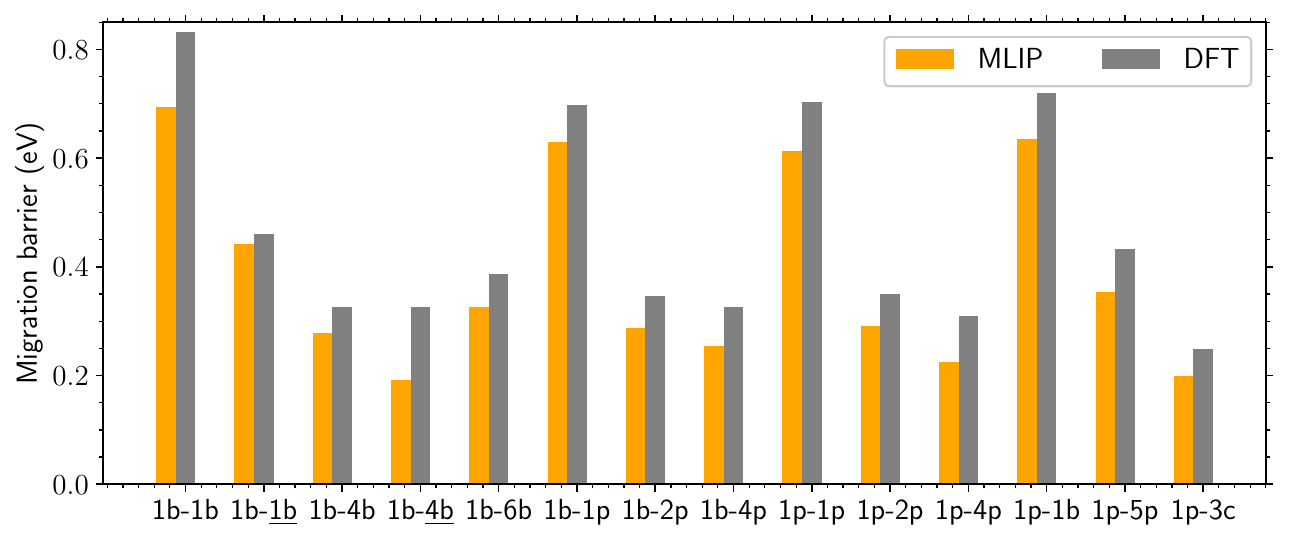}}
    \subfloat[\label{fig:hops_diagram}]{\includegraphics[width=0.3\linewidth]{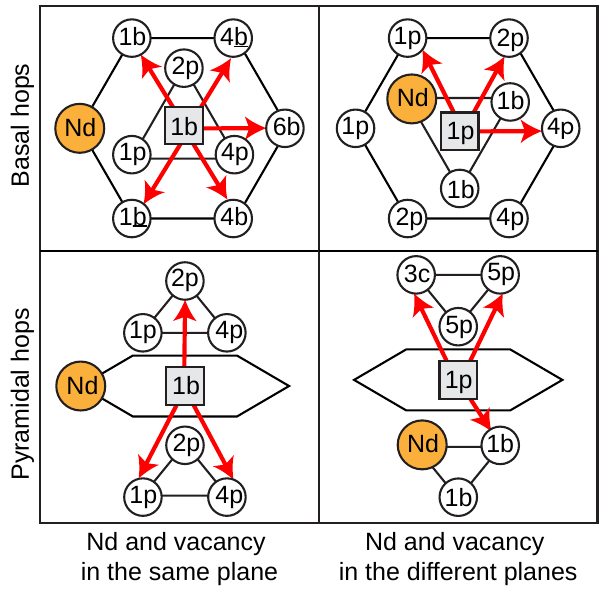}}
    \caption{(a) Comparison of migration energy barriers for vacancy hops away from a Nd atom in hcp-Mg computed with MLIP and DFT (DFT data from \cite{agarwal_exact_2017}). (b) Schematic of the hop geometry. The gray square and white circles indicate the vacancy and Mg atoms, respectively. Labels specify the neighbor shell (1st NN, 2nd NN, etc.) and geometric location ($b$: basal, $p$: pyramidal, $c$: column), following the convention in \cite{agarwal_exact_2017}.}
    \label{fig:solute_Mg_hops}
\end{figure*}

\setcounter{figure}{7}
\begin{figure}[!ht]
    \centering
    \includegraphics[width=0.7\columnwidth]{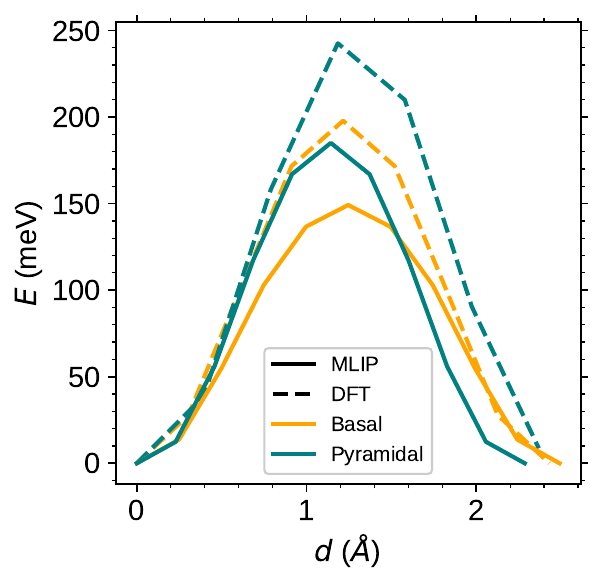}
    \caption{KRA energy profiles for Nd-vacancy exchange in hcp-Mg computed with the NEB method using DFT and MLIP.}
    \label{fig:Nd_vac_exchange}
\end{figure}

\subsubsection{Kinetic properties}

\Cref{fig:kra,fig:Nd_vac_exchange,fig:Mg_vacancy_exchange,fig:vacancy_hops_betap} compare kinetically resolved activation (KRA) barriers for several vacancy migration events. We computed vacancy exchanges with both Nd and Mg atoms in a large hcp cell containing predominantly Mg and within $\beta^{\prime}$. We used a $4\times4\times3$ supercell of hcp-Mg and a $2\times2\times3$ supercell of $\beta'$, computing all symmetrically distinct vacancy hops within $\beta^{\prime}$. We obtained vacancy migration energy profiles with the NEB method \cite{jonsson_nudged_1998}. Reference DFT values came from analogous DFT calculations or static DFT calculations on the MLIP configurations.

\Cref{fig:kra} shows KRA barriers as a function of Nd composition computed with the MLIP and DFT. The inset parity plot compares KRA barriers predicted by the MLIP against DFT values, showing close agreement. \Cref{fig:Nd_vac_exchange} compares NEB energy profiles obtained with DFT and the MLIP for the vacancy-Nd exchange in hcp-Mg. The DFT-computed barriers for basal and pyramidal jumps are 198 and 243 meV, respectively. The MLIP underestimates these values by approximately 50 meV, giving 150 and 185 meV. \Cref{fig:Mg_vacancy_exchange} shows migration barriers for vacancy jumps away from a Nd atom in hcp-Mg. The energy barriers are calculated relative to the starting vacancy configuration. \Cref{fig:hops_diagram} illustrates the vacancy jump geometry and details the labeling convention. We compare the MLIP results to DFT data available in \cite{agarwal_exact_2017}. Although the MLIP slightly underestimates the migration barriers, it correctly reproduces the overall trends.

\Cref{fig:vacancy_hops_betap} provides an overview of the 26 symmetry-distinct vacancy hops in $\beta'$. The crystal structure at the top of the diagram marks the different jumps with arrows, distinguishing basal (solid) and pyramidal (dashed) hops for Mg (red) and Nd (blue). The KRA barriers evaluated with the MLIP are plotted at Nd compositions around $x_{\text{Nd}} = 0.125$ in \cref{fig:kra}. The corresponding DFT values come from static calculations on structures obtained from NEB calculations with the MLIP. The plots in \cref{fig:vacancy_hops_betap} indicate good agreement between the MLIP and DFT. Interestingly, several vacancy hops in $\beta'$ have very small barriers or are unstable. This is the case for hops number 2, 6, 7, 10, 11, 15, 19, 20, 22, and 23.

\setcounter{figure}{9}
\begin{figure*}[p]
    \includegraphics[width=0.8\linewidth]{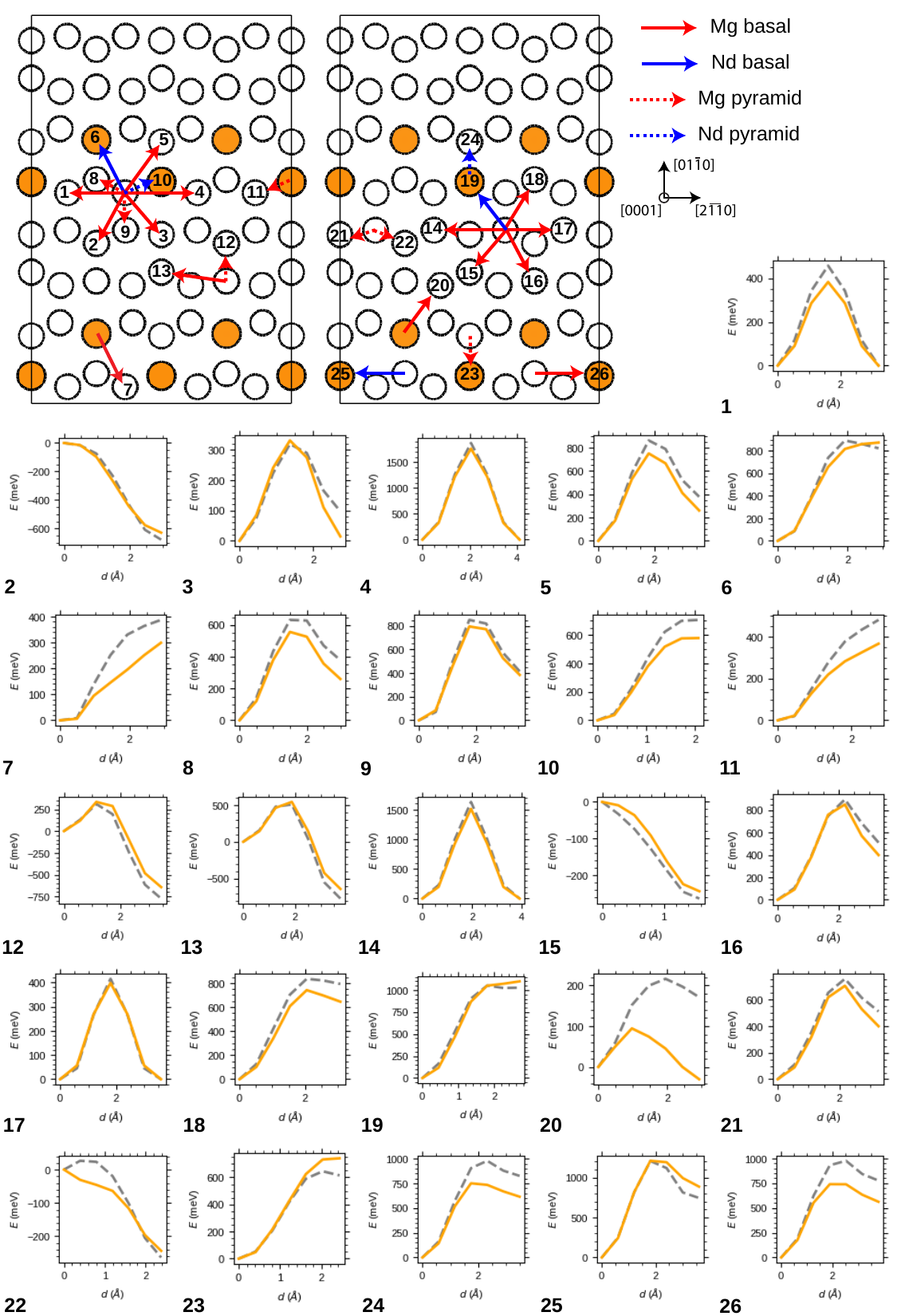}
    \caption{Energy profiles for 26 symmetry-distinct vacancy hops in $\beta'$. In the schematic crystal structure, arrows indicate vacancy jumps from the source (tail) to the numbered target atom (head).  Numbers in each energy profile correspond to these target atoms. Energies are referenced to the initial state. The orange solid line is for the MLIP, the gray dashed line is for DFT.}
    \label{fig:vacancy_hops_betap}
\end{figure*}

\bibliographystyle{unsrt}
\bibliography{references}